\documentclass[twocolumn,aps,prl,nofootinbib]{revtex4}

\usepackage{graphicx}
\usepackage{dcolumn}
\usepackage{epsfig}

\input babarsym

\long\def\ignore#1{\relax}
\newcommand{\CPviolate}{\CP\put (-13,-0.7){\line(5,3){12}}}

\begin{document}

\title {\begin{flushright}
{\small
CALT 68-2447\\
August 2003\\}
\end{flushright}
\Large\bf Sensitivity of CKM fits to theoretical uncertainties\\ and their
       representation\footnote{Work partially supported by Department of Energy under Grant DE-FG03-92-ER40701.}} 

\begin{abstract}

There is now a rapidly growing body of experimental data relevant to 
the question of whether the standard model CKM quark mixing matrix is 
a correct description of \CP-violation as well as of 
non--\CP-violating flavor decay processes.
In the detailed comparisons with theoretical predictions that are 
required to investigate this, a key challenge
has been the representation of non-statistical uncertainties, 
especially those arising in theoretical calculations.
The analytical procedures that have been used to date require 
procedural value judgments on this matter that color the
interpretation of the quantitative results they produce.
Differences arising from these value judgments in the results obtained
from the various global CKM fitting techniques in the literature are
of a scale comparable to those arising from the other
uncertainties in the input data and therefore cannot be ignored.

We have developed techniques for studying and visualizing the
sensitivity of global CKM fits to non-statistical uncertainties and
their parameterization, 
as well as techniques for visual evaluation of the consistency of
experimental and theoretical inputs that minimize the implicit use
of such value judgments, while illuminating their effects.
We present these techniques and the results of such studies using 
recently updated theoretical and experimental inputs,
discuss their implications for the interpretation of global CKM fits,
and illustrate their possible future application as the uncertainties
on the inputs are improved over the next several years.

\end{abstract}

\author{G. P. Dubois-Felsmann}
\author{D. G. Hitlin}
\author{F. C. Porter}
\affiliation{California Institute of Technology, Pasadena, CA 91125 USA}
\author{G. Eigen}
\affiliation{University of Bergen, Bergen, Norway}

\maketitle

\section{Introduction}

The three-generation Cabibbo-Kobayashi-Masakawa (CKM) quark mixing matrix 
is a key feature of the standard model,
with rich phenomenological consequences. Testing the consistency of 
different ways of measuring the matrix elements and the consistency of this
matrix with unitarity in three dimensions has become an industry, and, as the
precision of such tests improves, may lead to the observation of phenomena
outside of the standard model.
The recent experimental observation of \CP-violating asymmetries in $B$ decays
presents us with the hope to investigate the \CP-violating phase of the CKM 
matrix from many perspectives, and further presents a new source of constraints
on the unitarity aspect.
A detailed answer to the consistency question requires a comprehensive 
comparison of a great deal of experimental
data with corresponding theoretical predictions. We present here 
an approach to 
investigate and visualize the state of our knowledge on this matter.

Motivated by rapidly improving experimental knowledge, 
there has been substantial effort towards
deriving statistics for describing the
self-consistency of the standard model in the quark sector.
A number of methods for performing such global fits have been presented in the literature over the past decade,
including the so-called ``scanning method''~\cite{bib:BABAR}, which delineates a region of
consistency of parameters, but does not attempt to ascertain a ``best value'' of the CKM
parameters.
Several of the more recently developed techniques~\cite{bib:ciuchini,bib:ckmfitter}
aim to produce quantitative measures of the overall consistency of the experimental
data and theoretical model, and to put forward preferred values for the CKM matrix
elements.

A key challenge in the development of these procedures has been the
representation in the fits of non-statistical uncertainties,
especially on theoretical parameters such as the $B$ meson
pseudoscalar decay constant $f_B$ and the ``bag factor'' $B_K$ arising
in $K$ meson decay calculations.
Calculations of these parameters are generally published with 
``uncertainty ranges'' reflecting the authors' degree of belief in the
correctness of model assumptions or the effects of ignoring
higher-order terms in expansions. 
These do not in general have a precise statistical definition
(except for those contributions arising from statistics-limited Monte Carlo
or lattice calculations), and there is no clear consensus in the
community for the meaning to be attached to their precise numerical
value.

Nevertheless, in order to perform a ``global fit'' that incorporates
inputs with such uncertainties, in the framework of a standard
minimization-of-deviations fitting procedure, the goodness-of-fit
metric to be used must somehow be constructed to give them
quantitative effect.
Because of the lack of a clear statistical meaning for these
uncertainties or a consensus on how to interpret them, the
schemes used to do this have been a subject of considerable debate in
the literature, and the procedural value judgments involved 
inevitably color the interpretation of the quantitative results
obtained from such analyses.

Differences arising from these value judgments in the results obtained
from the various global CKM fitting techniques in the literature have been
shown to be of a scale comparable to the effects of the other
uncertainties in the input data, and therefore cannot be 
ignored~\cite{bib:CKM2002}.

In the fitting procedures published to date, these choices tend to be
embedded in the details of the analysis, and their effects are not readily
made manifest in a study of the outputs of the fit.
For example, Ciuchini {\it et~al.}~\cite{bib:ciuchini} 
take a Bayesian approach, choosing a uniform probability density
function (p.d.f.) when 
``the parameter is believed to be (almost) certainly in a given interval, 
and the points inside this interval are 
\emph{considered equally probable}'' (italics ours).
The fact is that values of a theoretical parameter within an uncertainty
range of this type may or may not be equally probable: 
their distribution is generally {\it a priori} unknown. 
Thus their choice of {\it a priori} p.d.f.\ is in fact a considerably
stronger statement than saying ``the value should lie within this range'',
and than the original theoretical author may have intended.
It is when this p.d.f.\ is then convolved with others in the course of the
fit that particular problems can arise.

This is easily seen by considering the case of a single hypothetical
positive-definite parameter $\zeta$.  Imagine a theoretical prediction
that $\zeta$ should lie between $\zeta_1$ and $\zeta_2$, based perhaps
on the range of values yielded by taking a few different approaches to
an approximate calculation.
One would expect such a prediction to have the same intellectual content
as one stating that $\zeta^2$ should lie between $\zeta_1^2$ and $\zeta_2^2$;
yet in the Bayesian method, this is not so.
The meaning of a flat p.d.f.\ depends on the choice of the actual form 
of the parameter in which its distribution is expected to be uniform.  
Whether this is chosen to be in terms of $\zeta$ or $\zeta^2$---a question
to which there may not be a principled answer---can clearly introduce 
a differential bias in what values of $\zeta$ are preferred in the resulting 
fits.

This illustrates what we believe to be a serious and intrinsic weakness in
the Bayesian approach.

H\"ocker, {\it et al.}'s approach~\cite{bib:ckmfitter} to treating
theoretical uncertainties is that of frequentist statistics.
In their ``RFit'' scheme, they explicitly do not impute a uniform p.d.f.\ to 
theoretical uncertainties. 
Rather, they include in their likelihood fits an unnormalized penalty 
function, not interpretable as a p.d.f., which entirely bars values of 
theoretical parameters outside their allowed ranges, while having no effect 
inside the range.
This effectively replaces the Bayesian method's convolution integral over
a theoretical parameter's p.d.f.\ with a logical OR\@.
The consequence is that when an overall likelihood is formed the 
results at a point {\it a} in the parameter space should be interpreted as an 
``upper bound of the confidence level one may set on {\it a},
which corresponds to the best possible set of theoretical parameters''.
This avoids problems of the sort mentioned above.

It retains the disadvantage that, in the graphical outputs generated by the
method, points lying on the same contour---say, 95\%---of 
the ``upper bound of the confidence level'' can appear as somehow 
similar to each other.  It requires considerable discipline to avoid
reading such a graph as stating that points on the same contour are in
fact ``equally likely''.  Yet, to the extent that the best-fit results for
two such points arise from different values of the theoretical parameters,
they are not.  Such a statement, we maintain, simply cannot be made for
non-probabilistic theoretical uncertainties.

This critique of the RFit frequentist analysis primarily relates to
the interpretation of results and the need for care in presentation.  While
this may seem an unduly psychological point, in practice it can be observed
that independent readers of its results frequently fail to retain the 
qualification that its results are ``upper bounds'' on confidence levels. 

Our principal concern, however, which applies to both approaches, is
the degree to which experimental uncertainties, probabilistic and typically
approximately gaussian, are intertwined with theoretical uncertainties 
with no principled probabilistic interpretation.
We believe that this obscures the importance of distinguishing among sources 
of uncertainty and lends an undue patina of statistical precision to the 
numerical results obtained.
This may result in inappropriate conclusions on the significance of
results obtained from CKM fits, and on questions of such central importance
as the consistency of data with the standard model.

We present herein techniques for ascertaining the sensitivity of
global CKM fits to non-probabilistic uncertainties and their
representation.  These include novel techniques for visual evaluation of 
the consistency of experimental and theoretical inputs, which minimize the
implicit use of value judgments concerning such uncertainties, while
illuminating their effects.

These ideas are of general applicability; in this paper, however, we
concentrate on the information available for the determination of the
point $(\bar\rho,\bar\eta)$, the apex of the conventional Unitarity Triangle.
Our approach relies on a careful separation of the types of
uncertainties in the inputs to the analysis.
We use a fitting procedure that avoids the need for Bayesian p.d.f.\ or
likelihood representations of non-probabilistic uncertainties, and, 
in a frequentist framework,
defers the application of the theoretical inputs until the
stage of presentation and interpretation of the fit outputs.

This allows us to develop a clearer understanding of the role of these
inputs in determining our present and future knowledge of $\bar\rho$
and $\bar\eta$ and the consistency of theory with experiment.

\section{Theoretical and Experimental Background}

For the purposes of this analysis, 
we initially assume the standard model CKM matrix to be a comprehensive 
description of weak flavor physics (and therefore unitary), 
and the only source of \CP-violation.
We are then able to adopt a standard parameterization of it that extends 
the one of Wolfenstein~\cite{bib:wolfenstein} to higher order in 
$\lambda$~\cite{bib:burasLO}, 
as shown in Eqn.~\ref{eqn:CKMmatrix} below.  
This parameterization is sufficiently precise for present purposes.
As is conventional, we define $\bar\rho \equiv \rho\cdot(1-\lambda^2/2)$ 
and $\bar\eta \equiv \eta\cdot(1-\lambda^2/2)$.

We take the parameter $\lambda$, which determines $|V_{us}|$ and
the rest of the light-quark sector of the CKM matrix, as an input, 
$\lambda = 0.2241 \pm 0.0033$,
as reported in a recent combined analysis of light-quark sector 
data \cite{bib:CKM2002}.
We combine a variety of other measurements to constrain the matrix 
elements involving the third generation, 
$V_{cb}$, $V_{ub}$, $V_{td}$, and $V_{ts}$.

The magnitude of the $V_{cb}$ CKM matrix element can be measured using either
inclusive or exclusive techniques.
It has been measured in the exclusive decay $B\to
 D^*\ell\nu$, where $\ell = e$ or 
$\mu$~\cite{bib:CLEO0126,bib:BELLE-vcb,bib:LEP-vcb}:
\begin{equation}
{d{\cal B}\over dw}(w) = {G_F^2\over 48\pi^3}{1\over \tau_B}
 {\cal K}(w)\left[|V_{cb}|{\cal F}_{D^*}(w)\right]^2.
\end{equation}
$G_F$ is the Fermi constant, 
${\cal K}(w)$ is a precisely known kinematic factor, 
and ${\cal F}_{D^*}(w)$ is a theoretically uncertain form factor. 
The measured quantities which are used in determining $|V_{cb}|$ 
are the branching fraction and the $B$ lifetime $\tau_B$. 
Extrapolating the fit to $w=1$ results in the
measurement of $|V_{cb}|{\cal F}_{D^*}(1)$.

\begin{widetext}
\begin{equation}
\label{eqn:CKMmatrix}
\pmatrix{V_{ud} & V_{us} & V_{ub} \cr V_{cd} & V_{cs} & V_{cb} \cr
         V_{td} & V_{ts} & V_{tb} \cr} 
       = \pmatrix{1-{\lambda^2\over 2} - {\lambda^4\over 8} &
         \lambda & A\lambda^3(\rho-i\eta) \cr
         -\lambda+A^2\lambda^5({1\over 2}-\rho-i\eta) &
           1-{\lambda^2\over 2}-{\lambda^4\over 8}(1+4A^2) & A\lambda^2 \cr
         A\lambda^3(1-\bar\rho-i\bar\eta) & -A\lambda^2+A\lambda^4({1\over2}
 -\rho-i\eta) & 1-{1\over2}A^2\lambda^4 \cr}+O(\lambda^6).
\end{equation}
\end{widetext}

The form factor ${\cal F}_{D^*}(1)$, which is calculated at zero recoil and
in the heavy quark limit is identical to one, 
contributes an estimated theoretical uncertainty corresponding to a range in 
$V_{cb}$ of $(40.2-43.8) \times 10^{-3}$~\cite{bib:LEP-vcb}.

For $B \rightarrow D^{*} \ell \nu$ the hadronic form factor at zero recoil is 
approximately given by~\cite{bib:BABAR}
\begin{equation}
\label{eq:ffds}{\cal F}_{D^*}(1) = \eta_A \eta_{QED} (1 + \delta_{1/m^2_b} + 
\delta_{1/m^3_b} +...),
\end{equation}
where $\eta_A$ represents a short-distance correction arising from the 
finite QCD renormalization of the flavor-changing  axial current at zero 
recoil, $\eta_{QED} \simeq 1.007$ denotes QED corrections in 
leading-logarithmic order, and the $\delta_{1/m^n_b}$ are higher order
corrections in powers of the $b$-quark mass.
Note that the first order is missing due to Luke's theorem~\cite{bib:luke}. 
The exact two-loop expression yields $\eta_A = 0.96 \pm 0.007$~\cite{bib:czar}.

For the higher-order power corrections,
Ref.~\cite{bib:fds1} calculated a range of 
$-0.08 < \delta_{1/m^2_b} < -0.03$ which agrees with the result of 
$\delta_{1/m^2} = -0.055$ given in Ref.~\cite{bib:fds2}.
Only at order $\delta_{1/m^3_b}$ do the predictions differ. 
While in Ref.~\cite{bib:fds1} 
this is accounted for in $\delta_{1/m^2_b}$, 
an extra contribution is estimated from sum rules in Ref.~\cite{bib:fds2},
yielding  $\delta_{1/m^3_b}= -0.03$. 
Plugging these values into Eqn.~\ref{eq:ffds} yields
\begin{equation}
\label{eq:ffds2}{\cal F}_{D^*}(1)=0.913 \pm 0.007_{pert} \pm 0.024_{1/m^2_b} \pm 0.011_{1/m^3_b}.
\end{equation}
The errors denote theoretical uncertainties 
from perturbative QCD,  
from $1/m^2_b$, and from $1/m^3_b$ terms, respectively. The central values 
differ by $\sim 3\%$, while the theoretical uncertainties added in quadrature 
amount to $2.7\%$ and $5\%$, for the two predictions. Both predictions are lower than, but
consistent with, a recent quenched lattice gauge calculation of 
${\cal F}_{D^*}(1)=0.935 \pm 0.033$~\cite{bib:fds3}.
According to the recommendation of the LEP working group, we use ${\cal F}_{D^*}(1)=0.91 \pm 0.04$.

%

The magnitude of the $V_{ub}$ CKM matrix element can also be measured either 
via inclusive or exclusive techniques. 
The measurement of the exclusive branching fraction for 
$B\to\rho\ell\nu$ is related to
$|V_{ub}|$ according to Ref.~\cite{bib:rholnuCLEO}:
\begin{equation} 
\label{eq:vubexcl}
{\cal B}(B\to\rho\ell\nu) =
|V_{ub}|^2 \cdot \widetilde \Gamma_{\rho\ell\nu} \cdot \tau_{B}.   
\end{equation}
A similar relation holds for the extraction of $V_{ub}$ from inclusive
branching fraction measurements.
The predicted reduced exclusive decay rate, $\tilde \Gamma_{\rho\ell\nu}$, 
depends on form factors which have been estimated in various 
models~\cite{bib:vub-thex}. The predicted inclusive decay rates are
calculated in the heavy quark expansion~\cite{bib:vub-thin}. 
The extraction of $V_{ub}$ from exclusive (inclusive) branching fraction 
measurements introduces a theoretical uncertainty of  $15\%$ ($10\%$).
In the $(\bar \rho, \bar \eta)$ plane a range of $|V_{ub}/ V_{cb}|$ 
appears as a circular band centered at (0,0).

%

The CKM element $V_{td}$ may be extracted from the $B^0 \bar B^0$ oscillation 
frequency 
\begin{equation}
\label{eq:dmd}
\Delta m_{B_d} = {G_F^2 \over 6 \pi^2} \eta_B m_{B_d} m_W^2 
S_0(x_t) f^2_{B_d} B_{B_d} \mid V_{td} V_{tb}^* \mid^2,
\end{equation}
where $\eta_B = 0.55 \pm 0.01$ is a QCD factor, 
$m_{B_d}$ the $B$~meson mass, $m_W$ the $W$ boson mass, 
$S_0(x_t)$ the Inami-Lim function~\cite{bib:inami} for the box diagram, 
$x_t = m^2_t / m^2_W$ the ratio of top-quark 
mass to $W$~boson mass squared, $f_{B_d}$ the $B$~meson decay constant, 
and $B_{B_d}$ the so-called bag factor.
The estimated theoretical uncertainty introduced through 
$f_{B_d} \sqrt{B_{B_d}}$ is
of the order of $20\%$. In the  $(\bar \rho, \bar \eta)$ plane 
a range of $|V_{td}/ V_{cb}|$ appears as a circular band centered at (1,0).

The \CP-violation parameter $| \epsilon_K |$ in the $K^0 \bar K^0$  system
provides another constraint in the $(\bar{\rho}, \bar{\eta})$ plane:
\begin{eqnarray}
| \epsilon_K | & \propto & B_K {\cal I}m (V_{td} V^*_{ts}) \nonumber \\
 &  & \mbox{} \times \{ {\cal R}e (V_{cd} V^*_{cs}) [\eta_1 S_0(x_c)- \eta_3 S_0(x_c,x_t)] \nonumber \\
 &  & \mbox{} - {\cal R}e (V_{td}V^*_{ts}) \eta_2 S_0(x_t)\ .
\end{eqnarray}
Here the geometrical representation is a hyperbolic band
in the $(\bar \rho, \bar \eta)$ plane.  $\eta_1, \eta_2, \eta_3$ are 
QCD parameters~\cite{bib:eta1,bib:etab,bib:eta3};  
$x_c= m^2_c / m^2_W$ is the ratio of 
charm-quark mass to $W$~boson mass squared, and $S_0(x)$ are the
Inami-Lim functions \cite{bib:inami} for the electroweak box diagrams with charm and top quarks in the loop. For the central values of the running top and charm quark masses used herein, $S_0(x_t) = 2.45$,
$S_0(x_t,x_c) = 2.31 \times 10^{-3}$, and
$S_0(x_c) = 2.62 \times 10^{-4}$.

Mixing in $B_s$ mesons provides a means to measure the CKM quantity 
$|V_{tb}^*V_{ts}|$. 
Similarly to Equation~(\ref{eq:dmd}) for $\Delta m_{B_d}$, the
$B_s$ oscillation frequency is related to $|V_{tb}^*V_{ts}|$ by:
\begin{equation}
\Delta m_{B_s} = \frac{G_F^2}{6 \pi^2} \eta_B m_{B_s} m_W^2
f_{B_d}^2 B_{B_d}  \xi ^2\  S_0(x_t)|V_{tb}^*V_{ts}|^2,
\end{equation}
where $\xi \equiv (f_{B_s} \sqrt{B_{B_s}})/(f_{B_d} \sqrt{B_{B_d}})$.
Experimentally, a lower limit on $\Delta m_{B_s}$
has been determined by combining analyses of different experiments using the 
amplitude method~\cite{bib:moser}. For given values 
of $\xi$ and $f_{B_d}^2B_{B_d}$, this leads to a
lower limit on $|V_{tb}^*V_{ts}|$. 
A measurement of $|V_{ts}/V_{td}|$ would yield
another circular band in the $(\bar \rho, \bar \eta)$ plane centered at (1,0).
We can, at present, constrain only the upper side of this band.

The measurements considered up to now are related to the sides of the
Unitarity Triangle. Additional measurements, such as of the angles $\alpha,
\beta, \gamma$ of the Unitarity Triangle,
will provide additional constraints on the CKM parameters. 
Measurements of $\sin 2 \beta$ are now rather precise~\cite{bib:BABAR-sb,bib:BELLE-sb,bib:CDF-sb,bib:ALEPH-sb}:
the world average is
$\sin 2 \beta = 0.731 \pm 0.055$. 
This angle is related to the CKM elements by~\cite{bib:dib}:
\begin{equation}
\beta\equiv \hbox{arg}\left(-{V_{cd}V^*_{cb} \over V_{td}V^*_{tb}}\right).
\end{equation}
$\sin 2 \beta$ appears as a set of rays that cross the point
(1,0) in the $(\bar \rho, \bar \eta)$ plane.



\section{Fitting the data}


We first describe our approach to the global fit in conceptual terms.
The fit begins with a set of experimental observables $E_i$, which can be 
predicted on theoretical grounds.  
The inputs to these predictions are:
the CKM matrix, assumed $3\times3$ and unitary, represented in terms of
the Wolfenstein parameters $W_j$; 
a list of quantities $C_k$, such as masses and lifetimes, with 
experimentally derived or other probabilistic uncertainties $\sigma_{C_k}$; 
and a set of ``theoretical parameters'' $T_\ell$.
The parameters $T_\ell$ are, in principle, exactly calculable from the
theory and the $C_k$, but this is in practice difficult to do, and the
choices of approximation methods, cutoffs, and the like introduce
non-probabilistic uncertainties $\Delta_{T_\ell}$ which arise, in 
essence, from human judgment.  
As it becomes possible to use lattice
calculations and other quasi-statistical tools to evaluate the $T_\ell$,
these uncertainties will be replaced by probabilistic ones.

The predictions for the $E_i$ take the functional form
\begin{equation}
\langle E_i\rangle = {\cal E}_i(W_j; C_k; T_\ell).
\end{equation}
These predictions are to be compared with experimental measurements of the
observables, $E_i \pm \sigma_{E_i}$.

In order to test the consistency of theory and experiment and 
determine numerical values for the CKM matrix elements, then,
we wish to perform a fit to the ensemble of data above.
Notionally we would allow all the input parameters to vary in the fit,
minimizing a consistency statistic (a $\chi^2$ or likelihood)
constructed from the deviations of the predictions $\langle E_i \rangle$
from the measurements $E_i$, including their uncertainties, 
and from the deviations of the
input parameters from their nominal values, including their
uncertainties as well.
The $W_j$, on which there is no {\it a priori} information, would be
entirely free in the fit.

The results of the fit would be an overall optimum set of values for
the fit variables, including the CKM parameters $W_j$, a confidence level 
for the best fit, and a correlation matrix or set of confidence level
contours characterizing the possible deviations from the optimum point.

The problem, described previously, is that the uncertainties $\Delta_{T_\ell}$
on the theoretical parameters do not in fact have a principled statistical
interpretation.  
Thus, constructing the consistency statistic for the full fit would
require the use of a value judgment on how to proceed in the face of this.
In a Bayesian analysis, we would impute an {\it a priori} p.d.f.\ to
each of the $T_\ell$, somehow based on the $\Delta_{T_\ell}$, and then 
proceed with a conventional fit.  In the frequentist RFit approach, the
contributions of the $T_\ell$ would be represented as unnormalized ``penalty
function'' factors in the overall likelihood function, having value zero
outside the intervals $\mbox{}\pm\Delta_{T_\ell}$ and one inside.  These
constrain the fit to the theoretically allowed intervals without introducing
any statistical preferences within them.  The fit can then proceed, but
the use of these penalty functions requires the re-interpretation of the
results as yielding not confidence levels but upper bounds on them.

In order to avoid the need to convolute probabilistic and 
non-probabilistic uncertainties, we take a different approach.  
We acknowledge that we simply do not have
the necessary information on the $T_\ell$ to be able to perform a global fit.
We choose instead to treat each point in the space $T$ independently,
and defer the consideration of non-probabilistic uncertainties 
until the stage of interpretation of the fit results.
The space $T$ can be taken as extending well beyond its ``theoretically 
allowed'' bounds; this is useful in clarifying the interplay between
theoretical and experimental constraints.

For any given point in this space, which we call a {\em model,} we 
fix its $T_\ell$ values,
and then we perform the fit, allowing only the $W_j$ and $C_k$ to vary.
The result of this procedure is, for each model in $T$, an optimum 
value of the fit variables, a value of the consistency statistic (a
confidence level), and a measure of the uncertainties around the
optimum point.  
This information can then be put to a variety of uses which we have found 
to shed interesting light on our understanding of the CKM data.  
We discuss these below after the presentation of the details of the
fits as actually performed.

A prerequisite to this approach is a detailed analysis of the
uncertainties on each input to the fit, separating those which are
statistical in nature from those which are not and for which there is
no principled way to determine a probability density function.
In some cases this has already been done carefully by the authors of
the inputs we use, but in others we have had to try to separate them
out ourselves.  That isn't always possible---sometimes uncertainties
have been irreversibly convolved---and we hope by our work to
demonstrate the importance of keeping these different sources of
uncertainty distinct in published results.

We follow the established convention in high energy physics for
treating experimental systematic uncertainties as if they were
statistical and approximately Gaussian.
This convention is typically justified by noting that experimental 
systematics tend to be built up from many individual contributions that 
are not strongly correlated.  
Applying a heuristic descended from the Central Limit Theorem, this 
leads to the expectation that it is reasonable to approximate their sum 
as having a statistical, Gaussian distribution.
Compared to the case of theoretical uncertainties, there is a longer
history of doing this and a more uniform consensus on calculating systematic
uncertainties so that their scale can be interpreted as a standard deviation.

Nevertheless, if in the future we find that a particular experiment has a 
large systematic uncertainty dominated by a single contribution, we
can choose to break it out for treatment similar to that for theoretical 
uncertainties.

Once the uncertainties contributing to the fit have been identified and
cataloged in this manner, we then select those non-probabilistic 
uncertainties that are expected to have the largest effect on the overall 
fit results.
This is necessary because the methods described below are limited in
the number of these sources of uncertainty that can be studied
simultaneously: the computational load is exponential in this number.
As a result, we must compromise and adopt approximate statistical
interpretations for the remaining, smaller, non-statistical uncertainties.  
Iterative application of our method can be used to explore the validity of 
these choices.

\subsection{Separate Treatment of Probabilistic and Non-probabilistic Uncertainties}

Consider a quantity $x$ having both probabilistic and non-probabilistic 
uncertainties.  We may represent this as
\begin{equation}
x = \langle x\rangle \pm \sigma_x \pm \Delta_x,
\end{equation}
where $\langle x\rangle$ is the nominal central value of $x$,
$\sigma_x$ represents the sum in quadrature of all probabilistic
uncertainties on $x$, and $\Delta_x$ represents a non-probabilistic
uncertainty, typically a theoretical one.  While $\sigma_x$ has its usual
interpretation as the width of a normal distribution, at this point we 
need not commit to any particular interpretation of $\Delta_x$---it is merely
a placeholder.

For the purposes of the construction of the fit we restate this schematically 
as:
\begin{equation}
x = ( \langle x\rangle \pm \sigma_x ) + ( 0 \pm \Delta_x ).
\end{equation}
That is, we represent it as the sum of a normally distributed variable
carrying the probabilistic uncertainties, and an additive shift with 
purely non-probabilistic uncertainties.  
We could also use an equivalent multiplicative definition, where appropriate.

The value of $x$ actually used in the fit is then computed as 
$x = x_{\rm P} + x_{\rm NP}$, 
the sum of particular values of the two terms above, respectively.
The value of $x_{\rm P}$ is allowed to vary in the fit, with a contribution
to the goodness-of-fit statistic representing its probabilistic 
uncertainty.  For a $\chi^2$-based fit, this contribution would be
\begin{equation}
\chi^2|_x = (x_{\rm P} - \langle x\rangle)^2 \over \sigma^2_x
\end{equation}

The value of $x_{\rm NP}$ is scanned over a range related to the uncertainty
$\Delta_x$, chosen as appropriate to the interpretation of that uncertainty,
with an independent fit for $x_{\rm P}$ (and any other fit variables) 
performed for each value of $x_{\rm NP}$.

When discussing the results of the scan analysis, in order to improve the
accessibility of the presentation, we compute and display the value
\begin{equation}
\widetilde{x} \equiv \langle x\rangle + x_{\rm NP},
\end{equation}
thus recentering the scanned values on the nominal central value.

In some cases, again for the purpose of clarity of presentation, we 
present the results not in terms of the fit variable $x$, but in terms
of a transformation of that variable.

Thus, for instance, from Equation \ref{eq:vubexcl} we see that we can
compute $|V_{ub}|$ from exclusive data as:
\begin{equation}
|V_{ub}| = \left( { { {\cal B}(B\to\rho\ell\nu) } \over 
                    { \Gamma_{\rho\ell\nu} \cdot \tau_{B} } } \right)^{1\over2}.
\end{equation}
$\Gamma_{\rho\ell\nu}$ is the quantity on which the theoretical uncertainties
actually arise, and the one treated as the ``$x$'' in the fit according to
the procedure above, with $\widetilde{\Gamma}_{\rho\ell\nu}$ representing
the scanned theoretical value.
For the purposes of presentation, however, this variable is somewhat 
unattractive; readers will be more familiar with considering the theoretical
uncertainties as applied to $|V_{ub}|$, 
and so will more readily understand the scanned values when
presented in that form.

These are \emph{inputs} to fits, and must be displayed as such.
We therefore define $|\widetilde{V}^{\rm ex}_{ub}|$ as the value of 
$|V_{ub}|$ computed from the above equation using 
$\widetilde{\Gamma}_{\rho\ell\nu}$ 
and the \emph{input} values for ${\cal B}(B\to\rho\ell\nu)$ and $\tau_{B}$.
When this appears in the results below, it should not be confused with 
the value of $|V_{ub}|$ resulting from the fits, which will in general be
different.

The variables 
$|\widetilde{V}^{\rm in}_{ub}|$, 
$|\widetilde{V}^{\rm ex}_{cb}|$, and 
$|\widetilde{V}^{\rm in}_{cb}|$ 
are defined similarly and used in the presentation of results below.

\vskip0.3cm

\subsection{Details of the Fit}



Following the procedure described above for
dealing with the non-probabilistic uncertainties on 
the theoretical parameters $T_\ell$, we define the notion of a
``model'', a specific set of these parameters
\begin{eqnarray}
{\cal M} & \equiv & \bigl \{{\cal F}_{D^*}(1),  
\widetilde{\Gamma}_{\rho\ell\nu}, \widetilde{\Gamma}_{u\ell\nu},\widetilde{\Gamma}_{c\ell\nu},
\widetilde{f_{B_d} \sqrt{B_{B_d}}},\nonumber\\
&& \phantom{\{} \widetilde{B}_K, 
\widetilde{\xi}, \widetilde{\eta}_1, 
\widetilde{\eta}_2, \widetilde{\eta}_3, \widetilde{\eta}_B \bigr \}.
\end{eqnarray}

The fit for a given ``model'' ${\cal M}$ by definition
incorporates no contribution from the non-probabilistic uncertainties
on these parameters.
The parameters are then scanned within designated ranges, performing a
fit for each resulting model.
We may define the range of the scan as the 
theoretically preferred intervals shown in Table~\ref{tab:ckm-th-input},
or, for the sensitivity studies present here, substantially broader one.

In detail, for each model ${\cal M}$, the $\chi^2$ expression which is minimized is:
\begin{widetext}
\begin{eqnarray}
 \chi_{\cal M}^2 (A,\bar \rho ,\bar \eta ) &=& 
   \left( {\frac{{\left\langle {\left| {V_{cb} {\cal F}_{D^*}(1)} \right|} \right\rangle  - A^2 \lambda ^4 \left| {{\cal F}_{D^*}(1)} \right|^2 )}}{{\sigma _{V_{cb} {\cal F}_{D^*}(1)} }}} \right)^2 
    + \left( {\frac{{\left\langle {B_{c\ell\nu } } \right\rangle  - \widetilde{\Gamma}_{c\ell\nu } A^2 \lambda ^4 \tau _b }}{{\sigma _{B_{c\ell\nu } } }}} \right)^2 
    + \left( {\frac{{\left\langle {B_{\rho \ell\nu } } \right\rangle   - \widetilde{\Gamma}_{\rho \ell\nu } A^2 \lambda ^6 \tau_{B^0} (\rho ^2  + \eta ^2 )}}{{\sigma _{B_{\rho \ell\nu } } }}} \right)^2   \nonumber\\ 
  &&+ \left( {\frac{{\left\langle {B_{u\ell\nu } } \right\rangle  - \widetilde{\Gamma}_{u\ell\nu } A^3 \lambda ^6 \tau _b (\rho ^2  + \eta ^2 )}}{{\sigma _{B_{u\ell\nu } } }}} \right)^2 
    + \left( {\frac{{\left\langle {\Delta m_{B_d } } \right\rangle  - \Delta m_{B_d } (A,\bar \rho ,\bar \eta )}}{{\sigma _{\Delta m} }}} \right)^2 
    + \chi^2_{\Delta {m_{B_s}}}(A,\bar \rho,\bar \eta)
 \nonumber\\ 
  &&+ \left( {\frac{{\left\langle {a_{\psi K_s } } \right\rangle  - \sin 2\beta (\bar \rho ,\bar \eta )}}{{\sigma _{\sin 2\beta } }}} \right)^2 
    +\left( {\frac{{\left\langle {f_B \sqrt {B_B } } \right\rangle  - f_B \sqrt {B_B } }}{{\sigma _{f_B \sqrt {B_B } } }}} \right)^2  
    + \left( {\frac{{\left\langle {B_K } \right\rangle  - B_K }}{{\sigma _{B_K } }}} \right)^2 
    + \left( {\frac{{\left\langle {\left| {\varepsilon _K } \right|} \right\rangle  - \left| {\varepsilon _K } \right|(A,\bar \rho ,\bar \eta )}}{{\sigma _\varepsilon  }}} \right)^2 
 \nonumber\\ 
  &&+ \left( {\frac{{\left\langle \xi  \right\rangle  - \xi }}{{\sigma _\xi  }}} \right)^2    
    + \left( {\frac{{\left\langle \lambda  \right\rangle  - \lambda }}{{\sigma _\lambda  }}} \right)^2
    + \left( {\frac{{\left\langle {m_t } \right\rangle  - m_t }}{{\sigma _{m_t } }}} \right)^2 
    + \left( {\frac{{\left\langle {m_c } \right\rangle  - m_c }}{{\sigma _{m_c } }}} \right)^2
    + \left( {\frac{{\left\langle {m_W } \right\rangle  - m_W }}{{\sigma _{M_W } }}} \right)^2  \nonumber\\ 
  &&+ \left( {\frac{{\left\langle {\tau _{B^0 } } \right\rangle  - \tau _{B^0 } }}{{\sigma _{\tau _{B^0 } } }}} \right)^2
    + \left( {\frac{{\left\langle {\tau _{B^ +  } } \right\rangle  - \tau _{B^ +  } }}{{\sigma _{\tau _{B^ +  } } }}} \right)^2
    + \left( {\frac{{\left\langle {\tau _{B_s } } \right\rangle  - \tau _{B_s } }}{{\sigma _{\tau _{B_s } } }}} \right)^2
    + \left( {\frac{{\left\langle {\tau _{\Lambda_b} } \right\rangle  - \tau _{\Lambda_b} }}{{\sigma _{\tau _{\Lambda _b } } }}} \right)^2
     \nonumber \\  
 && + \left( {\frac{{\left\langle {f_{B^{+ } } } \right\rangle   - f_{B^{+ } } }}  {{\sigma _{f_{B^{+ } } } }}} \right)^2
    + \left( {\frac{{\left\langle {f_{B_s } } \right\rangle      - f_{B_s } }}     {{\sigma _{f_{B_s } } }}} \right)^2
    + \left( {\frac{{\left\langle {f_{B^{+, 0 } } } \right\rangle  - f_{B^{+, 0 } } }} {{\sigma _{f_{B^{+, 0 } } } }}} \right)^2.  
\label{eqn:chisq}
\end{eqnarray}
\end{widetext}
In this expression, the notation $\langle \rangle$ is used to denote the experimental input, averaged over experiments.

The minimization solution $(\lambda, A, \bar \rho, \bar \eta)_{{\cal M}}$ 
for a particular model now depends only
on measurement errors and other probabilistic uncertainties. 
We have included in the $\sigma_{E_i}$ any probabilistic component of 
the uncertainties on the 
theoretical parameters relevant to each particular measurement. 
This is why terms for some of the model parameters appear in the $\chi^2$.
We have also treated the comparatively
small uncertainties arising from $\eta_2$, $\eta_3$, and
$\eta_B$ as probabilistic.

We use the experimental measurements
listed in Table~\ref{tab:ckm-th-input}.
While the CKM matrix element $|V_{td}|$ is obtained 
from the measurement of the $B^0_d \bar B^0_d$
oscillation frequency~\cite{bib:LEP-osc}, the
CKM matrix elements $|V_{cb}|$, and $|V_{ub}|$ can be extracted from both
inclusive and exclusive semileptonic decays.

\begin{table*}
\caption [ ] {Input parameters for unitarity triangle fits. 
Parameters cited as ``Statistical'' are allowed to vary in the fit,
with a probabilistic uncertainty incorporated into the $\chi^2$
function defined in Eqn.~\ref{eqn:chisq}.
Parameters marked ``Scanned'' are held fixed in the minimization
procedure, but varied from fit to fit over the non-probabilistic range
shown here.
Parameters with both designations are treated as discussed in the text;
those with neither are treated as precise constants.
The theoretical uncertainties on the QCD $\eta$ parameters are
in some cases scanned and in others incorporated into the fit as
statistical, as described in the text.
\label{tab:ckm-th-input}}
\begin{center}
\begin{tabular}{|l|l|c|c|c|}
\hline
Variable & Value &  Statistical & Scanned & Ref. \\
\hline
\hline
\multicolumn{4}{|l}{$B$- and $K$-meson decay properties} & \\
\hline
$\Delta m_{B_d}$ & $0.503 \pm 0.006~\rm ps^{-1}$ & $\times$ & & \cite{bib:LEP-osc}\\
$\Delta m_{B_s}$ & $> 14.4\ \rm ps^{-1}\  \ (95\%$ C.L.) & $\times$ & & \cite{bib:LEP-osc}\\
$\epsilon_K$ & $2.271 \pm 0.017 \times 10^{-3}$ & $\times$ & & \cite{bib:pdg}\\
$\sin 2 \beta$ ($B_d \rightarrow (c\bar c) K$) & $0.731 \pm 0.055$ & $\times$ & & \cite{bib:BABAR-sb,bib:BELLE-sb} \\
\hline
\multicolumn{4}{|l}{$b$-hadron lifetimes} & \\
\hline
$\tau_{B^0}$ & $(1.542 \pm 0.016)$ ps & $\times$ & & \cite{bib:pdg} \\
$\tau_{B^+}$ & $(1.674 \pm 0.018)$ ps & $\times$ & & \cite{bib:pdg} \\
$\tau_{B_s}$ & $(1.461 \pm 0.057)$ ps & $\times$ & & \cite{bib:pdg} \\
$\tau_{\Lambda_b} $ & $(1.229 \pm 0.08)$ ps & $\times$ & & \cite{bib:pdg} \\
\hline
\multicolumn{4}{|l}{$b$-hadron production parameters} & \\
\hline
$f_{B^+}=f_{B^0}$ & $0.388 \pm 0.013$ & $\times$ & & \cite{bib:pdg} \\
$f_{B_s}$         & $0.106 \pm 0.013$ & $\times$ & & \cite{bib:pdg} \\
$f_{B^{+,0}}$     & $1.04  \pm 0.08$  & $\times$ & & \cite{bib:pdg} \\ 
\hline
\multicolumn{4}{|l}{Meson masses} & \\
\hline
$m_{B_d}$ &  $ 5279.4   ~\rm MeV/c^2$ & & & \cite{bib:pdg}\\
$m_{B_s}$ &  $ 5369.0   ~\rm MeV/c^2$ & & & \cite{bib:pdg}\\
$m_{K^0}$ &  $ 497.7    ~\rm MeV/c^2$ & & & \cite{bib:pdg} \\
\hline
\multicolumn{4}{|l}{Running quark masses} & \\
\hline
$ m_t$ &  $169.3 \pm 5.1~\rm GeV/c^2$ & $\times$ & & \cite{bib:ckmfitter}\\
$ m_c$ &    $1.3 \pm 0.1~\rm GeV/c^2$ & $\times$ & & \cite{bib:pdg}\\
\hline
\multicolumn{4}{|l}{Standard model parameters} & \\
\hline
$m_W $ & $ 80.423 \pm 0.039~{\rm GeV/c^2}$ & $\times$ & & \cite{bib:pdg}\\
$ G_F$ & $(1.16639 \pm 0.00001) \times 10^{-5}~{\rm GeV} $ & &  & \cite{bib:pdg} \\
$\lambda$ & $0.2241 \pm 0.0033$ & $\times$ & &
\cite{bib:CKM2002} \\
\hline
\multicolumn{4}{|l}{$B$ factors and decay constants} & \\
\hline
$B_K$ &  $0.72$ -- $1.0 \pm 0.06$   & $\times$ & $\times$ &\cite{bib:lat00}\\
$f_{B_d} \sqrt{B_{B_d}}$ & $211 $ -- $235 \pm 33$~MeV &  $\times$ & $\times$ & \cite{bib:lat00}\\
$\xi$ & $1.18 $ -- $ 1.30 \pm 0.04$ & $\times$ & $\times$ & \cite{bib:lat00}\\
\hline
\multicolumn{4}{|l}{Form factors and reduced decay rates} & \\
\hline
${\cal F}_{D^*}(1) $ & $0.87$ -- $0.95$ &  & $\times$ & \cite{bib:BABAR} \\
$\widetilde{\Gamma}_{\rho\ell\nu}$ & 12.0 -- 22.2 ps$^{-1}$ & & $\times$ & \cite{bib:rholnuCLEO,bib:rholnuBaBar} \\
$\widetilde{\Gamma}_{u\ell\nu}$ & 54.8 -- 79.6 ps$^{-1}$ & & $\times$ & \cite{bib:ulnu} \\
$\widetilde{\Gamma}_{c\ell\nu}$ &38.0 -- 41.5 ps$^{-1}$ & & $\times$ & \cite{bib:clnu} \\
\hline
\multicolumn{4}{|l}{QCD parameters} & \\
\hline
$\eta_B$ & $0.55\pm  0.01$ & & $(\times)$ & \cite{bib:etab}\\
$\eta_1$ & $1.32\pm  0.32$ & & $(\times)$ & \cite{bib:eta1}\\
$\eta_2$ & $0.574\pm 0.01$ & & $(\times)$ & \cite{bib:etab} \\
$\eta_3$ & $0.47\pm  0.04$ & & $(\times)$ & \cite{bib:eta3}\\
\hline
\end{tabular}
\end{center}
\end{table*}

\subsubsection{Inputs for $|V_{cb}|$}

The measured
inclusive $B \rightarrow X_c \ell \nu$ decay rate is related to $|V_{cb}|$ by
\begin{equation}
|V_{cb}| = \sqrt{\frac{\Gamma(B \rightarrow X_c \ell \nu)}
{\widetilde\Gamma_{\scriptstyle\rm SL}}},
\end{equation} 
where $\widetilde \Gamma_{\scriptstyle\rm SL}$ is the predicted semileptonic 
$b \rightarrow c \ell\nu$ decay rate. 
We use the most recent branching fraction measurements 
from \babar, Belle, CLEO, and LEP 
 (see Table~\ref{tab:bcl}) 
that have been collected by the Heavy Flavor Averaging Group. 
At the $\Upsilon(4S)$ the semileptonic decay rate is measured to be 
 $\Gamma(B \rightarrow X_c \ell \nu) = 
0.446\cdot (1 \pm 0.023 \pm 0.007)\times 10^{-10}~\rm MeV$ while at LEP 
a value of $\Gamma(B \rightarrow X_c \ell \nu) = 
0.441\cdot (1 \pm 0.018)\times 10^{-10}~\rm MeV$ is obtained. Both values are 
averaged including the $B$ meson and $b$ quark life times,
respectively. In our maximum likelihood fits we scan the entire
theoretical uncertainty in $\widetilde \Gamma_{\scriptstyle\rm SL}$, which results by adding the 
individual theoretical uncertainties in quadrature. This amounts to a 
$5\%$ theoretical error in $|V_{cb}|$.

In exclusive $B \rightarrow D^* \ell \nu$ decays 
$|V_{cb}|$ may be extracted from the lepton spectrum at zero 
recoil $|V_{cb}| \times F(1)$~\cite{bib:LEP-vcb}. 
We use the average value for $|V_{cb}| \times F(1)$ provided by the 
Heavy Flavor Averaging Group, 
based on results from CLEO, Belle, ALEPH, DELPHI, and OPAL. 
We scan over the theoretical uncertainty of the prediction of ${\cal F}_{D^*}(1)$ which 
is presently $4.4\%$. 

\subsubsection{Inputs for $|V_{ub}|$}

The inclusive value of $V_{ub}$ is obtained from five branching fraction 
measurements at the $\Upsilon(4S)$ and the LEP average
(see Table~\ref{tab:bul}). 
Here the theoretical 
uncertainties enter in three ways, in the efficiency, in the branching 
fraction and in the extraction of $V_{ub}$. 
In order to incorporate these errors in the scan
we adopt the following procedure. We first determine a weighted average of the
five $\Upsilon(4S)$ measurements using statistical and systematic errors only.
Then we determine weighted averages for the upper and lower bounds determined 
by the theoretical uncertainties ($B_i + \Delta_i, B_i -\Delta_i$), 
where the weights are determined from statistical and systematic errors only. 
The difference between the averaged upper and lower bounds and the average of 
the central value yield the theoretical uncertainty in the averaged 
branching fraction $B \rightarrow X_u \ell \nu$ measured at the
$\Upsilon(4S)$. We then average the $\Upsilon(4S)$ results and the LEP results
using the $B$-meson and $b$ quark life times. From this value 
$V_{ub}$ is extracted using~\cite{bib:vub-thin}
\bigskip
\begin{eqnarray}
|V_{ub}| = &  0.0445 \left[ {\frac{B(b \rightarrow u \ell \nu)}{0.002} }\cdot
{\frac{1.55 ps}{\tau_b}}\right]^{1/2} \nonumber \\
 &    \times\ (1 \pm0.02_{QCD}
\pm0.052_{m_B}).
\end{eqnarray}
The theoretical uncertainties from the branching fraction and those from 
the extraction of $|V_{ub}|$ are added in quadrature and converted into a 
factor that is scanned.

Though CLEO and Belle have measured $|V_{ub}|$ in various exclusive decay modes
we use the branching fraction of $B^0 \rightarrow \rho^-\ell \nu$ by averaging
the CLEO and  \babar\ measurements (also shown in Table~\ref{tab:bul}). 
To incorporate the different
theoretical uncertainties we also average the $\pm 1 \Delta$ upper and lower
bounds using only statistical and systematic errors. The theoretical 
uncertainty is incorporated into the predicted reduced rate that is scanned
over in the maximum likelihood fit.


\begin{table*}[t]
\caption [ ] {Inclusive branching fraction measurements of 
$B \rightarrow X_c \ell \nu$. 
\label{tab:bcl}}
\begin{center}
\begin{tabular}{@{\extracolsep{2pt}}|@{ }l|l|l|}
\hline
Experiment &Branching Fraction [$\%$] & Reference \\
\hline
\hline
$\Upsilon(4S)$ average & $10.70 \pm 0.28$ & \cite{bib:pdg03} \\
\hline
LEP average & $10.42 \pm 0.26$ & \cite{bib:LEP-vcb} \\
\hline
\end{tabular}
\end{center}
\end{table*}

\begin{table*}[htb]
\caption [ ] {Inclusive branching fraction measurements of $B \rightarrow X_u \ell \nu$
and selected exclusive measurements of $B(B^0 \rightarrow \rho^- e^+ \nu_e)$
\label{tab:bul}}
\begin{center}
\begin{tabular}{@{\extracolsep{2pt}}|@{ }l|l|l|l|}
\hline
Experiment & Method  & Branching Fraction [$10^{-3}]$ & Reference \\
\hline
\hline
CLEO & endpoint analysis & $1.77\pm0.115_{stat}\pm0.269_{sys}\pm
0.327_{fu-stat}\pm 0.20_{fu-sys}$  & \cite{bib:cleo-ulnu} \\
\babar & endpoint analysis & $2.054\pm0.189_{stat}\pm0.189_{sys}\pm0.388_{fu-stat}
\pm 0.25_{fu-sys} $ &  \cite{bib:babar-ulnu} \\
\babar & hadronic mass & $2.14 \pm0.29_{stat}\pm0.25_{sys}
\pm0.37_{b \rightarrow u}$ & \cite{bib:ulnu} \\
Belle & $D^{(*)} \ell \nu$ tags & $2.62\pm0.63_{stat}\pm0.23_{sys}\pm0.05_{b \rightarrow c} \pm 0.41_{b \rightarrow u}$ & \cite{bib:ulnu} \\
Belle & Improved $\nu$ reconstruction & 
$1.64\pm0.14_{stat}\pm0.36_{sys}\pm0.28_{b \rightarrow c}
\pm 0.22_{b \rightarrow u}$ & \cite{bib:ulnu} \\
\hline 
$\Upsilon(4S)$ average &  & $2.031\pm0.215_{sta+sys}\pm0.31_{th}$ & \\
\hline
LEP average & hadronic mass/neural network &
$1.71\pm 0.31_{stat+sys}\pm0.37_{b \rightarrow c} \pm 0.21_{b \rightarrow u}$ & \cite{bib:LEP-vub} \\
\hline
\hline
CLEO & $B^0 \rightarrow \rho^- e^+ \nu_e$ & $2.17 \pm 0.34_{stat}$
$^{+0.47}_{-0.54\ sys}\pm 0.01_{} \pm 0.041_{th}$ & \cite{bib:rholnuCLEO}\\
\babar & $B^0 \rightarrow \rho^- e^+ \nu_e$ &  $3.29\pm 0.42_{stat} \pm 
0.47_{sys} \pm 0.6_{th}$ & \cite{bib:rholnuBaBar}\\
\hline
$\Upsilon(4S)$ average &  & $2.68\pm0.43_{sta+sys}\pm0.50_{th}$ & \\
\hline
\end{tabular}
\end{center}
\end{table*}

\ignore{
\\caption [ ] {Measurement inputs used for unitarity triangle fits.
\label{tab:ckm-input}}
\begin{center}
\begin{tabular}{@{\extracolsep{2pt}}|@{ }l|l|l|l|l|}
\hline
Observable & Process & Measurement & Experimental Error & Reference \\
\hline
\hline
${\cal B}(B \rightarrow \rho \ell \nu)$ &   &
$ \vert V_{cb} \vert {\cal F}_{D^*}(1) $ & $ B \rightarrow D^* \ell \nu$ &

$ 0.03822$ & $\pm 0.0011$ & \cite{bib:LEP-vcb} \\
$ \Delta m_{B_d} $ & $B_d$--$\bar B_d$ oscillations & $ 0.503~ \rm ps^{-1}
$
& $\pm 0.006~\rm ps^{-1}$ &  \cite{bib:LEP-osc}\\
$\Delta m_{B_s} $ & $B_s$--$\bar B_s$ oscillations & $> 14.4\ \rm
 ps^{-1}\  \ (95\%$ C.L.)&  & \cite{bib:LEP-osc}\\
$\epsilon_K$ &  \CPviolate\ in $K^0$--$\bar K^0$ mixing
 & $ 2.271 \times 10^{-3}$ & $\pm 0.017 \times 10^{-3}$ &
 \cite{bib:pdg}\\
$\sin 2 \beta$ & $B_d \rightarrow (c\bar c) K$ & $0.731$ & $\pm 0.055$ & \cite{bib:BABAR-sb,bib:BELLE-sb} \\
\hline
\end{tabular}
\end{center}
\end{table*}
}

\subsubsection{Incorporating $\Delta m_{B_s}$ data}

Including the current information on $\Delta m_{B_s}$  in the fit poses some special problems.
In this case, a lower limit at 95\%~C.L. has been determined by combining 
analyses of different experiments using the amplitude method~\cite{bib:moser}.
In this approach, we describe
the $\bar B_s$ and $B_s$ decays with the p.d.f.:
\begin{equation}
{\cal P} = \frac{1}{\tau} \ e^{- t/\tau} \  \frac{1 \pm {\cal A}
\cos(\Delta m_{B_s} t)}{2},
\end{equation}
where the amplitude ${\cal A}$ has been introduced. 
Experimental measurements provide
${\cal A}$ and its uncertainty $\sigma_{\cal A}$
as a function of $\Delta m_{B_s}$.
Comparing the measured amplitude to the expected one, we can
add a term to the $\chi^2$ function~\cite{bib:BABAR}:
\begin{equation}
\label{eq:chi-bs}
\chi^2_{\Delta {m_{B_s}}}(A,\bar \rho,\bar \eta) =  - 2\ln {\cal L}_\infty (\Delta m_{B_s}),
\end{equation}
where $-2 \ln {\cal L}_\infty (\Delta m_{B_s}) = 
\max(\frac{(1 - 2 \cal{A})}{\sigma_{{\cal A}}^2}, 0)$. 
For the non-zero values we have in the past used a subroutine provided by F.~Parodi~\cite{bib:parodi}
that incorporates the available experimental information. 
We required our implementation of the function to approach zero as $\Delta_{m_{B_s}}\to\infty$,
reflecting our contention that the experimental data does not rule out very large oscillation frequencies. 
In addition, we constrained the function to be nowhere negative, 
as required for a $\chi^2$ interpretation.
This procedure may be criticized for its {\it ad hoc} construction.
In addition, when using it in fits we have in some cases encountered
numerical instabilities arising from its multiple minima and lack of 
smoothness in the resulting function.

\begin{figure}
\centerline{
 \epsfig{file=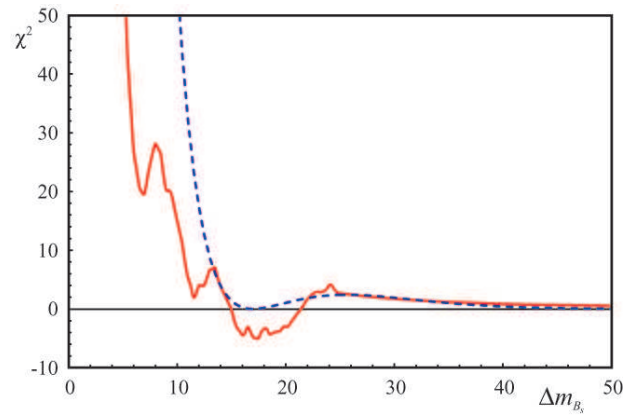,width=3.2in}  
 }
 \caption{The dashed blue curve is the empirical $\chi^2$ expression 
  used to represent the experimental data on $\Delta m_{B_s}$ in the
  fits described in this paper.
  The solid red curve is an implementation based on the amplitude 
  analysis method; when used in the role of a $\chi^2$ term, the
  function is truncated to non-negative values (see Eqn.~(\ref{eq:chi-bs})).
} 
\label{fig:deltamschisq}
\end{figure}

Given the difficulties with this approach, we now use a somewhat different method 
for incorporating the $\Delta m_{B_s}$ information into the fit. 
We start with the formula for significance~\cite{bib:willocq}:
\begin{equation}
 S = \sqrt{{N\over 2}} f_{B_s} (1-2w) e^{-{1\over 2}(\Delta m_s \sigma_t)^2},
\end{equation}
where $N$ is the sample size, $f_{B_s}$ is the $B_s$ purity, 
$w$ is the mistag fraction, and $\sigma_t$ is the resolution. 
We use this formula in an empirical approach, 
and substitute symbol $C$ for the expression $\sqrt{{N\over 2}} f_{B_s} (1-2w)$. 
Our use of this formula begins with interpreting $S$ 
as the number of standard deviations by which $\Delta m_{B_s}$ differs from zero, 
$S=\Delta m_{B_s}/\sigma_{\Delta {m_{B_s}}}$ 
(a similar result would be obtained with the 
interpretation applied instead to $1/\Delta {m_{B_s}}$). 
With this interpretation, we may express a contribution to the $\chi^2$ 
from the $\Delta m_{B_s}$ measurements as:
\begin{equation}
 \chi^2_{\Delta {m_{B_s}}} = C^2\left(1-{\Delta\over\Delta {m_{B_s}}}\right)^2
   e^{-(\Delta {m_{B_s}}\sigma_t)^2},
\end{equation}
where $\Delta$ is the best estimate according to experiment. 
The values of $(\Delta, C^2, \sigma_t)$ are chosen to give a minimum at 
$17~ \rm ps^{-1}$, 
and a $\chi^2$ probability of 5\% at $\Delta {m_{B_s}}=14.4~\rm ps^{-1}$. 
This function is plotted in Fig.~\ref{fig:deltamschisq}, 
with the $-2\ln {\cal L}_\infty(\Delta {m_{B_s}})$ curve described earlier superimposed. 
It may be noted that, in the region of small $\chi^2$, 
the two functions exhibit similar general features.
Deviations in the region where both curves have large values don't matter much --- 
in that region the fit is poor in either case. 
We have checked the sensitivity of our results to how rapidly the $\chi^2$ rises 
at low values of $\Delta {m_{B_s}}$, and find very little effect. 
Thus, we have some confidence that this empirical treatment is providing dependable answers.

\subsection{Fit results}

A ``model'' ${\cal M}$ and its best-fit solution are kept only
if the probability of the fit satisfies
$P(\chi^2_{\cal M}) > P_{min}$, which is typically chosen to be $5\%$. 
For each ``model'' ${\cal M}$ accepted, we draw a 95\%~C.L. contour 
in the $(\bar \rho, \bar \eta)$ plane. 
The fit is repeated for other ``models'' ${\cal M}$ by scanning 
through the complete parameter space specified by a region broad compared with
the estimated theoretical 
uncertainties. 

The $\chi^2$ minimization thus serves two purposes:

\begin{itemize}

\item[1.)] If a ``model'' ${\cal M}$ is consistent with the data, 
we obtain the best estimates for the 
three CKM parameters, and 95\%~C.L. contours are determined.

\item[2.)] If a ``model'' ${\cal M}$ is inconsistent with the data the 
probability ${\cal P}(\chi^2_{\cal M})$ will be low. Thus, the
requirement of ${\cal P}(\chi^2_{\cal M})_{min} > 5\%$ provides 
a test of compatibility between data and its theoretical 
description.

\end{itemize}

If no ``model'' were to survive we would have evidence of a consistency problem
between data and theory, independent of the calculations of the
theoretical parameters or the choices of their uncertainties.

\subsubsection{Single-parameter sensitivity studies}

\begin{figure*}
 \centerline{
   \epsfig{file=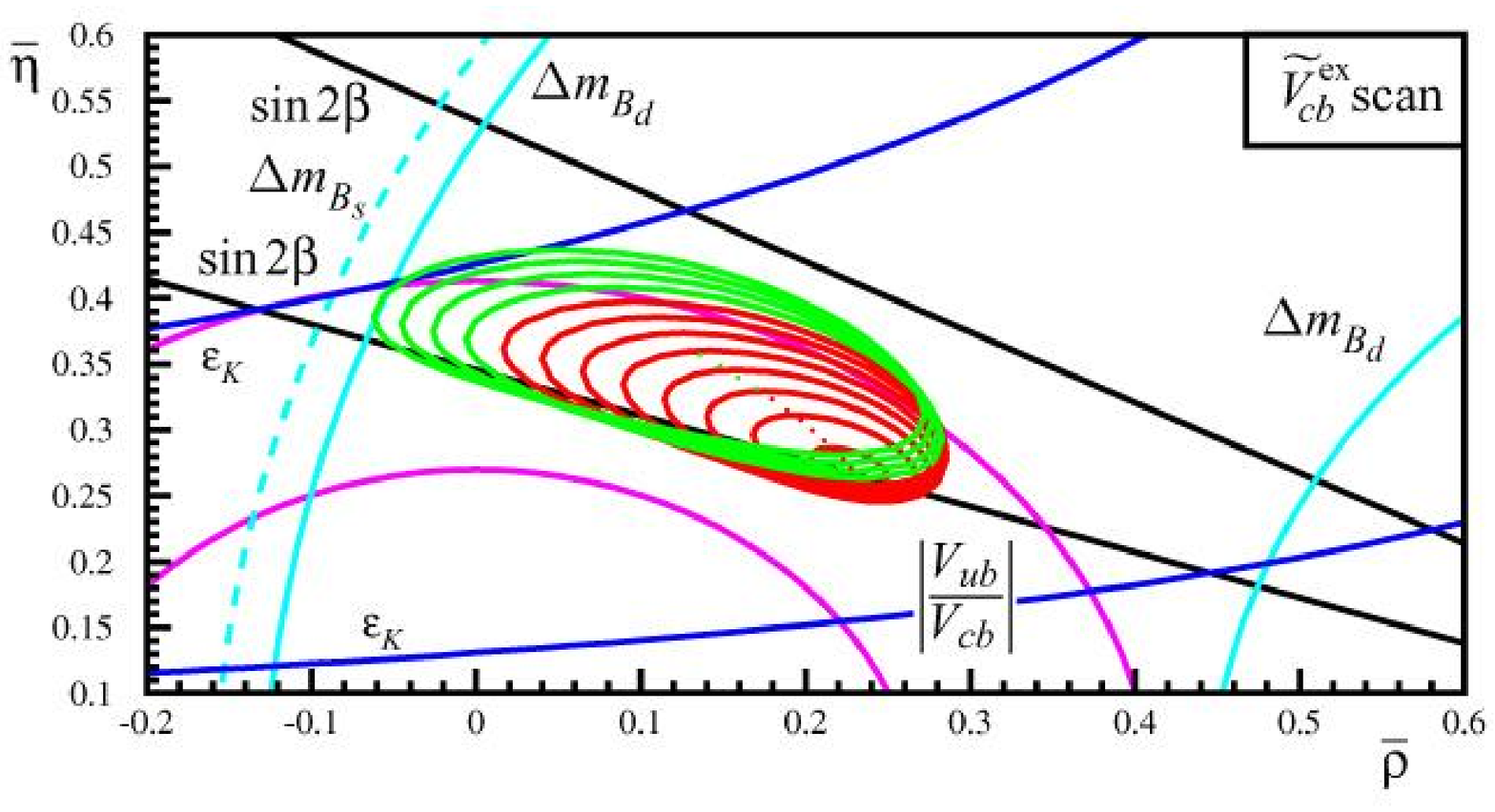,width=3.2in} 
  \hspace{1cm}
  \epsfig{file=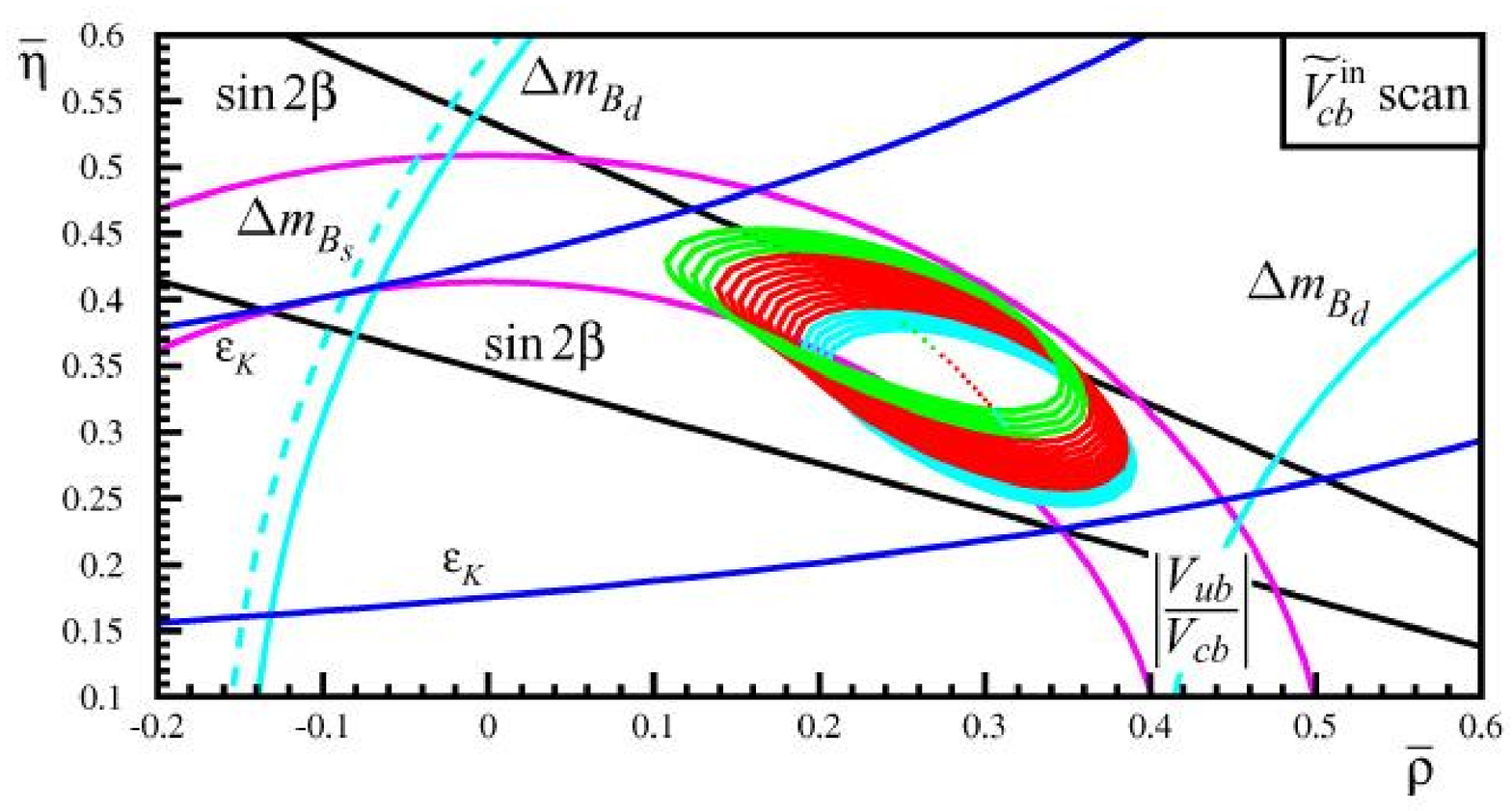,width=3.2in}
 }
\centerline{
  \epsfig{file=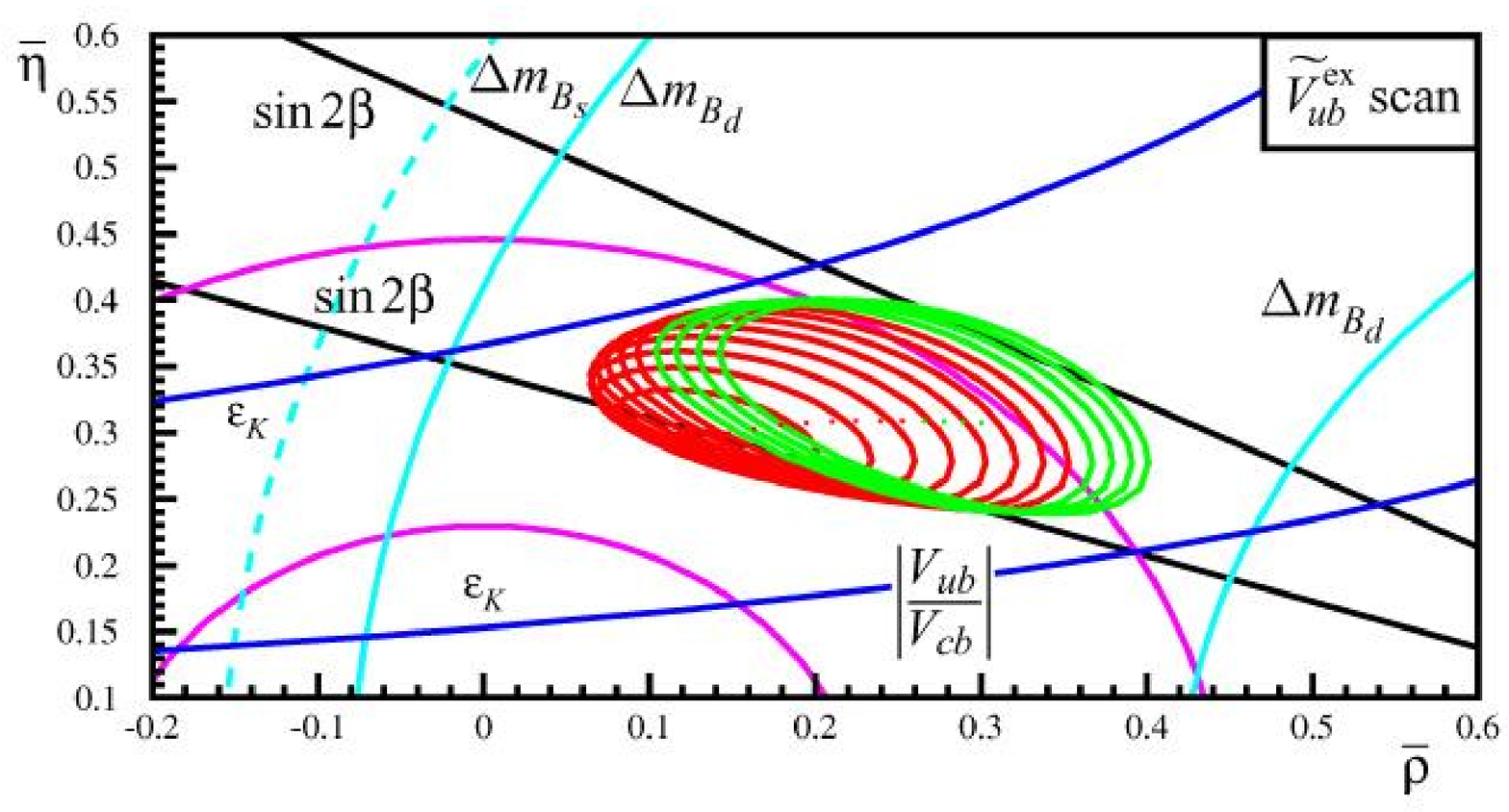,width=3.2in}  
  \hspace{1cm}
  \epsfig{file=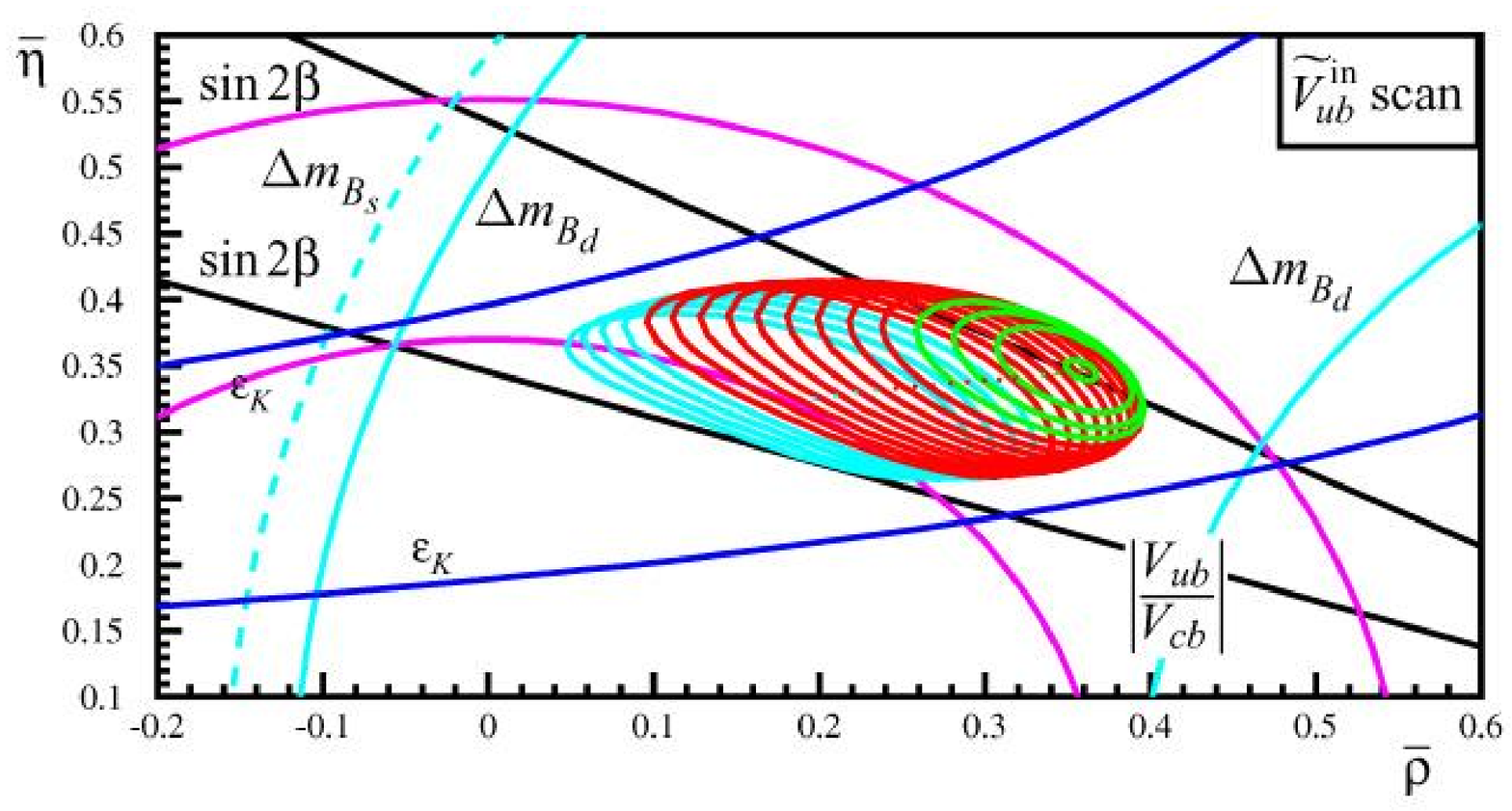,width=3.2in} 
 }
\centerline{
  \epsfig{file=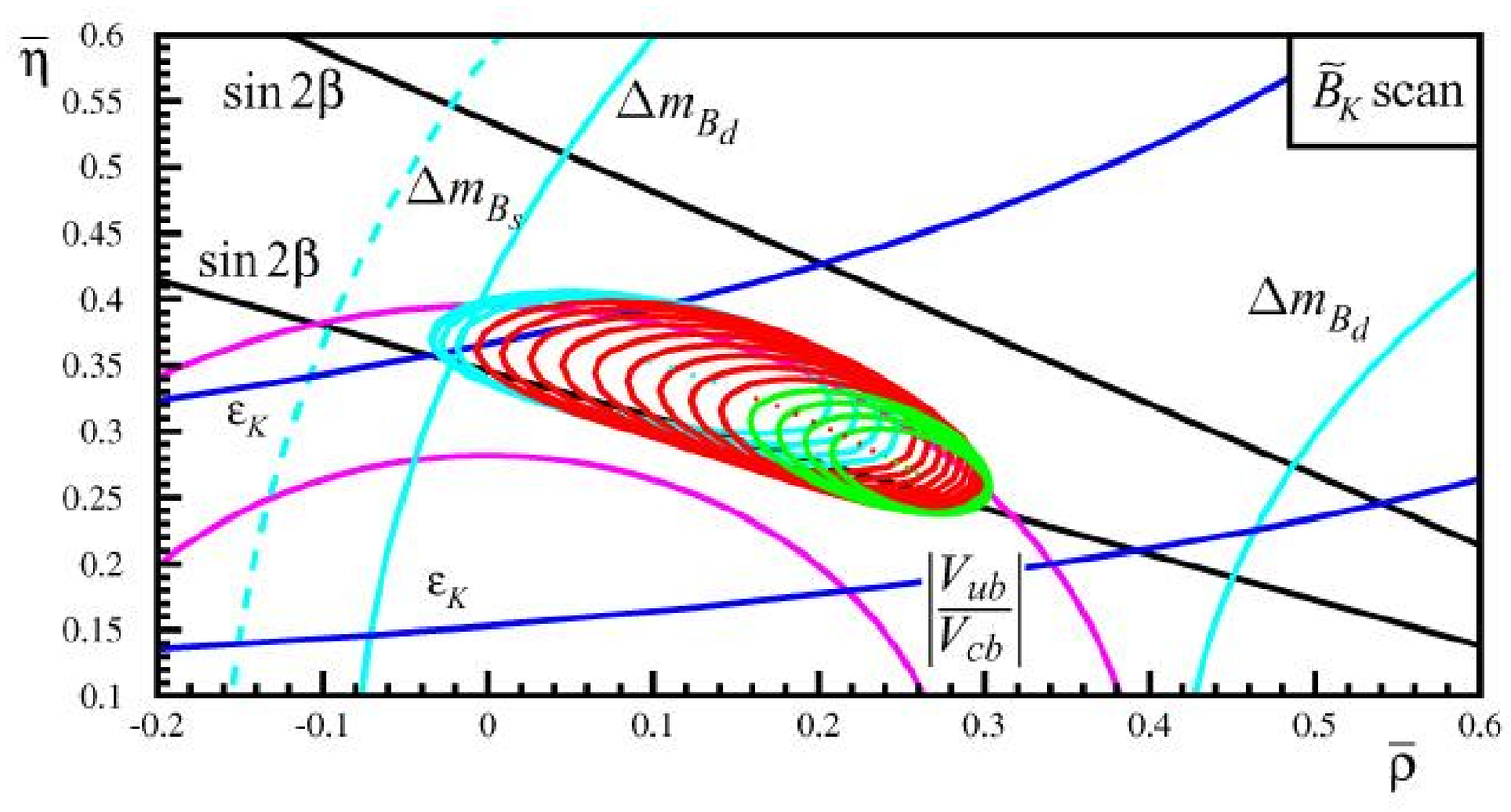,width=3.2in}  
  \hspace{1cm}
  \epsfig{file=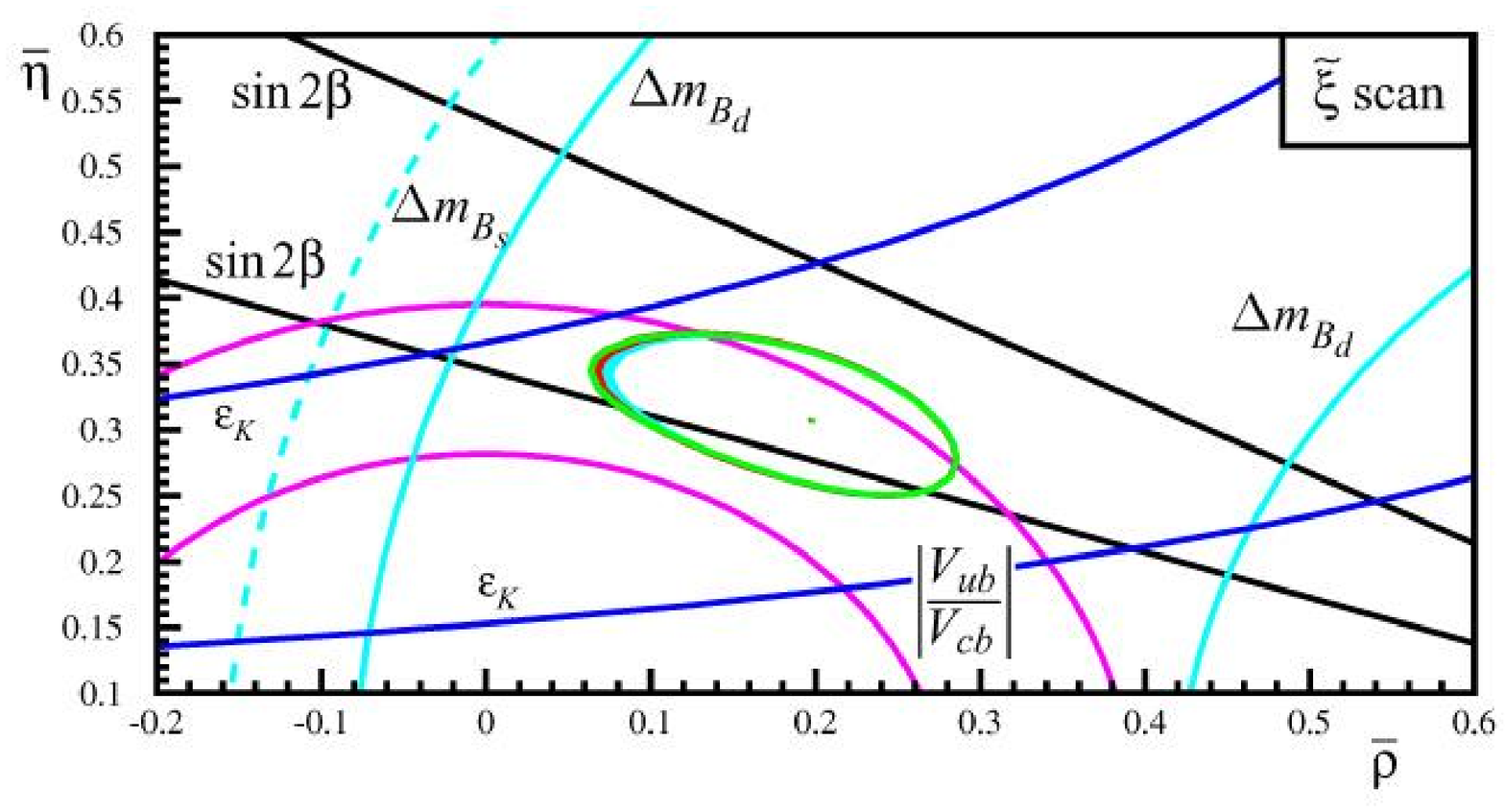,width=3.2in}
 }
\centerline{
  \epsfig{file=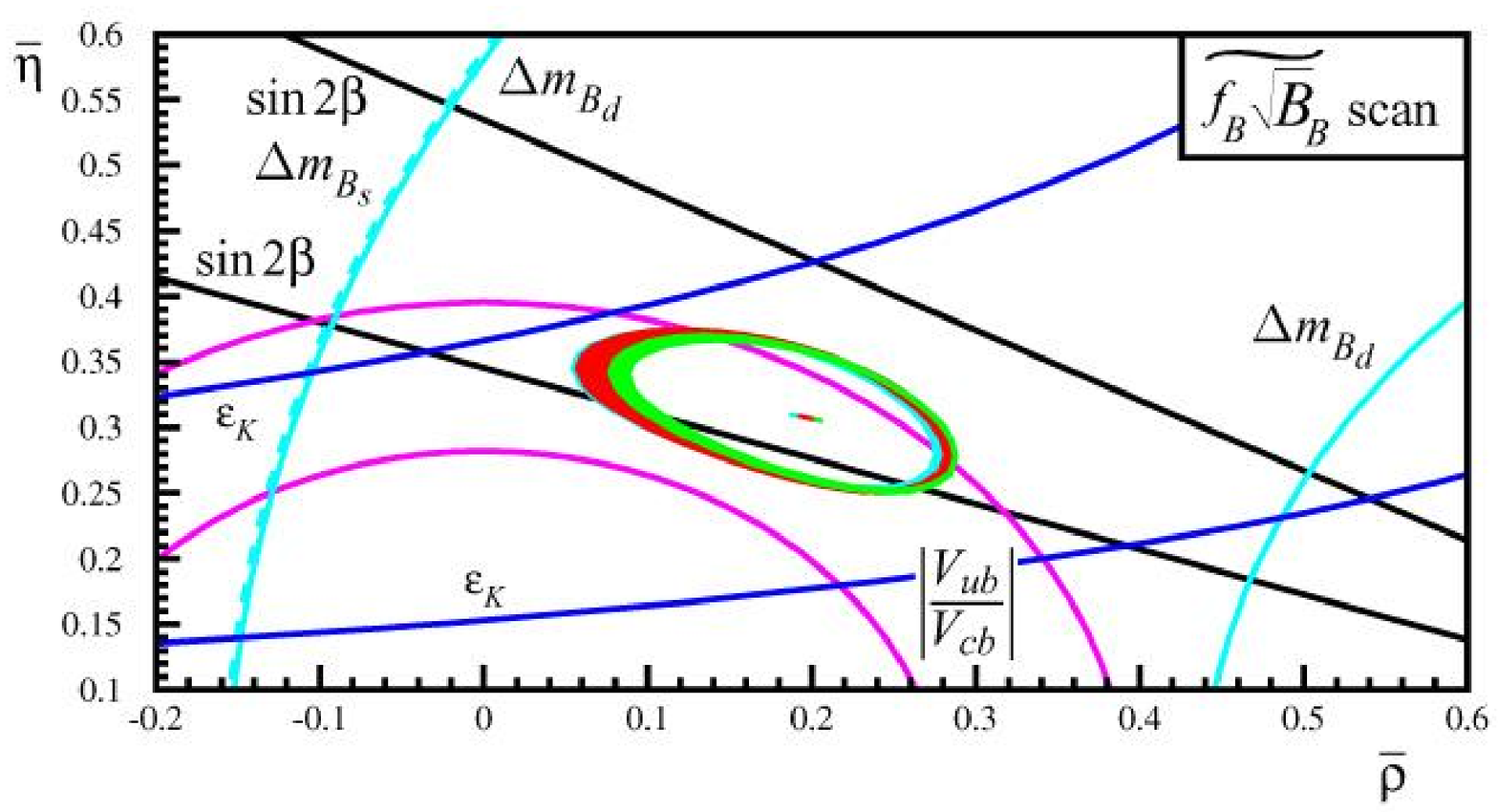,width=3.2in}
 }
 \caption{Contours of different models in the $(\bar \rho, \bar \eta)$
plane, by varying only one theoretical parameter at a time, left to right, 
top to bottom: 
$\widetilde{V}_{cb}^{\rm ex}$, $\widetilde{V}_{cb}^{\rm in}$,
$\widetilde{V}_{ub}^{\rm ex}$, $\widetilde{V}_{ub}^{\rm in}$, 
$\tilde B_K$, $\tilde\xi$, 
$\widetilde{f_B \sqrt{B_B}}$.
In each case, a total of 17 values were scanned: 
nine values (red contours) spanning the region $\pm\Delta_{\rm th}$ 
around the central theoretical value, 
where $\Delta_{\rm th}$ is the estimated theoretical uncertainty
of the parameter; 
four values (cyan contours) spanning the range from 
$-2\Delta_{\rm th}$ to $-\Delta_{\rm th}$; 
and four values (green contours) spanning the range from 
$\Delta_{\rm th}$ to $2\Delta_{\rm th}$.
Note that sometimes fewer than 17 contours appear --- 
this occurs when some of the values of the scanned parameter do not give an 
acceptable fit (less than 5\% $\chi^2$ probability).}
\label{fig:ut_par_1}
\end{figure*}

Demonstrating the impact of the different theoretical parameters on the
fit results in the $(\bar \rho, \bar \eta)$ plane, 
Figures~\ref{fig:ut_par_1} and~\ref{fig:ut_par_2} show contours for
fits in which only one parameter was scanned while the others were kept at 
their central values. 
These demonstrate the impact of the model dependence in each of the
theoretically uncertain parameters.

In these figures, the red contours correspond to variation of
the scanned parameter within its theoretically ``allowed'' range, 
while the green and cyan contours result from doubling this range, 
symmetrically to lower and higher values.
The scanned values of the parameter are spaced equidistantly, with nine
values covering the central theoretically preferred range and four each the 
upper and lower extensions.
If a fit fails the ${\cal P}(\chi^2_{\cal M})_{min} > 5\%$ requirement, 
no contour is drawn.

For qualitative comparison, we show the boundaries of the
four bands for $|V_{ub}/V_{cb}|$,  $|V_{td}/V_{cb}|$, $\sin2\beta$, and $\epsilon_K$.
Since the theoretical parameters are kept at their central values except for
the one being varied, the computation of these bands corresponding to the 
non-varied parameters reflects only experimental uncertainties.  

\ignore{
\begin{figure*}
\centerline{
   \epsfig{file=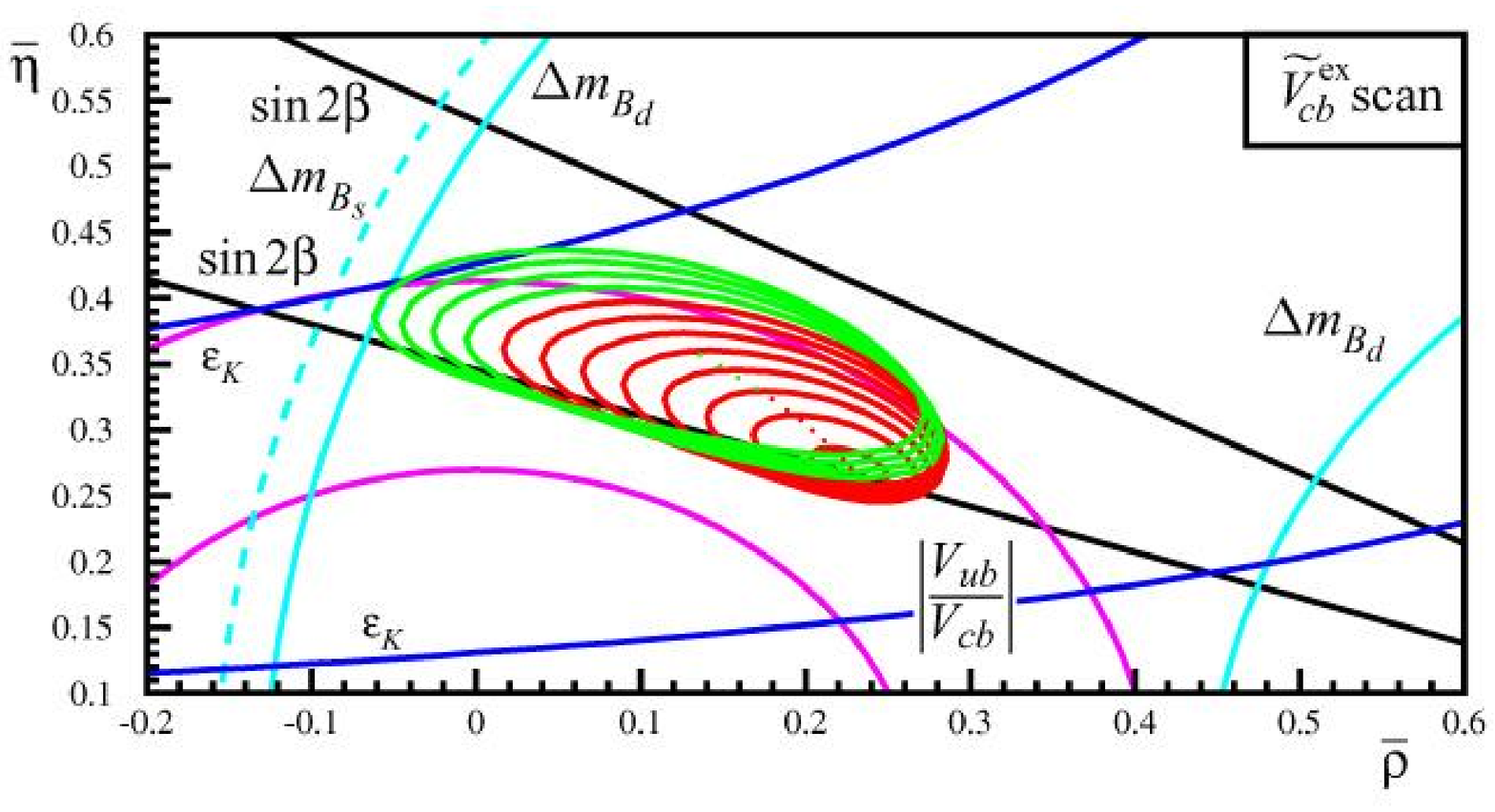,width=3.2in} 
  \hspace{1cm}
  \epsfig{file=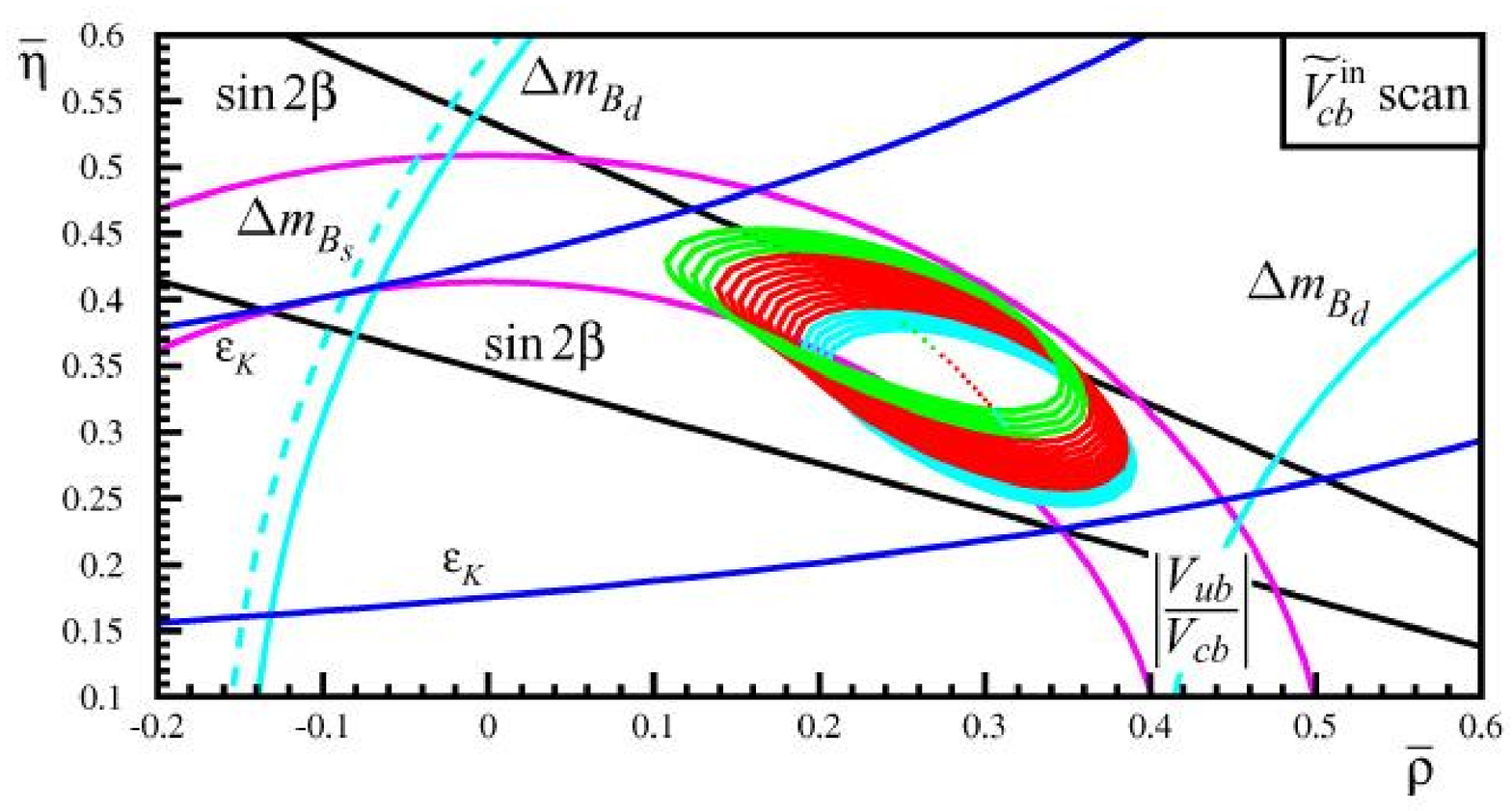,width=3.2in}
 }
\centerline{
  \epsfig{file=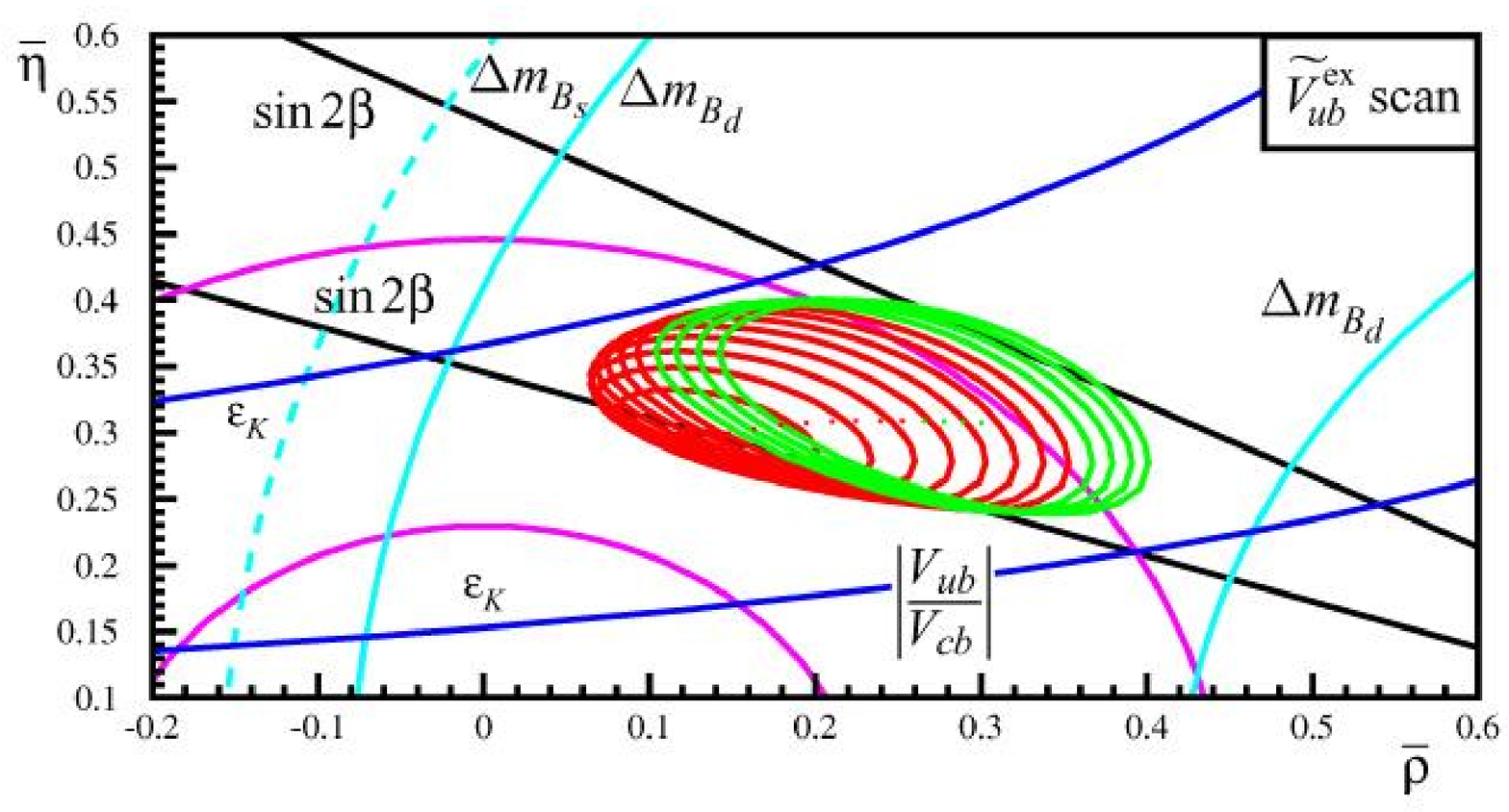,width=3.2in}  
  \hspace{1cm}
  \epsfig{file=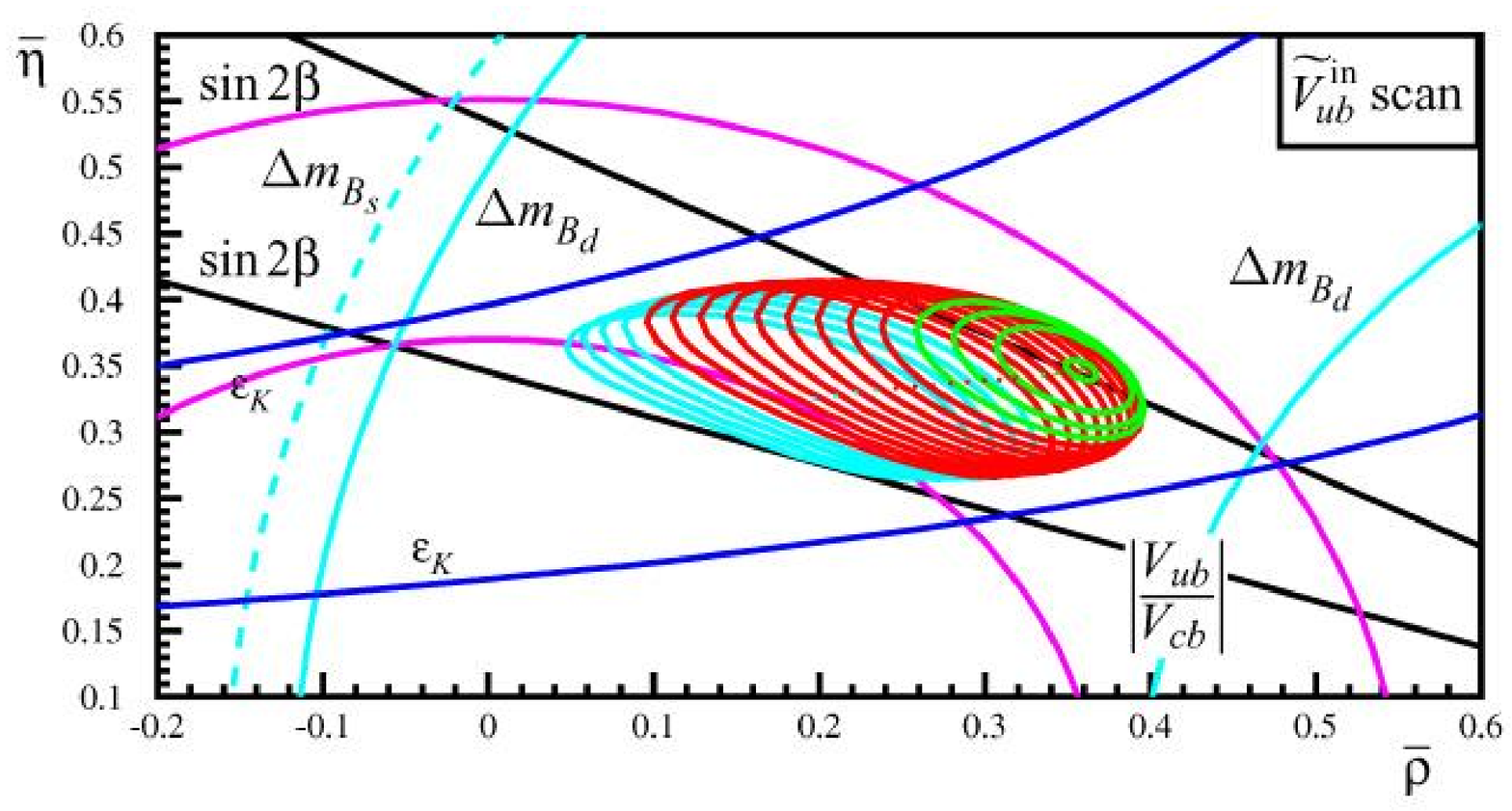,width=3.2in} 
 }
 \caption{Contours of different models in the $(\bar \rho, \bar \eta)$
plane, by varying only one theoretical parameter at a time, 
 $f_{B_d} \sqrt{B_{B_d}}$,  $B_K$, and
 $\xi$ (where $\Delta M_{B_s}$ is included in the fit). In each plot 9 different models are considered by varying
the theoretically-allowed range from the minimum value to the maximum value.
The figures are arranged with a) in the upper left, b) in the upper right, {\it etc.} 
 }
\label{fig:ut_par_2}
\end{figure*}  
}

\begin{figure*}
\centerline{
   \epsfig{file=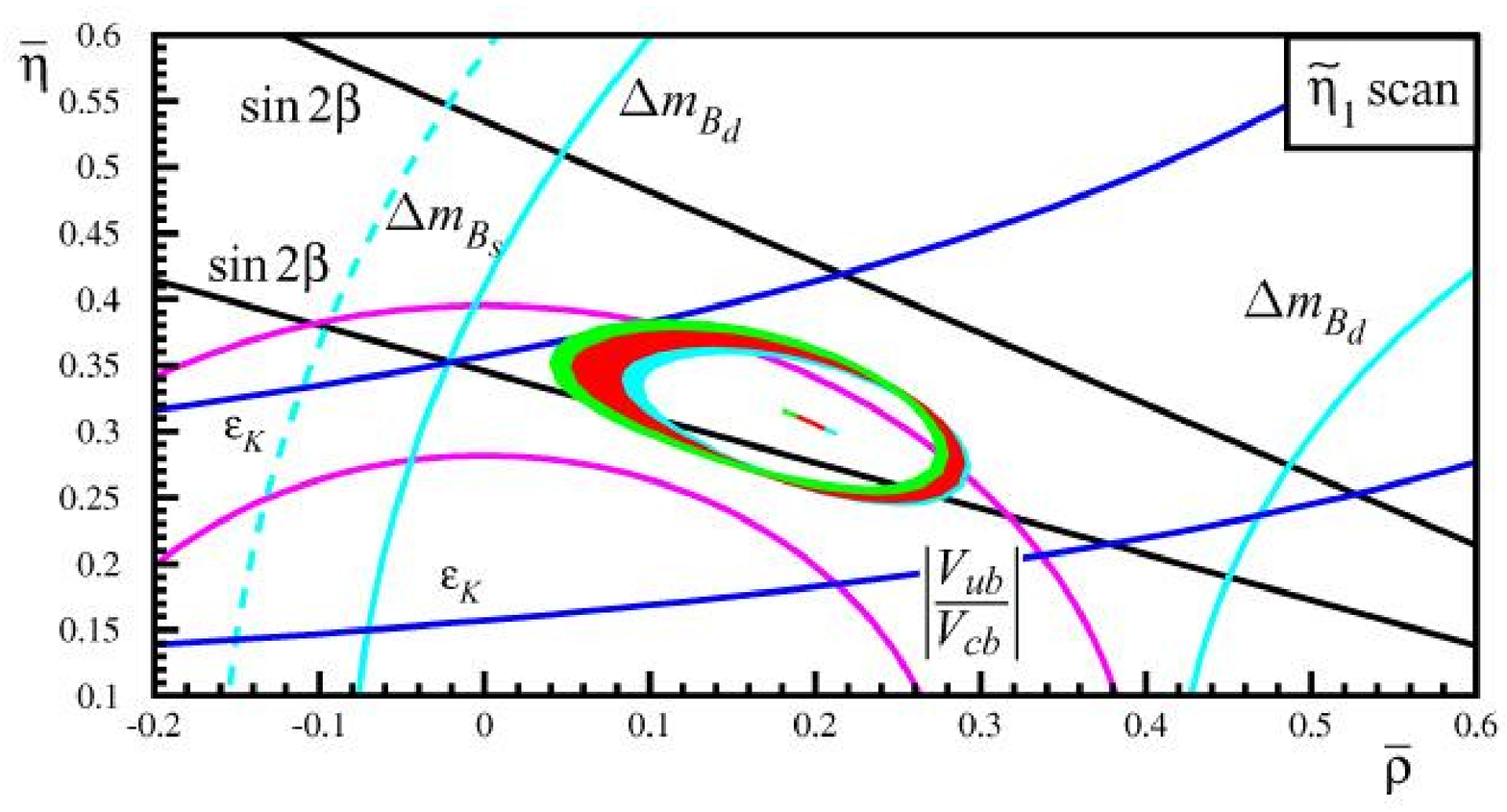,width=3.2in} 
  \hspace{1cm}
  \epsfig{file=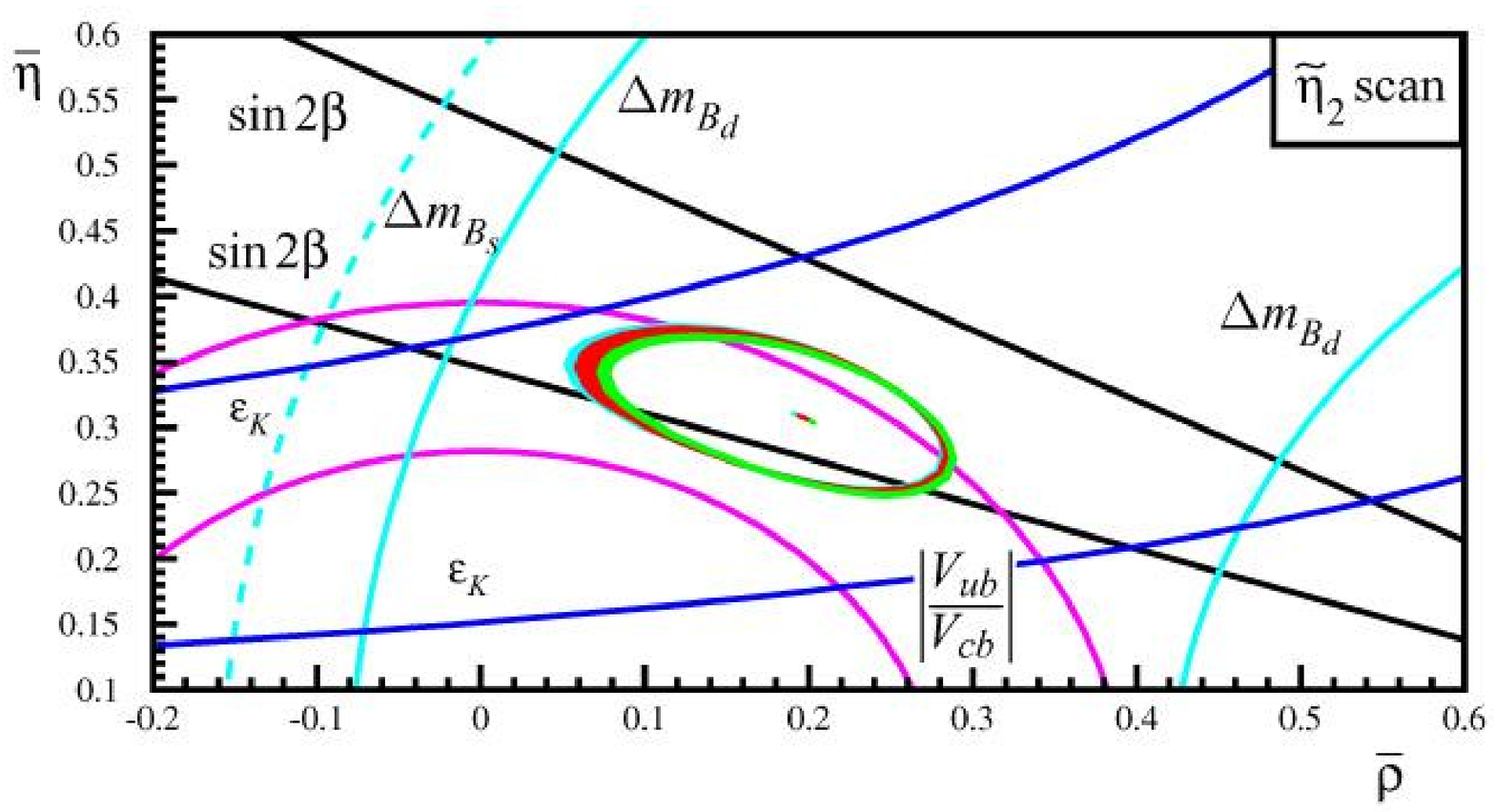,width=3.2in}
 } 
\centerline{
  \epsfig{file=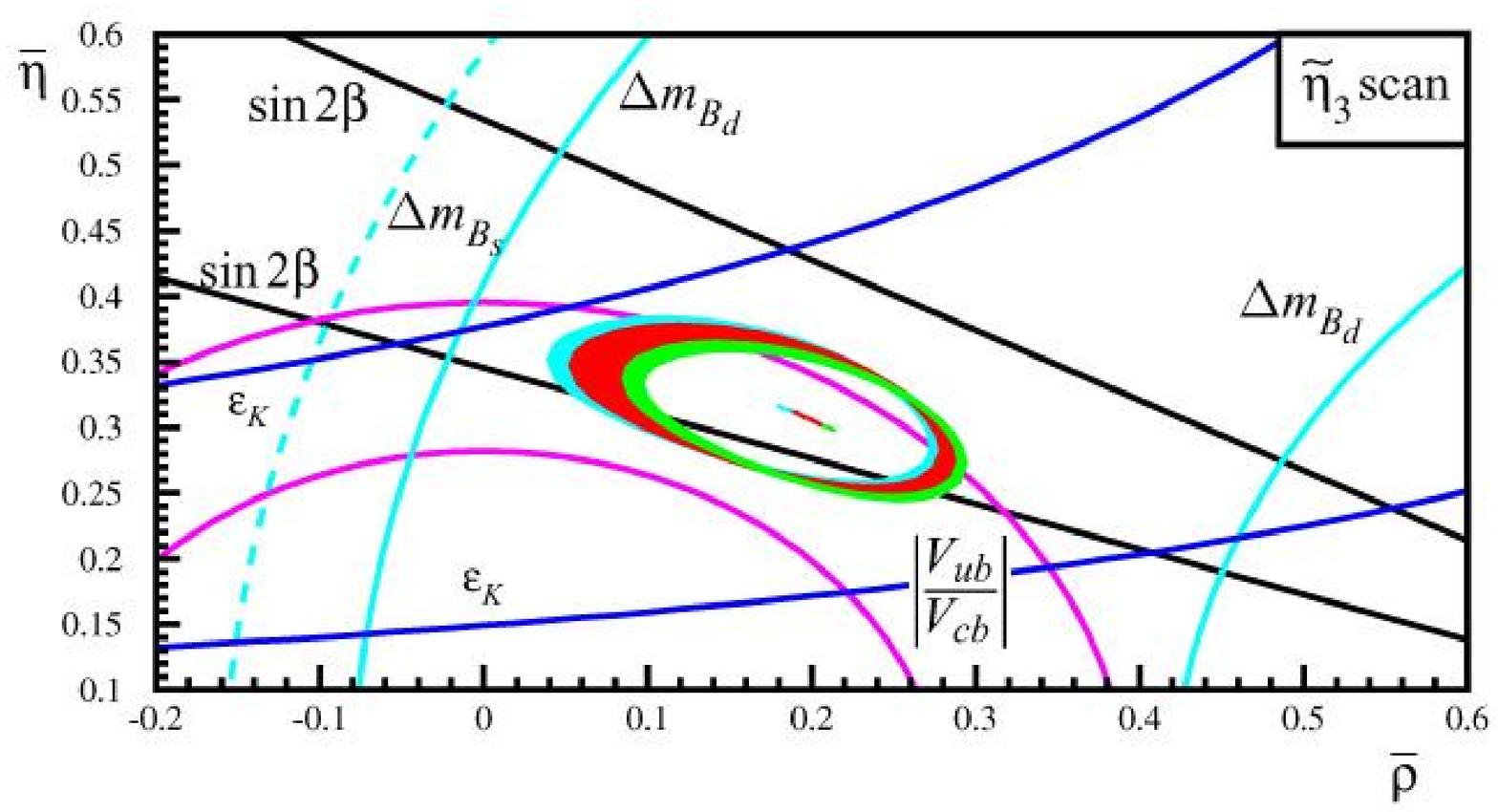,width=3.2in}  
  \hspace{1cm}
  \epsfig{file=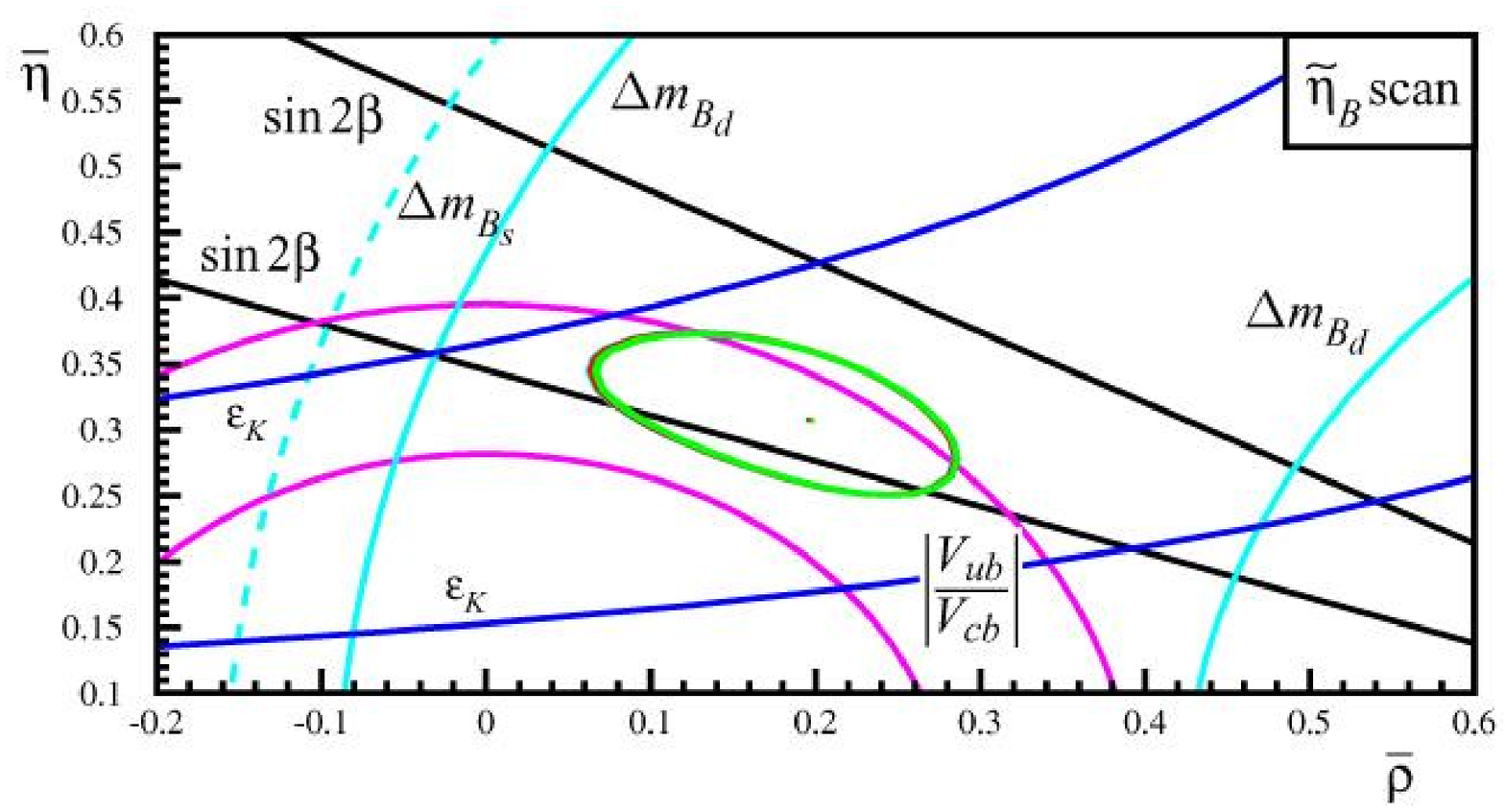,width=3.2in}
 } 
 \caption{Contours of different models in the $(\bar \rho, \bar \eta)$
plane, by varying only one theoretical parameter at a time, left to right, top to bottom: $\widetilde{\eta}_1$, $\widetilde{\eta}_2$, $\widetilde{\eta}_3$, 
and $\widetilde{\eta}_B$. The interpretation of the various contours is the same as in
Fig.~\ref{fig:ut_par_1}.
 }
\label{fig:ut_par_2}
\end{figure*}  

\begin{figure*}
\centerline{
  \epsfig{file=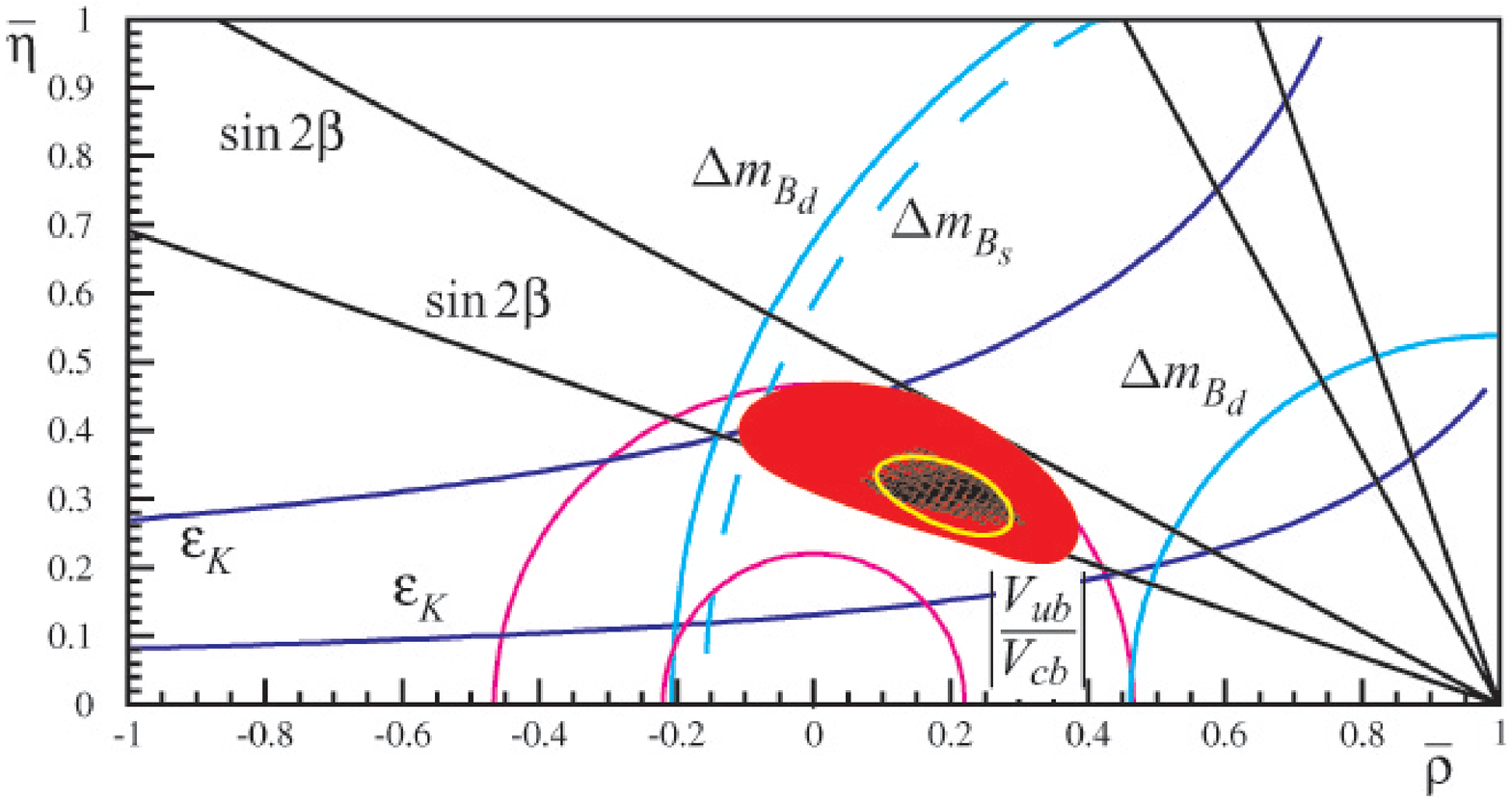,width=3.2in}  
  \hspace{1cm}
  \epsfig{file=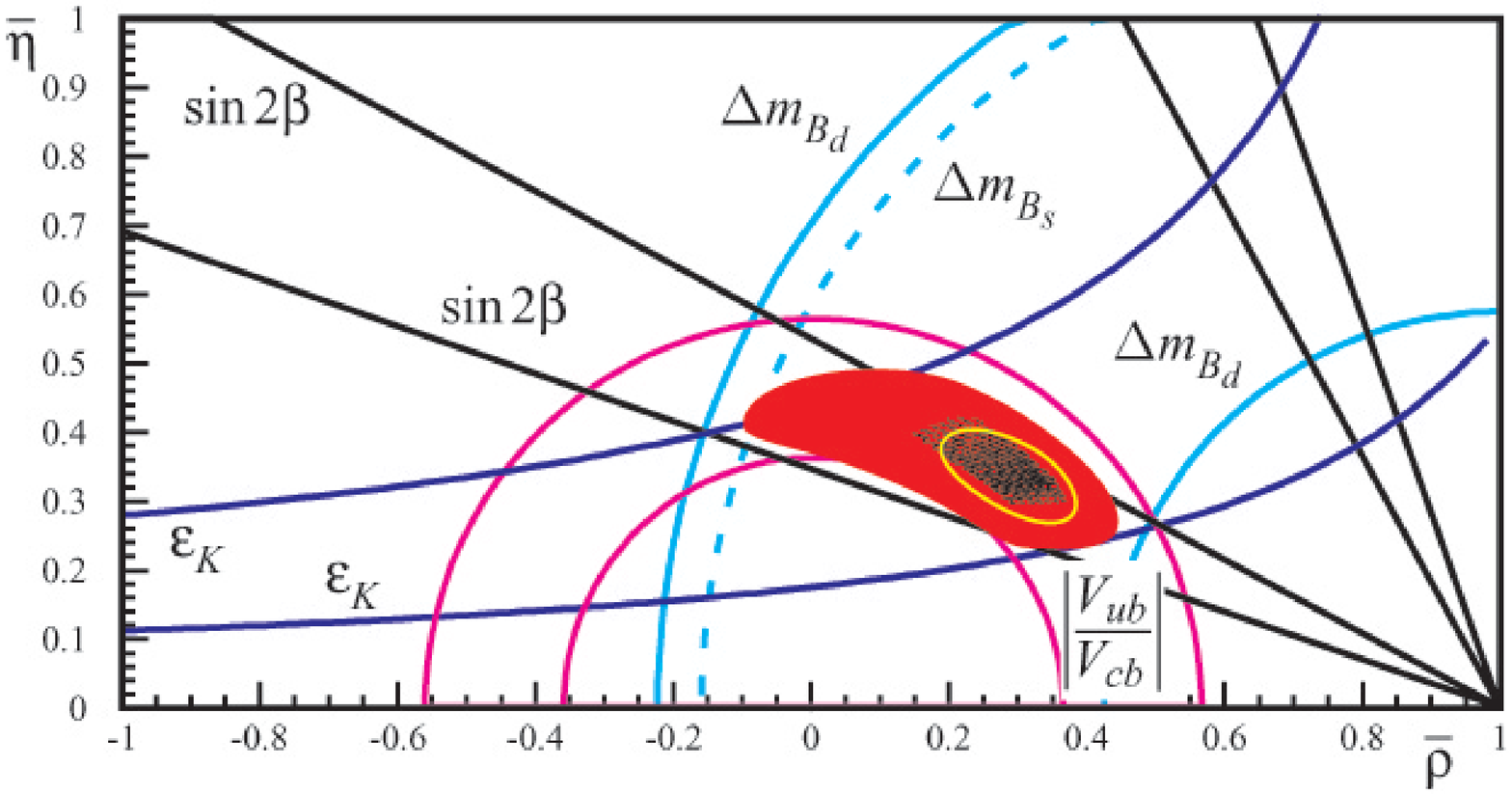,width=3.2in}
 }
\caption{The result of scanning over all theoretically uncertain parameters
where: the data from inclusive $|V_{ub}|$ and $|V_{cb}|$ 
measurements are excluded (left plot), and
the data from exclusive $|V_{ub}|$ and $|V_{cb}|$ 
measurements are excluded (right plot). 
The black dots show the central values from the fits, 
and a red contour for each fit indicates the 95\% $\chi^2$ probability region,
according to the experimental (and statistical theoretical) uncertainties.
 }
\label{fig:ut_par_4}
\end{figure*} 

\begin{figure*}
\centerline{
  \epsfig{file=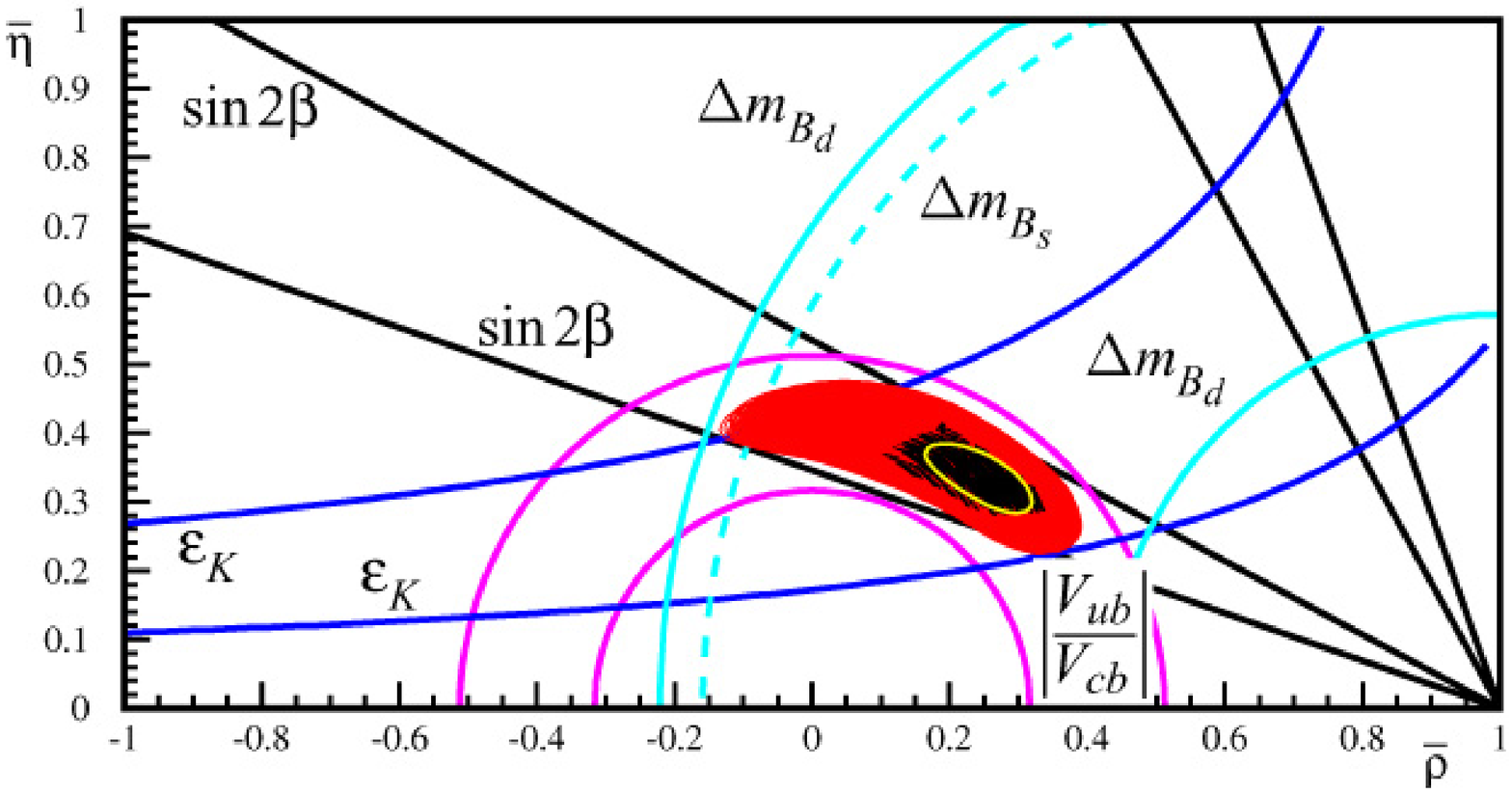,width=3.2in}  
  \hspace{1cm}
  \epsfig{file=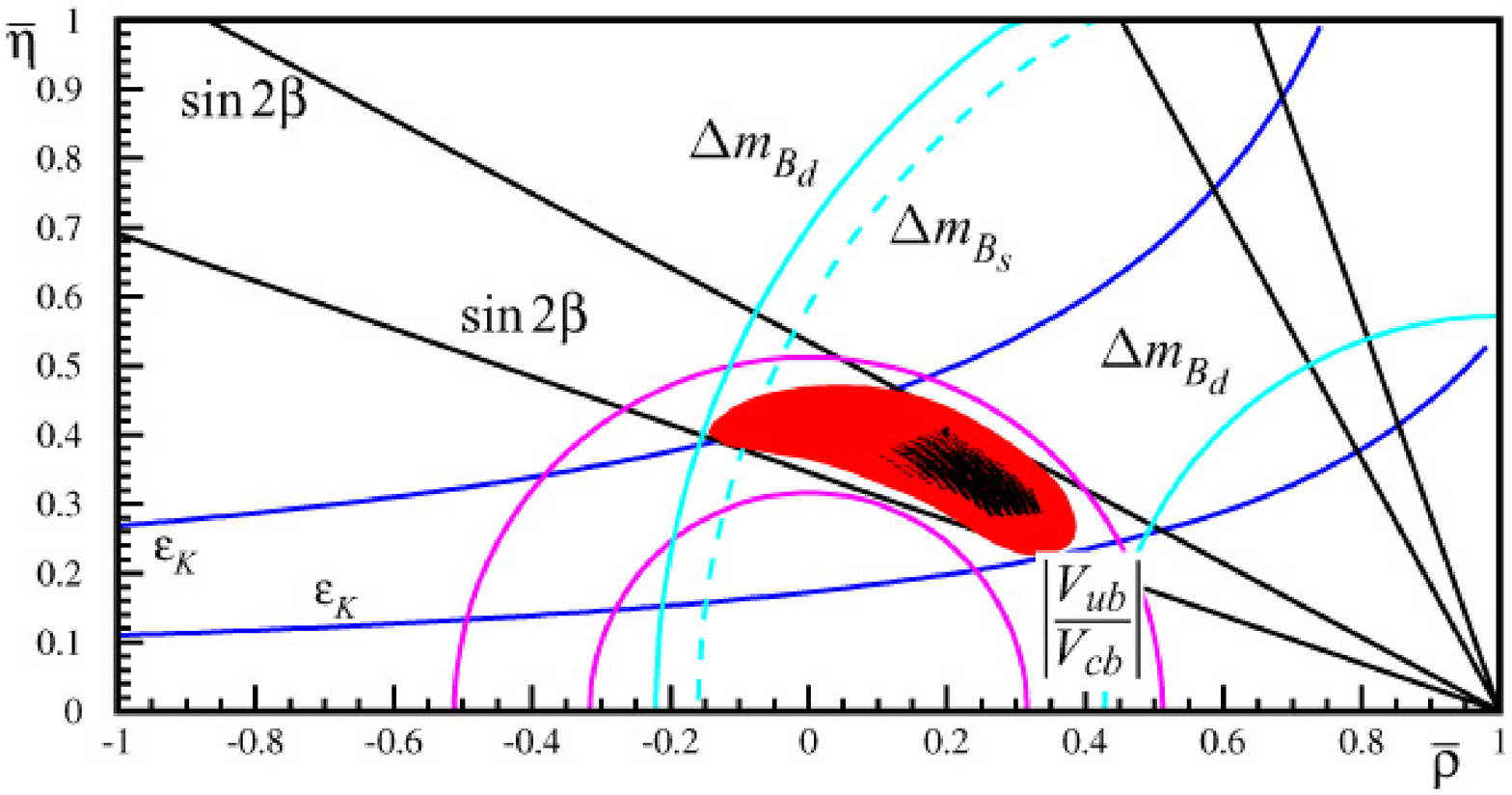,width=3.2in}
 }
\caption{The result of scanning over all theoretical parameters,
including information from both inclusive and exclusive measurements
of $|V_{ub}|$ and $|V_{cb}|$, 
with the $\Delta m_s$ constraint included in the fit (left) 
and excluded from the fit (right).
 }
\label{fig:ut_par_5}
\end{figure*} 

\begin{figure*}
\centerline{
  \epsfig{file=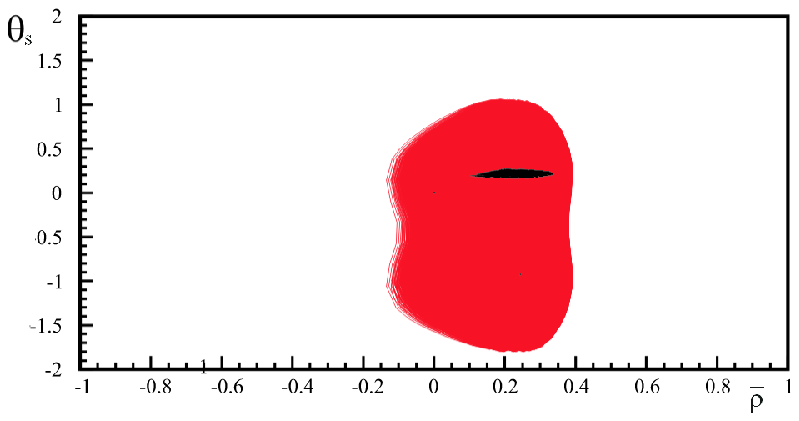,width=3.2in}  
  \hspace{1cm}
  \epsfig{file=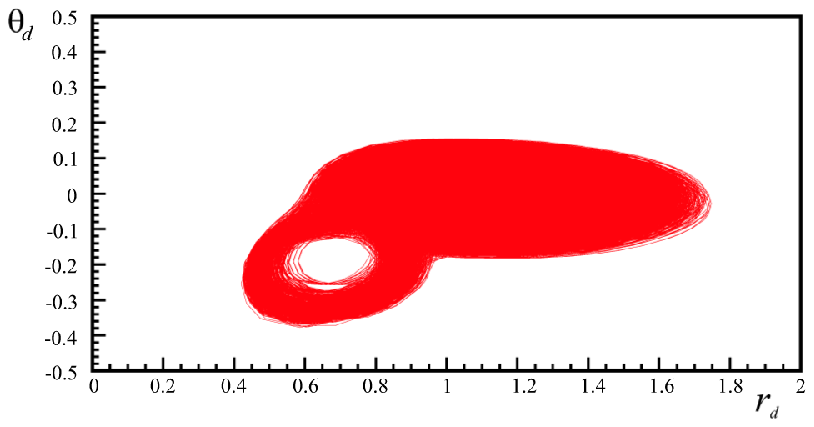,width=3.2in}
 }
\caption{The results for fits that include an extra phase, $\theta_s$, 
for the \CP asymmetry of $B \rightarrow \phi K^0_S$ (left plot),
and for a model-independent analysis that includes an extra scale parameter 
for $B^0_d \bar B^0_d$ mixing, $r_d$, 
and an extra phase for the \CP asymmetry of $B \rightarrow J/\psi K^0_s$, 
$\theta_d$. 
Shown are contours in the $\theta_s$--$\bar\rho$ plane and 
$\theta_d$--$r_d$ plane, respectively.
 }
\label{fig:ut_par_6}
\end{figure*} 

\begin{figure}
\centerline{
  \epsfig{file=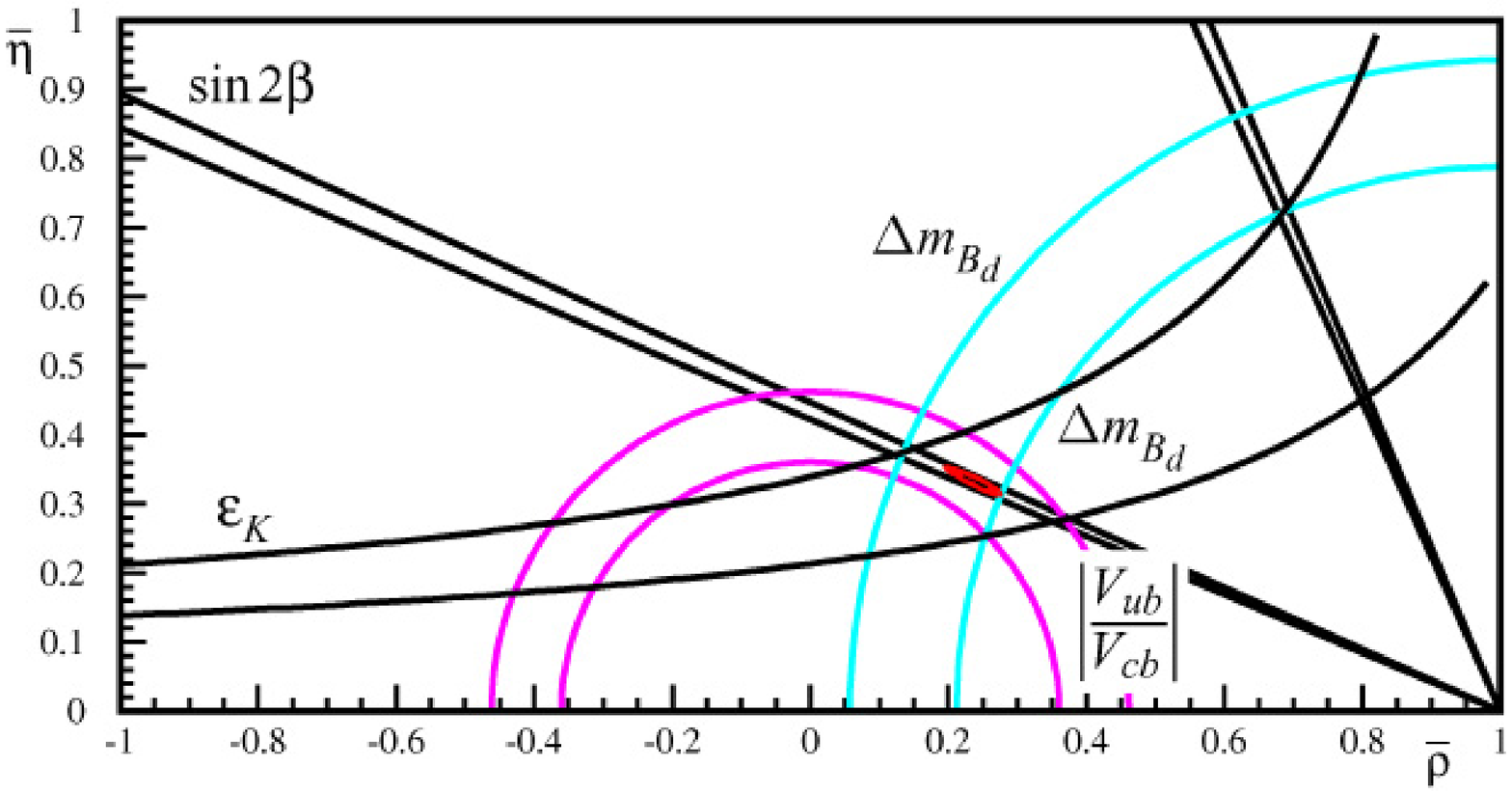,width=3.2in}  
 }
\caption{
A scan over the allowed range of fits for an envisioned future set of 
improved experimental and theoretical results.
 }
\label{fig:ut_par_7}
\end{figure}

\subsubsection{Scan results in the $\bar\rho$--$\,\bar\eta$ plane}

We now turn to scanning all parameters simultaneously within their
theoretically ``allowed'' ranges.
Figures~\ref{fig:ut_par_4} and \ref{fig:ut_par_5} show the resulting contours
for scans in the multi-dimensional model space with some selections on the
included experimental information.
Note that there is no frequency interpretation for comparing which models are
to be ``preferred'', other than the statement that at most one model is 
correct. In this analysis we cannot, and do not, give any relative 
probabilistic weighting among the contours, or their overlap regions.
If we wished to do so, we would proceed to a Bayesian analysis.

From Figure~\ref{fig:ut_par_5} 
we can determine ranges for the CKM parameters
$\bar\rho$, $\bar\eta$, $A$, and $\lambda$ that are consistent with
at least one set of scanned parameters. 
Table~\ref{tab:ranges} summarizes the present results. 
The $\pm1 \sigma$ asymmetric experimental errors are shown separately. 
Furthermore, we obtain ranges for the angles $\alpha$ and $\gamma$, 
as well as results for $m_t, \ m_c$, and $\beta$ 
that are consistent with the input values. 

\subsubsection{Extensions to the standard model}

\babar\ and Belle have measured the \CP asymmetry of 
$B \rightarrow \phi K^0_s$,
yielding a combined value of the sine term in the \CP asymmetry of 
$S_{\phi K^0_S}=-0.39 \pm 0.41$~\cite{bib:phiks}. 
In the standard model (SM) this should be equal to $\sin 2 \beta$ 
to within $\sim\!4\%$.
In models beyond the SM, however, it may differ. 
The present average deviates from $\sin 2 \beta$ by $2.7 \sigma$. 
We can use this result in our fits by adding an additional phase $\theta_s$.
The resulting set of contours in the $\theta_s$--$\bar\rho$ plane is shown in
Figure~\ref{fig:ut_par_6} (left plot). 
Presently, the phase is consistent with zero as expected in SM.

Physics beyond the SM may affect $B^0_d \bar B^0_d$ mixing and \CP violation 
in $B \rightarrow J/\psi K^0_s$ and $B \rightarrow \pi \pi$. 
Using a model-independent analysis we can add a scale parameter, 
$r_d$, for $B^0_d \bar B^0_d$ mixing and an additional phase, $\theta_d$, 
for parameterizing the \CP asymmetry in $B \rightarrow J/\psi K^0_s$. 
The SM is then represented by $r_d=1$ and $\theta_d=0$.
The plot on the right-hand side in Figure~\ref{fig:ut_par_6} 
shows the resulting contours in the $\theta_d$--$r_d$ plane. 
With present uncertainties $r_d$ and $\theta_d$ are consistent with 
the SM expectations.

In order to extrapolate the $\bar\rho$--$\bar\eta$ plane in the future, 
we have assumed that both experimental errors and non-probabilistic 
theoretical uncertainties are reduced according to projections given 
in \cite{bib:Snowmass2001}. 
In addition, we have tuned $V_{ub}$ determined from exclusive and inclusive
measurements to yield a central value that lies inside the $\sin 2 \beta$ 
band, as with the present central values no fit would survive. 
The resulting contours in the $\bar\rho$--$\bar\eta$ plane 
are shown in Figure~\ref{fig:ut_par_7}. 
In this case several sets of parameters are consistent with the SM. 
The main constraint is given by $\sin 2 \beta$.
Measurements of $\gamma$ and $\sin 2 \alpha$ are necessary to reduce the 
range along the $\beta$ rays.

\begin{table*}[htb]
\caption [ ] {Results for selected fit parameters and the angles of the
unitarity triangle from the fits in Figure~\ref{fig:ut_par_5}.
The second and third columns show a lower and upper bound obtained from
scanning all theoretical uncertainties. The fourth and fifth columns
show the asymmetric experimetal errors obtained from the fits.
\label{tab:ranges}}
\begin{center}
\begin{tabular}{@{\extracolsep{2pt}}|@{ }l|l|l|l|l|}
\hline
Variable & Mean value - $\Delta$ & Mean value + $\Delta$ & - $\sigma$ & $+ \sigma$\\
\hline
\hline
$\bar\rho$ & 0.1029  & 0.3372 & -0.0668  & +0.0264 \\
$\bar\eta$ & 0.281   & 0.409  & -0.0203  & +0.0337 \\
$A$        & 0.796   & 0.847  & -0.024   & +0.027 \\
$\lambda$  & 0.2231  & 0.2251 & -0.0032  & +0.0032 \\
$m_t$      & 168.4   & 170.2  & -5.04    & 5.08 \\
$m_c$      & 1.267   & 1.323  & -0.097   & 0.0969 \\
$\beta$    & 0.3619  & 0.4688 & -0.0458  & +0.108 \\
$\alpha$   & 1.451   & 2.0287 & -0.290   & +0.0941 \\
$\gamma$   & 0.70598 & 1.30   & -0.05655 & +0.144 \\
\hline
\end{tabular}
\end{center}
\end{table*}


\section{Visualizing the role of theoretical uncertainties}

Returning to the idealized scenario discussed above, recall that the
results of the scan of CKM fits over the space $T$ include a mapping
of the best-fit value of the fit's consistency statistic as a function
of the $T_\ell$.
This mapping, derived entirely without the use of theoretical
knowledge of these parameters, amounts to a summary of our 
{\it experimental} knowledge of them, assuming the general validity 
of the theoretical framework from which they arise.
It includes the effects of any correlations among the $T_\ell$ arising
from the experimental data.

We can now in principle combine probabilistic and non-probabilistic 
information in a useful way without any need for convolution over
{\it a priori} p.d.f.'s or an equivalent procedure:
the above mapping, can simply be compared directly with the 
theoretically calculated values and uncertainties $\Delta_{T_\ell}$
for the theoretical parameters.

\begin{figure}[hbtp]
 \centerline{
  \epsfig{file=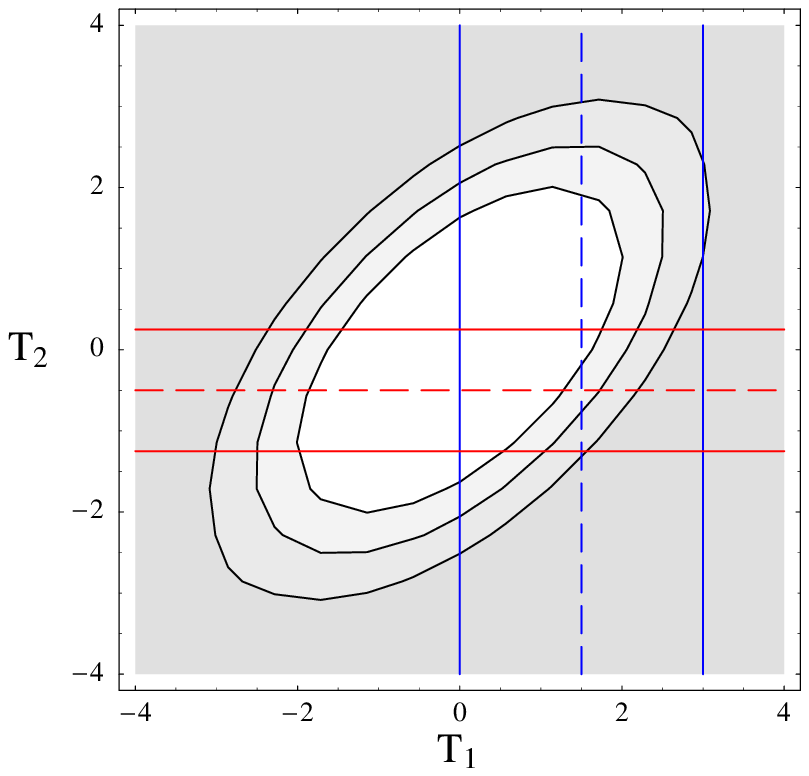,width=3.2in}
 }
 \caption{
   Ideogram of the overlap of the theoretical bounds for two parameters
   $T_1$ and $T_2$ with a set of confidence levels contours obtained by
   scanning a fitting procedure, as described in the text, over the 
   space of the parameters.  The dashed lines show the nominal central
   values of the parameters and the solid lines their bounds.
 }
 \label{fig:ideogram2D}
\end{figure}

This is illustrated conceptually, for the case of two parameters, 
in Figure~\ref{fig:ideogram2D}.  
The curves represent confidence level contours in the space
${T_1,T_2}$ arising from the fit inputs with experimental (and other
probabilistic) uncertainties.  
The straight lines represent the theoretical information and its
uncertainties.
Our central point is that we hold that such a diagram contains an
efficient summary of all the available information (in a two-variable
situation), and that one can go no further in distilling down the
overall consistency of this information without introducing 
additional {\it a priori} information or value judgments. 

The challenge is to find a comparable way to display and understand the
multi-dimensional data set resulting from the actual fits.
Ideally, if we could display the entire 
space, $T$, of the selected parameters at once, no {\it a priori}
assumptions about theoretical uncertainties would be required at all.  
The viewer of such a multi-dimensional analog to Figure~\ref{fig:ideogram2D}
could see at once the entire set of fits
consistent with the experimental data, the effects on this set of the
application of the theoretical bounds, and the correlations of these
effects among the various parameters.

By adding more dimensions to the display for the output parameters of
the fit ({\it e.g.,} for $\bar{\rho}$ and $\bar{\eta}$), 
the data could be displayed as a hyperspatial surface in
whose shape all the correlations of the fitted CKM parameters with the
theoretical inputs would be manifest.  Unfortunately, such a plot is
beyond normal human ability to visualize directly, so some form of
reduction to fewer dimensions is needed.

This makes it necessary, in effect, to integrate over the undisplayed
dimensions' parameters.
In order for this to take into account available theoretical information
on those parameters, as it must in order to be useful, it is here that
{\it a priori} bounds or p.d.f.'s must enter.  
In a Bayesian approach, the integration would be done over an 
{\it a priori} p.d.f.\ as a true convolution.  
In a classical frequentist approach the integration in effect becomes 
a logical OR over all values of the undisplayed parameters that are 
considered ``acceptable''---also an {\it a priori} judgement, 
of course, but a weaker and less specific one than for the Bayesian approach.
As noted above, in either approach once the integration is performed 
the effects of the {\it a priori} input tend to be hidden in 
conventional displays of the fit results.

We have searched for a method that maximizes the amount of
simultaneously accessible information while minimizing the use and
specificity of {\it a priori} inputs, deferring as nearly as possible
until the moment of graphical display any convolution of probabilistic and
non-probabilistic uncertainties, and thus maintaining a principled
separation between these that clearly distinguishes the effects on the
CKM fit of any {\it a priori} assumptions about the theoretical
parameters.
The one we have developed is essentially frequentist in its treatment
of the projections it makes, which we believe best meets our
objectives, though a Bayesian interpretation of projection could also
be applied in what follows, {\it mutatis mutandi}.

\subsection{Methodology}

\begin{itemize}

\item
Pick two of the parameters $T$ for display.  Call these the {\it
primary parameters}, $T_1$ and $T_2$.

\item
Pick a third $T$ parameter, the {\it secondary parameter} $T_s$.  This
parameter is singled out for special attention to the effects of
projecting over it.

\item
Call all the other T parameters the {\it undisplayed parameters}, $T_X$.

\item
For each point P in the grid of scanned values of $T_1 \otimes T_2$, 
a number of fits will have been attempted, 
covering all the scanned values of $T_s$ and the $T_X$.  
For each P, evaluate the following hierarchy of
criteria against the ensemble of results of these fits, deriving 
for the point a value we call the ``Level'', an integer between 0 and 5
inclusive:

\begin{enumerate}

\item 
Define a minimum acceptable value of $P(\chi^2)$.  Typically we
use 1\%, 5\%, or 32\% for this threshold.
Did any of the fits for P pass this cut?  If not, assign Level~=~0 and
stop; otherwise assign Level~=~1 and continue.

\item
Did any of the remaining fits lie within the theoretically preferred 
region for {\it all} the undisplayed parameters \({T_X}\)?  If not, stop;
if yes, assign Level~=~2 and continue.

\item
Did any of the remaining fits have the secondary parameter \({T_s}\) within its theoretically preferred region?  If not, stop;
if yes, assign Level~=~3 and continue.

\item
Did any of the remaining fits have \({T_s}\) equal to its nominal central 
value?  (That value must have been included in the scan grid for this to
make sense.)  If not, stop; if yes, assign Level~=~4 and continue.

\item
Did any of the remaining fits have {\it all} the undisplayed parameters 
\({T_X}\) also at their nominal central values?    If not, stop; 
if yes, assign Level~=~5 and stop.

\end{enumerate}

\item
Now display contours of the quantity Level over the grid in the
\({T_1}\otimes {T_2}\) plane.  We assign a unique color to each
parameter T, so the contours for \({T_s}\) at Level~=~3,4 are
drawn in the color corresponding to that parameter.  The contours for
Level~=~4,5, which correspond to restrictions of parameters
exactly to their central values, are also drawn distinctively, with
dashing.

The Level~3 contour (solid, colored), in particular, displays the
allowed region, at the selected confidence level, for \({T_1}\) and
\({T_2}\), based on the experimental data and on limiting all other
theoretical parameters to their preferred ranges.
Study of the relative spacing of the Level~2, 3, and~4 contours readily 
reveals the effects of the application of the \({T_s}\) bounds on the 
fit results.

\item
Overlay the contours with straight lines showing the theoretically
preferred ranges and nominal central values for \({T_1}\) and
\({T_2}\), in their respective unique colors, again with dashing for the
central value.  This allows the theoretical bounds on \({T_1}\) and
\({T_2}\) to be evaluated directly for consistency against all other 
available data, yet avoiding any convoluted use of priors for these two
parameters.  

Comparison of these theoretical bounds for \({T_1}\) and \({T_2}\)
with the Level~3 contour that shows the experimental information, 
constrained by the application of the theoretical bounds on
\({T_s}\) and the \({T_X}\), allows a direct visual evaluation of the
consistency of all available information, with the effects of the 
application of all theoretical bounds manifest, not obscured by
convolutions performed in the fit itself.

\item
Cyclically permute the set \{\({T_1}\), \({T_2}\), \({T_s}\)\} 
right twice and repeat
the previous procedure for the resulting two permutations.  
Taking all three of the resulting plots together, then, each pair of parameters
in the above set will have been used as the primary parameters once, 
and each parameter will have been used as the secondary once.

\item
Stitch the three resulting plots together, along their common edges,
onto three faces of a cuboid volume and display the result as a
three-dimensional image.  The consistent use of color allows the
effects of the theoretical bounds on all three parameters to be
understood together, despite the large amount of information in the
plot.

\end{itemize}

\subsubsection{The basic two-dimensional visualization}

\begin{figure}[hbtp]
\centerline{
  \epsfig{file=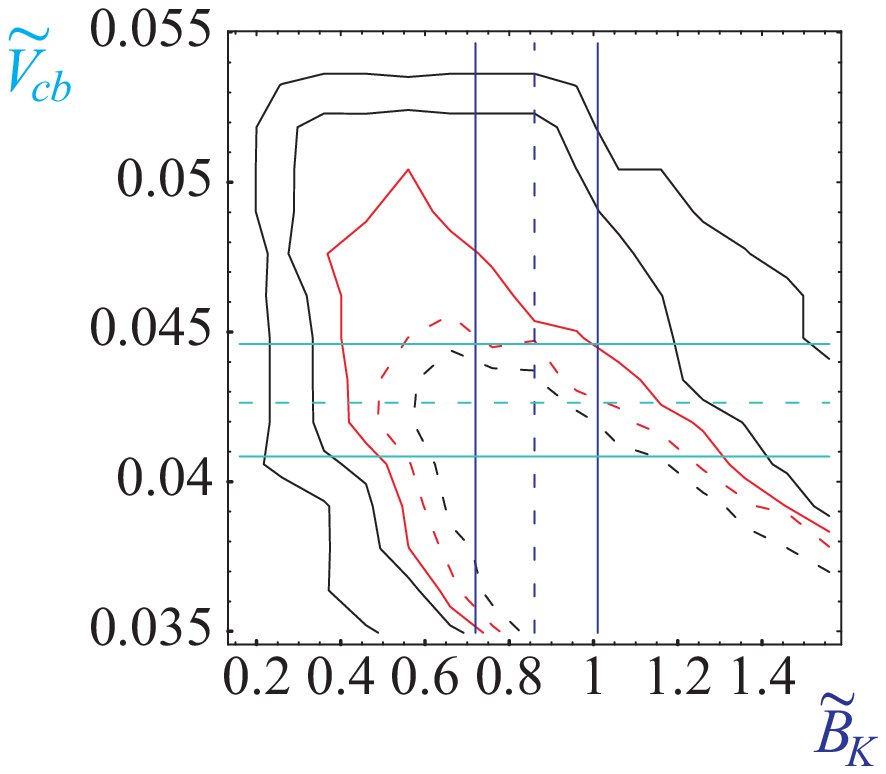,width=3.2in} 
 }
 \caption{Contours of the \emph{``Level''} quantity, defined in the text, 
  in $\widetilde{B}_K$ versus $\widetilde{V}_{cb}$, 
  calculated for a global CKM fit.
  From outside in, the contours are for 
  Level=1) the fit probability cut $P(\chi^2) > 0.32$ (in solid black), 
  2) restricting the ``undisplayed parameters'' $\widetilde\xi$ and 
     $\widetilde{f_B \sqrt{B_B}}$ to the ranges
     $[ 1.18, 1.30 ]$ and $[ 211, 235 ]$, respectively (solid black), 
  3) restricting the ``secondary parameter'' $\widetilde{V}_{ub}$ to the range
  $[ 0.0028, 0.0038 ]$ (solid red), 
  4) restricting it to its nominal central value of $0.00332$ (dashed red), and
  5) similarly restricting $\widetilde\xi$ to $1.24$
     and $\widetilde{f_B \sqrt{B_B}}$ to $0.223$ (dashed black).
  The orthogonal bands show the theoretically preferred ranges and 
  nominal central values for $\widetilde{B}_K$ (in blue) and
  $\widetilde{V}_{cb}$ (in green).
  The fit uses only information from exclusive measurements of
  $|V_{ub}|$ and $|V_{cb}|$.
 } 
\label{fig:vis-twod}
\end{figure}

We first illustrate the concept with a single two-dimensional plot
resulting from the procedure described above.
Figure~\ref{fig:vis-twod}
shows the results of this procedure for the standard inputs used
in this paper, for simplicity using only exclusive measurements for
$|V_{ub}|$ and $|V_{cb}|$.
The primary parameters \({T_1}\) and \({T_2}\) are
$\widetilde{B}_K$ and $\widetilde{V}_{cb}$, 
shown in blue and green, respectively.  The orthogonal bands show their
theoretically preferred ranges and nominal central values.

\ignore{
\begin{figure}[hbtp]
 \centerline{
  \epsfig{file=doe_ckm_twod.eps,width=3.2in}
 }
 \caption{
  Contours of the \emph{``Level''} quantity, defined in the text, 
  in $f_{B_d}\sqrt{B_{B_d}}$ versus ``pseudo-$V_{ub}$'', 
  calculated for a global CKM fit excluding $\sin 2\beta$ information.
  From outside in, the contours are for 
  Level=1) the fit probability cut $P(\chi^2) > 0.32$ (in solid black), 
  2) restricting the ``undisplayed parameter'' $\xi^2$ to the range 
     $[ 1.18, 1.43 ]$ (solid black), 
  3) restricting the ``secondary parameter'' $B_K$ to the range
  $[ 0.74, 1 ]$ (solid blue), 
  4) restricting it to its nominal central value of $0.87$ (dashed blue), and
  5) similarly restricting $\xi^2$ to $1.3$ (dashed black).
 }
 \label{fig:ckm_twod}
\end{figure}
} 

Recall that the symbols $\widetilde X$, as previously defined,
represent the non-probabilistic part of our lack of knowledge 
of the true parameters $X$.


The secondary parameter $T_s$ is $\widetilde{V}_{ub}$, 
and the undisplayed scanned parameters $T_X$ in this case are 
$\widetilde\xi$ and $\widetilde{f_B \sqrt{B_B}}$.
In this fit, and the others in this section, the QCD $\eta$ 
parameters are incorporated into the fit with probabilistic 
uncertainties, not scanned, as a simplification.

We have performed the fits scanned over a grid of all five of
these variables, expanded by a factor of five from their 
theoretically preferred ranges.
A strength of the present method is precisely that it permits the ready 
visualization of the results of going outside the nominal bounds.

The contours enclose successive values of the \emph{``Level''} quantity, 
as defined above.
The outermost, solid black contour surrounds the region of Level~1,
in which fits are required only to pass a $P(\chi^2) > 0.32$ cut;
the next, also solid black contour surroundsthe Level~2 region, which now
limits fits to those with the ``undisplayed parameters'' 
$\widetilde\xi$ and $\widetilde{f_B \sqrt{B_B}}$ within their nominal
theoretical ranges.  The third contour surrounds the region of 
Level~3, where the ``secondary parameter'' $\widetilde{V}_{ub}$
is limited to its nominal range, and is drawn in a solid color, in this
case red.

This contour represents our derived \emph{experimental} knowledge of 
$\widetilde{B}_K$ and $\widetilde{V}_{cb}$ given the theoretical
constraints on the other parameters.
It can be compared with the  \emph{theoretical} contraints on
$\widetilde{B}_K$ and $\widetilde{V}_{cb}$, show as the solid colored
orthogonal bands.
In this case, they overlap perfectly: thus we can see immediately that, 
given the inputs to these fits, our present experimental constraints on 
these parameters are strictly less restrictive than the 
present state of theoretical knowledge.

The dashed contours display the fit results when the theoretical inputs
are constrained to their nominal central values; first, at Level~4, 
for $\widetilde{V}_{ub}$ fixed to $0.00332$, drawn in dashed color,
and finally at Level~5 fixing all the remaining parameters 
($\widetilde\xi$ to $1.24$ and $\widetilde{f_B \sqrt{B_B}}$ to $0.223$), 
shown in dashed black.
Note the consistent use of dashing to represent central values of the 
parameters.

The fact that the dashed lines for $\widetilde{B}_K$ and $\widetilde{V}_{cb}$
cross within the innermost dashed contour is a representation of the fact
that a successful fit ($P(\chi^2) > 0.32$) is obtained even when all the 
scanned parameters are set to their central values---the set of all the
nominal central values is self-consistent.


\subsubsection{The three-dimensional triplet plot}

\begin{figure*}[p]
\centerline{
   \epsfig{file=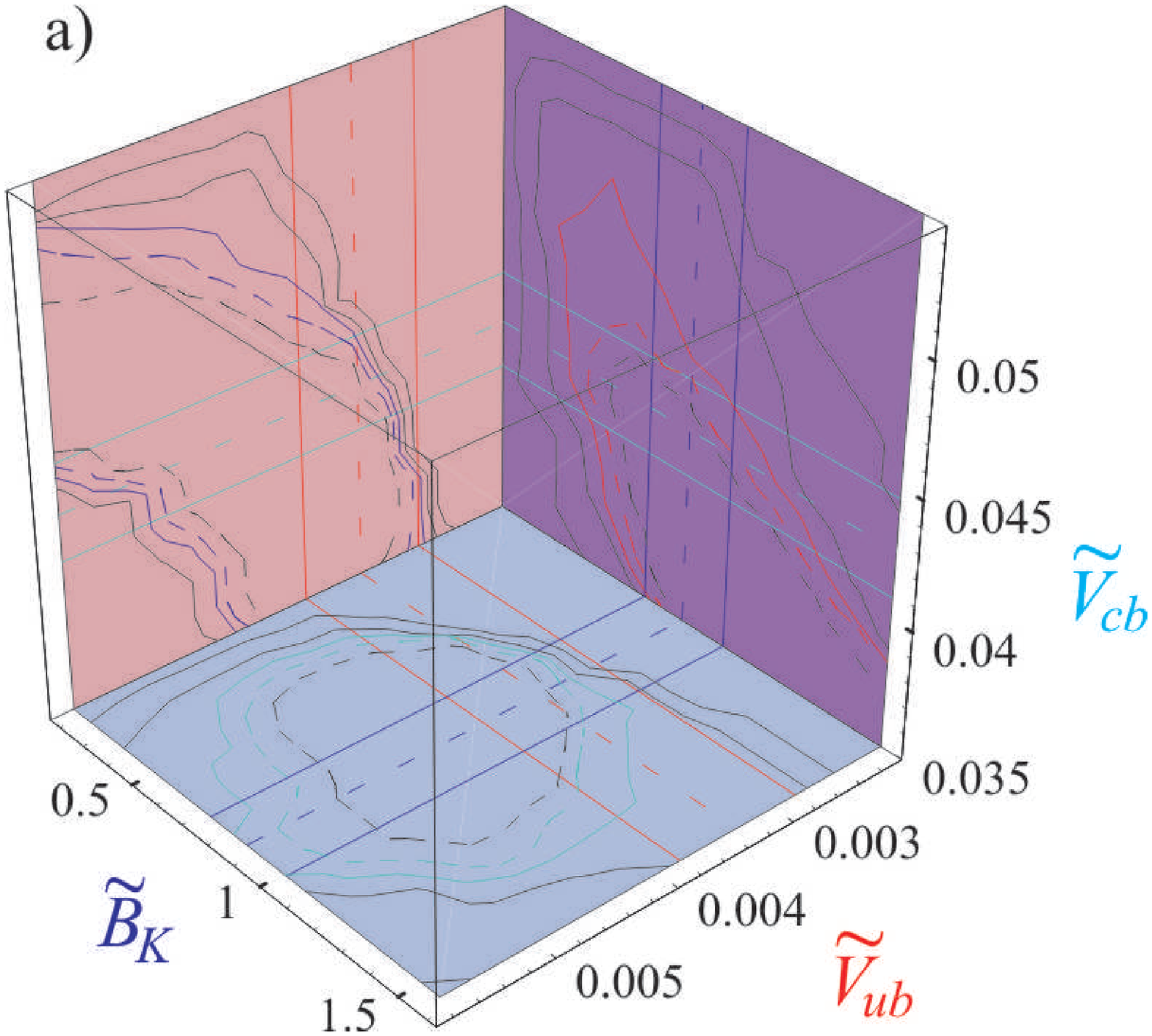,width=3.0in} 
  \hspace{0.5cm}
  \epsfig{file=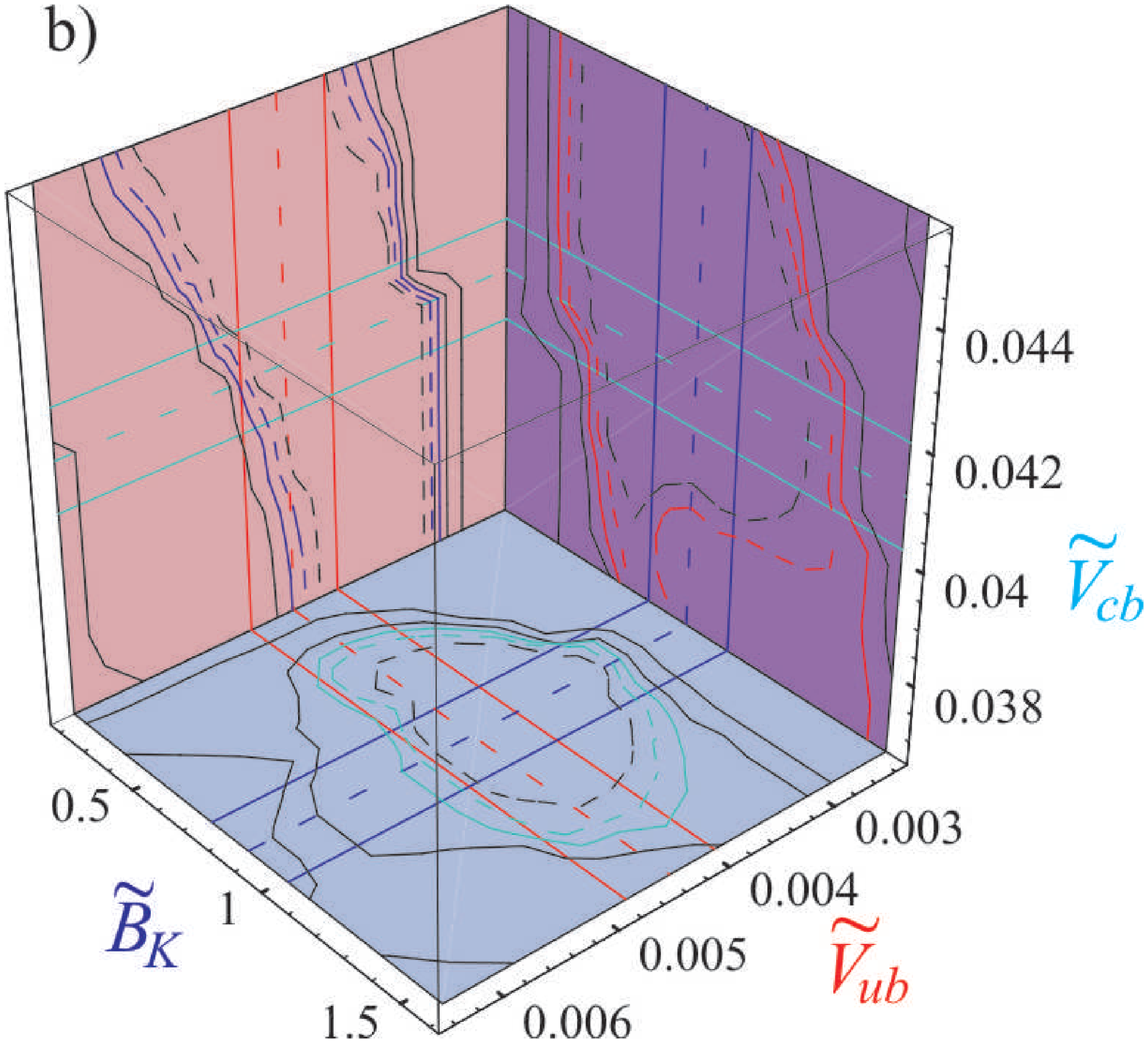,width=3.0in}
 }
\centerline{
  \epsfig{file=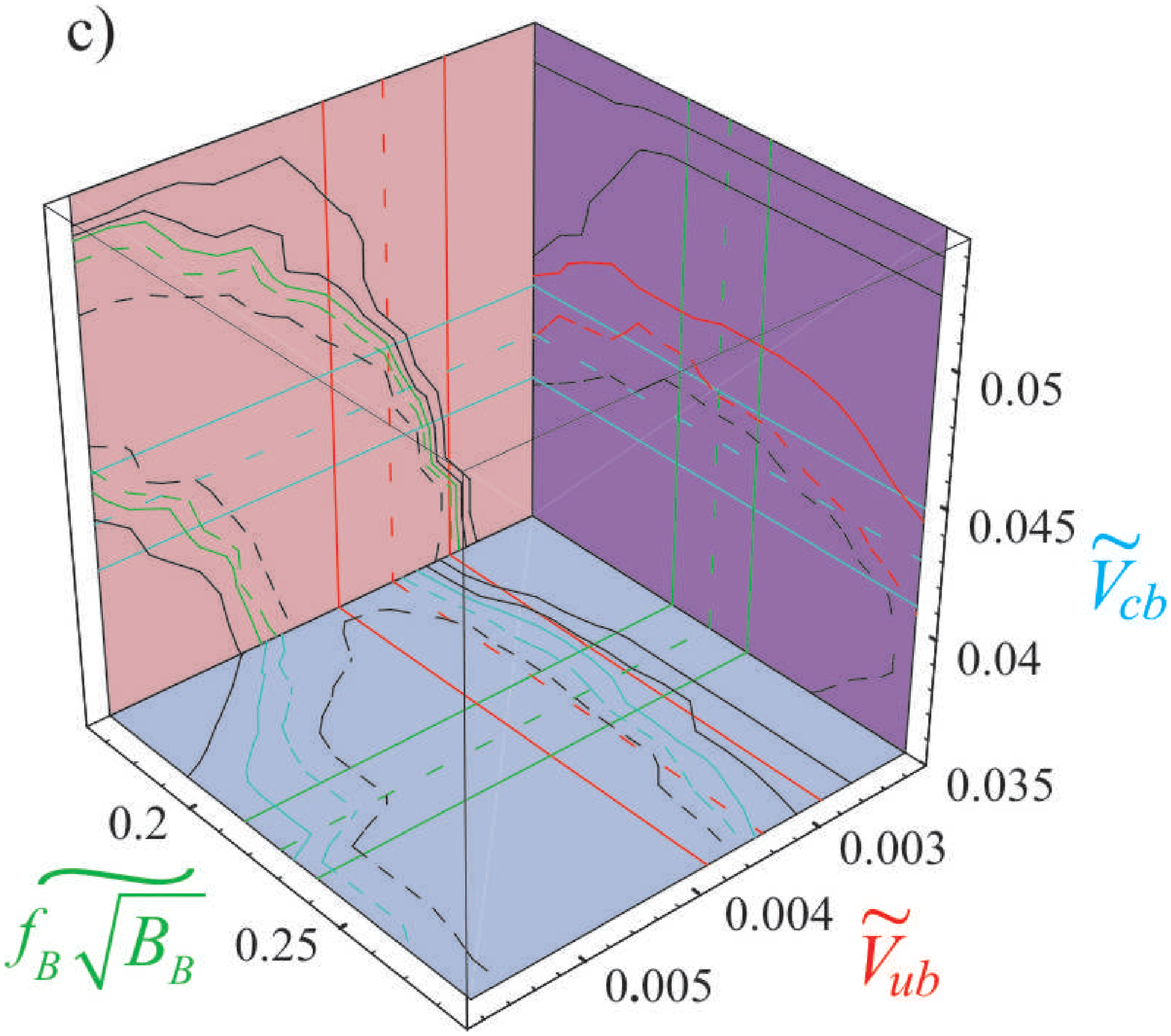,width=3.0in}  
  \hspace{0.5cm}
  \epsfig{file=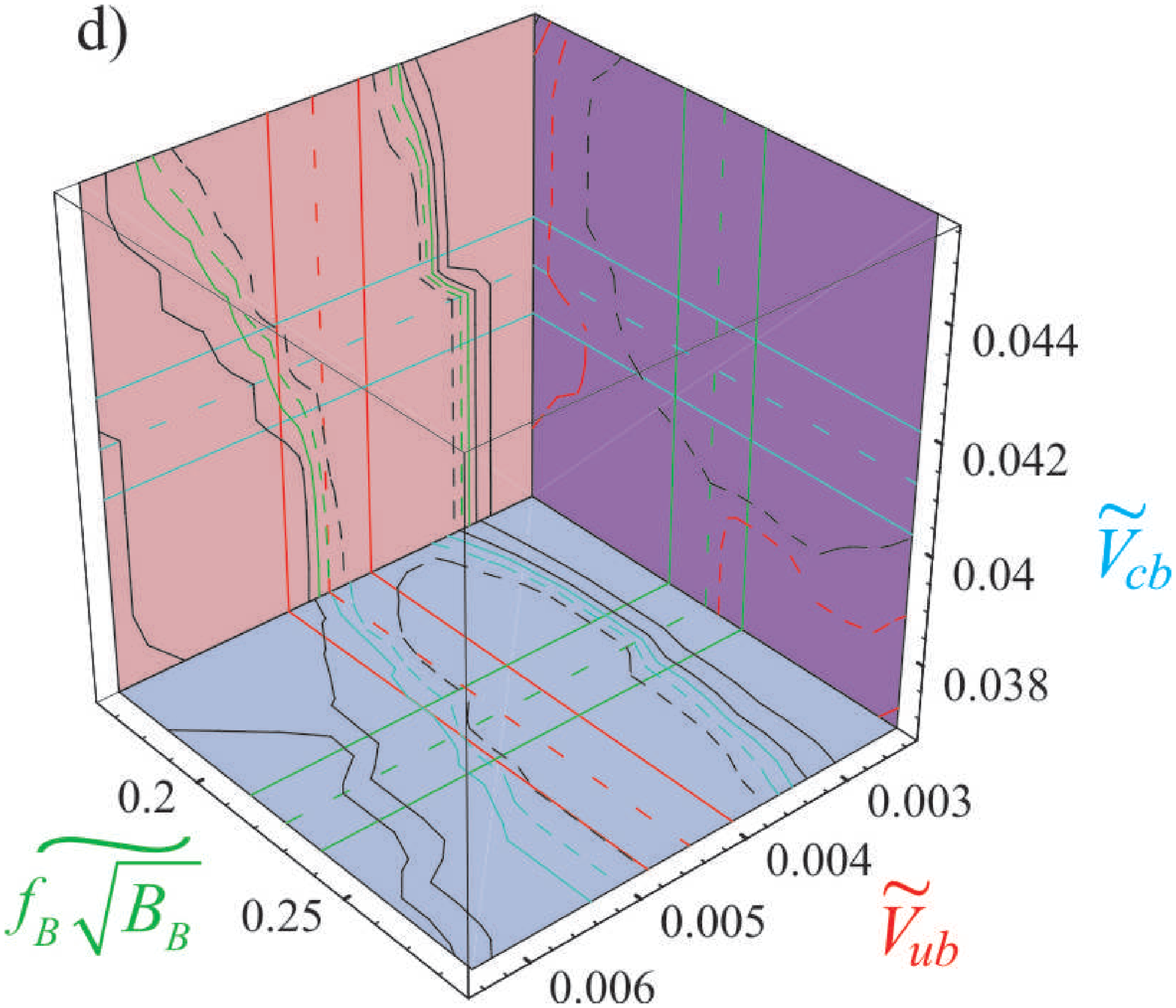,width=3.0in} 
 }
\centerline{
  \epsfig{file=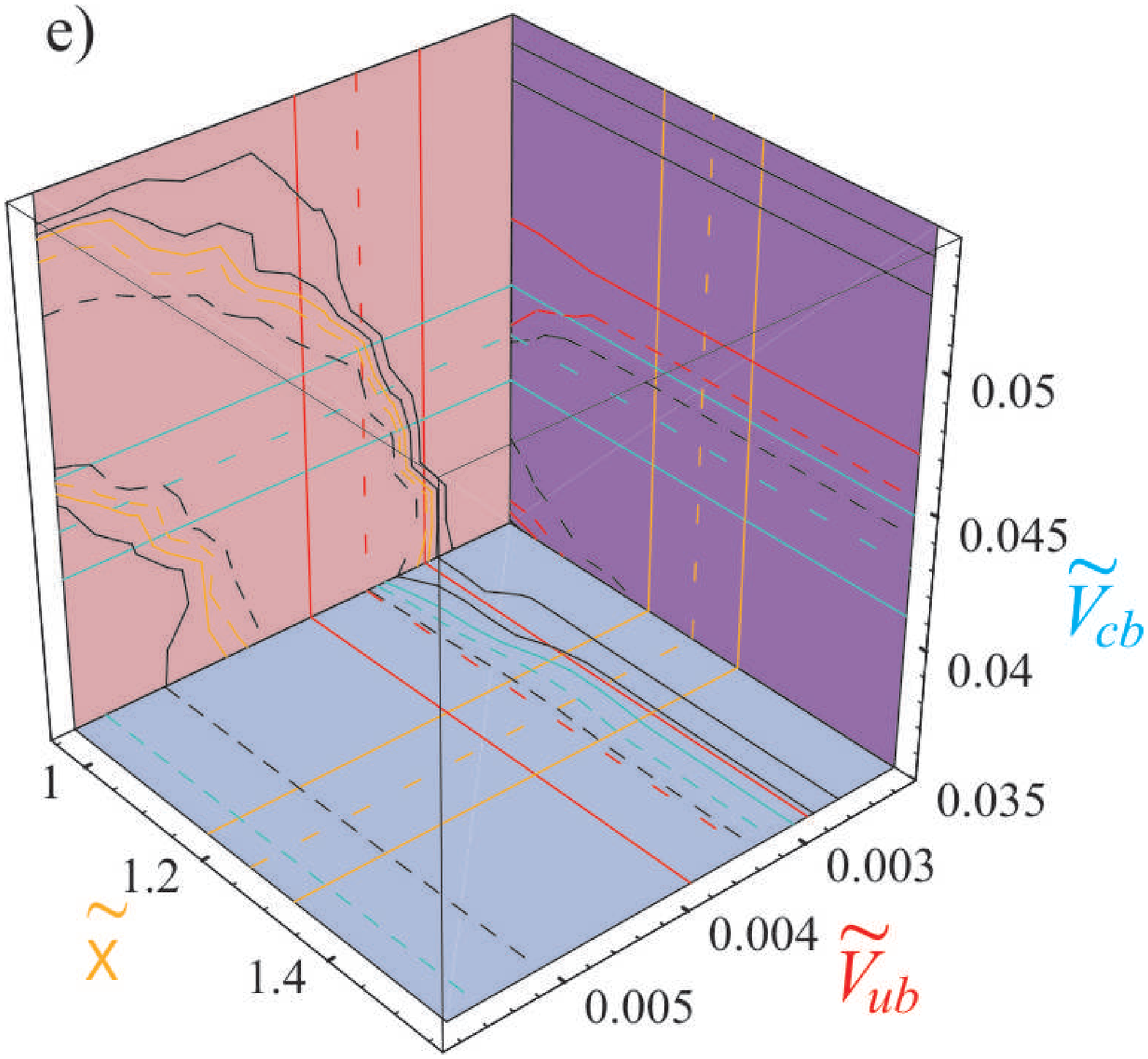,width=3.0in}  
  \hspace{0.5cm}
  \epsfig{file=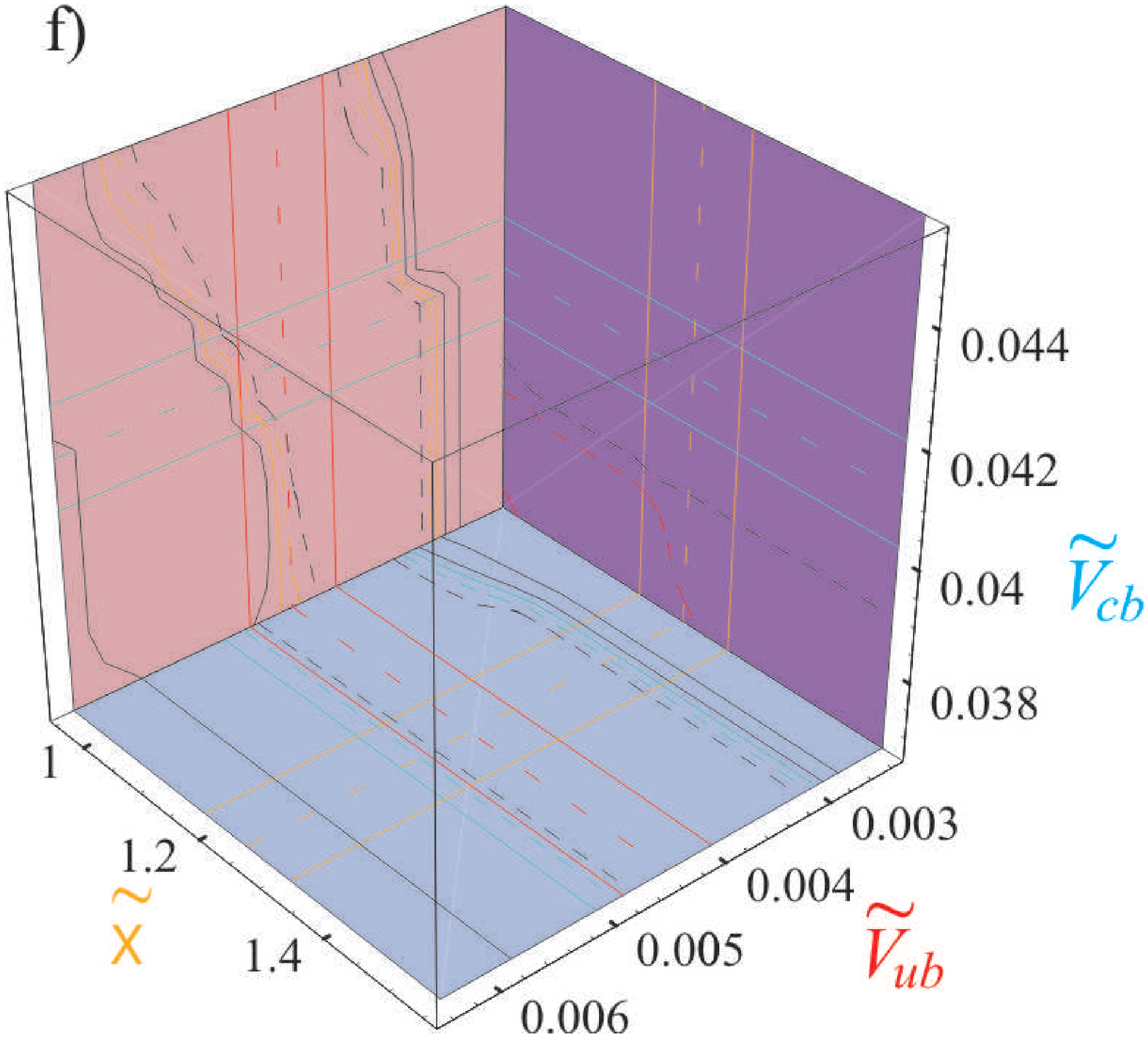,width=3.0in} 
 }
 \caption{
  ``Triplet plots'', as defined in the text, 
  for, top to bottom, various sets of three of the five scanned parameters and,
  left to right, for fits using only exclusive or only inclusive
  measurements of $|V_{ub}|$ and $|V_{cb}|$.
  Each triplet plot is a combination of graphs of the kind shown in 
  Figure~\ref{fig:vis-twod}, for all three combinations of a set of
  three theoretical parameters.
  The fit probability requirement remains $P(\chi^2) > 0.32$, and the
  theoretical uncertainty ranges and nominal values are those given
  elsewhere in the text.
 }
\label{fig:vis_2}
\end{figure*}

The full power of the method becomes apparent when carrying out the last
step in the program described above and considering the
three-dimensional plots constructed by assembling the three two-dimensional
plots for the pairwise cyclic permutations of a set of three variables.
Figure~\ref{fig:vis_2}a
shows the results of this for the same set of variables as in 
Figure~\ref{fig:vis-twod};
thus, the latter figure appears as one of the faces of the cuboid in
Figure~\ref{fig:vis_2}a.

The consistent use of color and texture coding is important to the 
comprehension of the plots.  Most importantly, in each face of a plot,
the colored contours show the effect of the theoretical restrictions 
on the parameter displayed perpendicularly to it --- and the same
line styles are used to display the restriction itself on the perpendicular
planes.

\ignore{
\begin{figure}[hbtp]
 \centerline{
  \epsfig{file=doe_ckm_threed.eps,width=3.2in}
 }
 \caption{
  Combination of graphs of the kind shown in Figure~\ref{fig:ckm_twod}
  for all three combinations of the variables ``pseudo-$V_{ub}$'',
  $f_{B_d}\sqrt{B_{B_d}}$, and $B_K$, showing correlated effects of 
  restrictions on each variable and on the ``undisplayed parameter'' $\xi^2$.
  The upper plot is for the global CKM fit excluding $\sin 2\beta$;
  in the lower plot the February 2002 world average of $0.79 \pm 0.17$
  has been included.  The theoretical bounds and nominal central values
  used for the parameters, in the form ``[lower, central, upper],'' are:
  ``pseudo-$V_{ub}$:'' [0.0027, 0.00325, 0.0038], 
  $f_{B_d}\sqrt{B_{B_d}}$: [0.21, 0.23, 0.25] GeV,
  $B_K$: [0.74, 0.87, 1],
  $\xi^2$: [1.18, 1.3, 1.43].  
  The fit probability requirement remains $P(\chi^2) > 0.32$.
 }
 \label{fig:ckm_threed}
\end{figure}
} 

This graphical presentation clearly exhibits
the correlations among parameters and measurements.  
For example, it is possible to tell by inspection in 
Figures~\ref{fig:vis_2}a,c,e that experimental data from exclusive
measurements is coming
to constrain the low side of $|V_{ub}|$ as much as or more than
the inputs from theory, with inclusive data similarly constraining 
the high side.

Further, there is now a region of low $|V_{ub}|$ versus high $|V_{cb}|$ 
that is excluded on experimental grounds alone, without the need for 
any of the theoretical inputs here considered.
In general, the correlation forced on $|V_{ub}|$ versus $|V_{cb}|$ by
the experimental data is readily apparent.

We can observe that $\widetilde\xi$ is entirely unconstrained by
experimental data within a range more than five times larger than its
present theoretical uncertainty.  Thus, the theoretical input on $\xi$
is seen to be of particular importance.

$B_K$ is seen to be constrained both from above and below by the
combination of experimental data and the theoretical input for 
$|V_{cb}|$, to a region two to three times larger than the
present theoretical uncertainty.  This can be seen as a measurement of
this quantitity, albeit one still dependent on the theoretical
input for $|V_{cb}|$.

These are not altogether novel observations --- they follow from the 
structure of the underlying theory, though the details are 
a quantitative matter --- but we believe that this visualization method
demonstrates them particularly clearly.  It seems a very useful tool
in understanding the other CKM analyses, with their typically opaque
convolution over the theoretical uncertainties.

It also permits the use of physical judgement in evaluating the meaning
of theoretical uncertainties: should a future version of one of these
plots reveal an inconsistency of model and data, readers concerned about
the role of theoretical uncertainties in that inconsistency
could simply read it off a figure, noting that, say, a modest shift in
a particular parameter could resolve the discrepancy.

\subsection{Extending to $\bar{\rho}$ and $\bar{\eta}$}

The method can be extended to include the Wolfenstein parameters
$\bar{\rho}$, $\bar{\eta}$, and $A$, or any other results of the fit,
in the visualization of the results.
Presentation in terms of $\bar{\rho}$ and $\bar{\eta}$, in particular,
permits making a connection with the
most common displays resulting from other CKM fitting techniques in
the literature, and allows direct visualization of the
effects of various theoretical bounds on the values of 
the Wolfenstein parameters.

However, for each of those parameters that appears in a graph
we must remove one of the theoretical parameters into the less readily
understood ``undisplayed parameters'' set and project over it, using
some sort of convolution over its assumed distribution.
This conflicts with the original rationale for our method, which is to
minimize as much as possible the use of hidden {\it a priori}
information, as in these projections.  Thus the extension of these
visualization methods to the Wolfenstein parameters represents a
compromise between objectives.

With that in mind, 
the methods we have developed nonetheless can be used to make triplet plots
including one, two, or even all three of the Wolfenstein parameters.
Since they are outputs of the fit, they are continuous variables
and do not form a grid the way the scanned parameters do.  
We simulate this by rounding them off to a grid---in effect, by binning
them---before generating the contours for display.

Because the Wolfenstein parameters do not have independent theoretical
bounds that can be used as cuts or displayed in a graphic, they cannot
serve exactly as the ``secondary parameters'' in the scheme as described above.
So, when one of them does appear in that position, we treat it as if
it were a secondary parameter without any cuts to be applied to it, and
show only the Level~1, 2, and~5 contours in the plane to which it is
orthogonal.  While we also cannot show the solid-color-edged band that we
use for theoretical bounds, but we do show a nominal value for each as
a colored, dashed line, purely as a guide to the eye.

\begin{figure*}[p]
\centerline{
   \epsfig{file=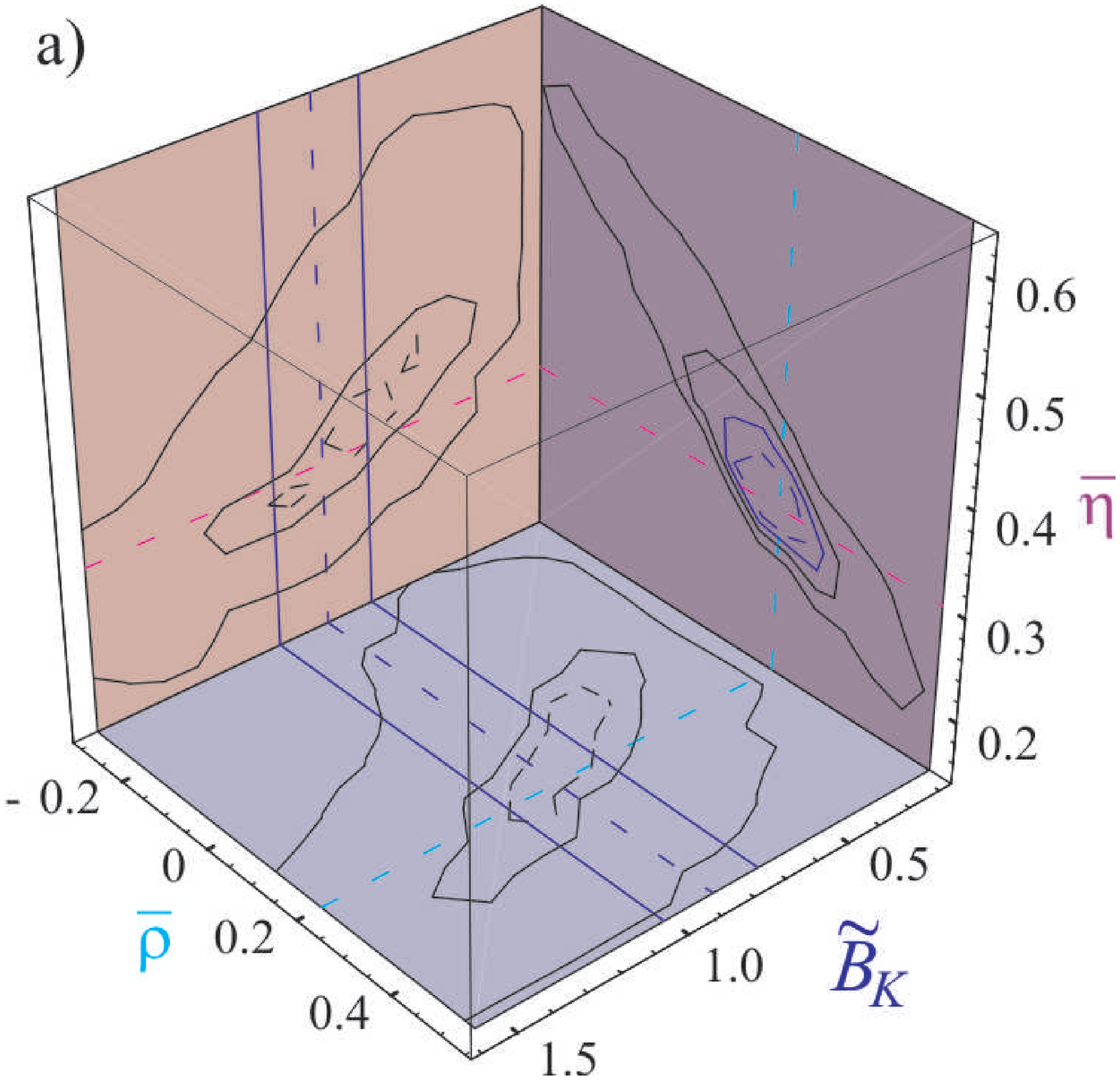,width=3.0in} 
  \hspace{0.5cm}
  \epsfig{file=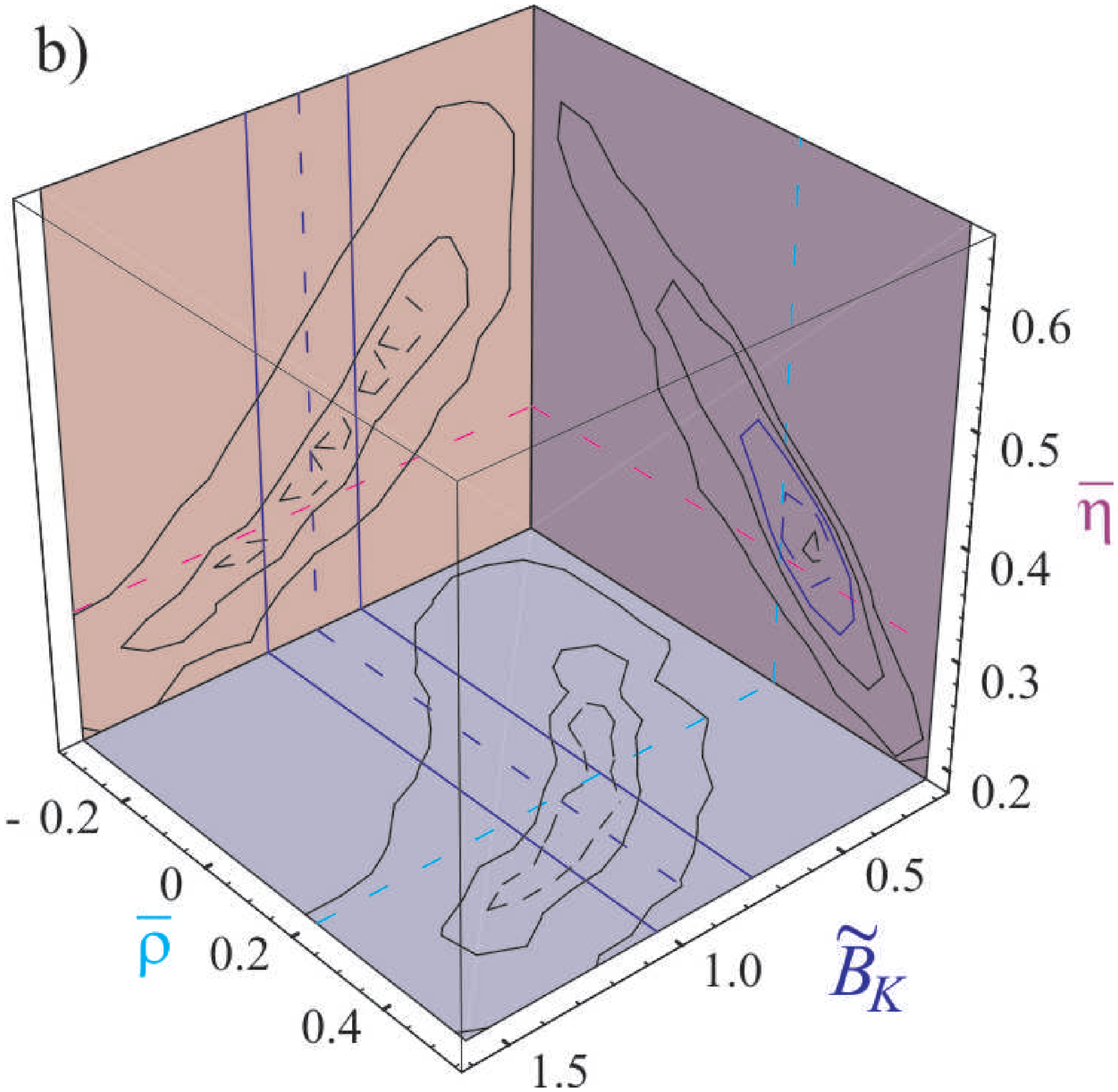,width=3.0in}
 }
\centerline{
  \epsfig{file=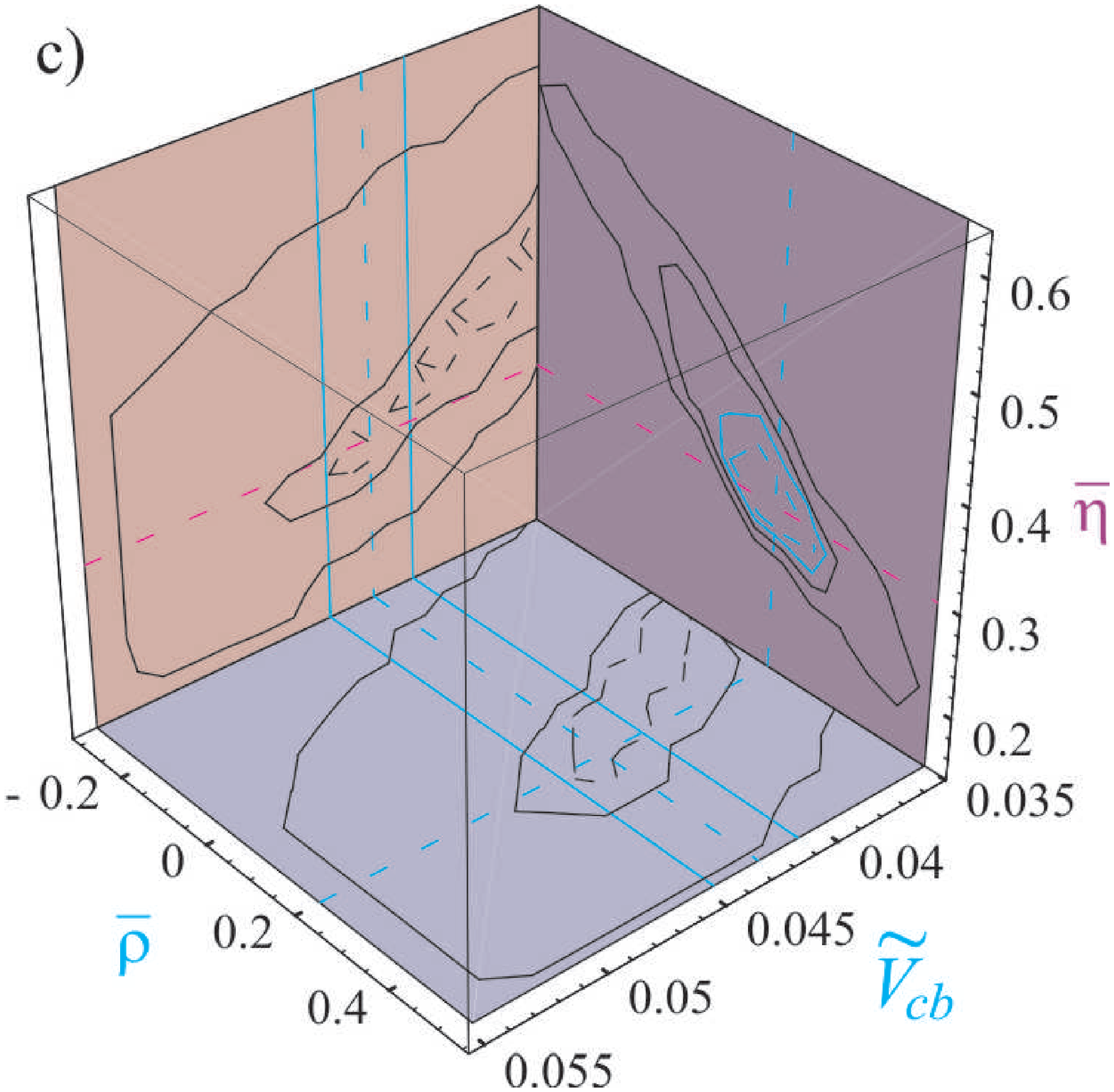,width=3.0in}  
  \hspace{0.5cm}
  \epsfig{file=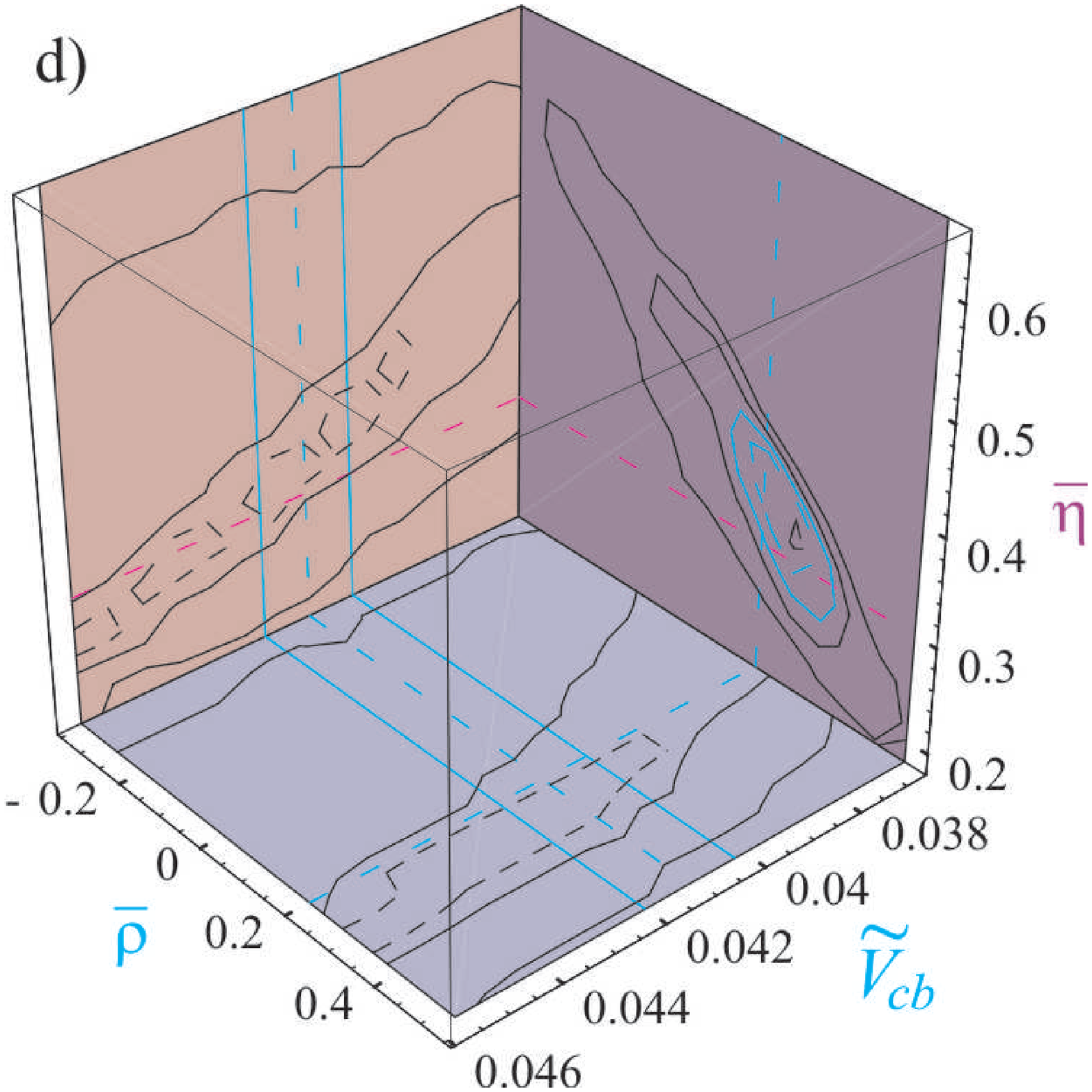,width=3.0in} 
 }
\centerline{
  \epsfig{file=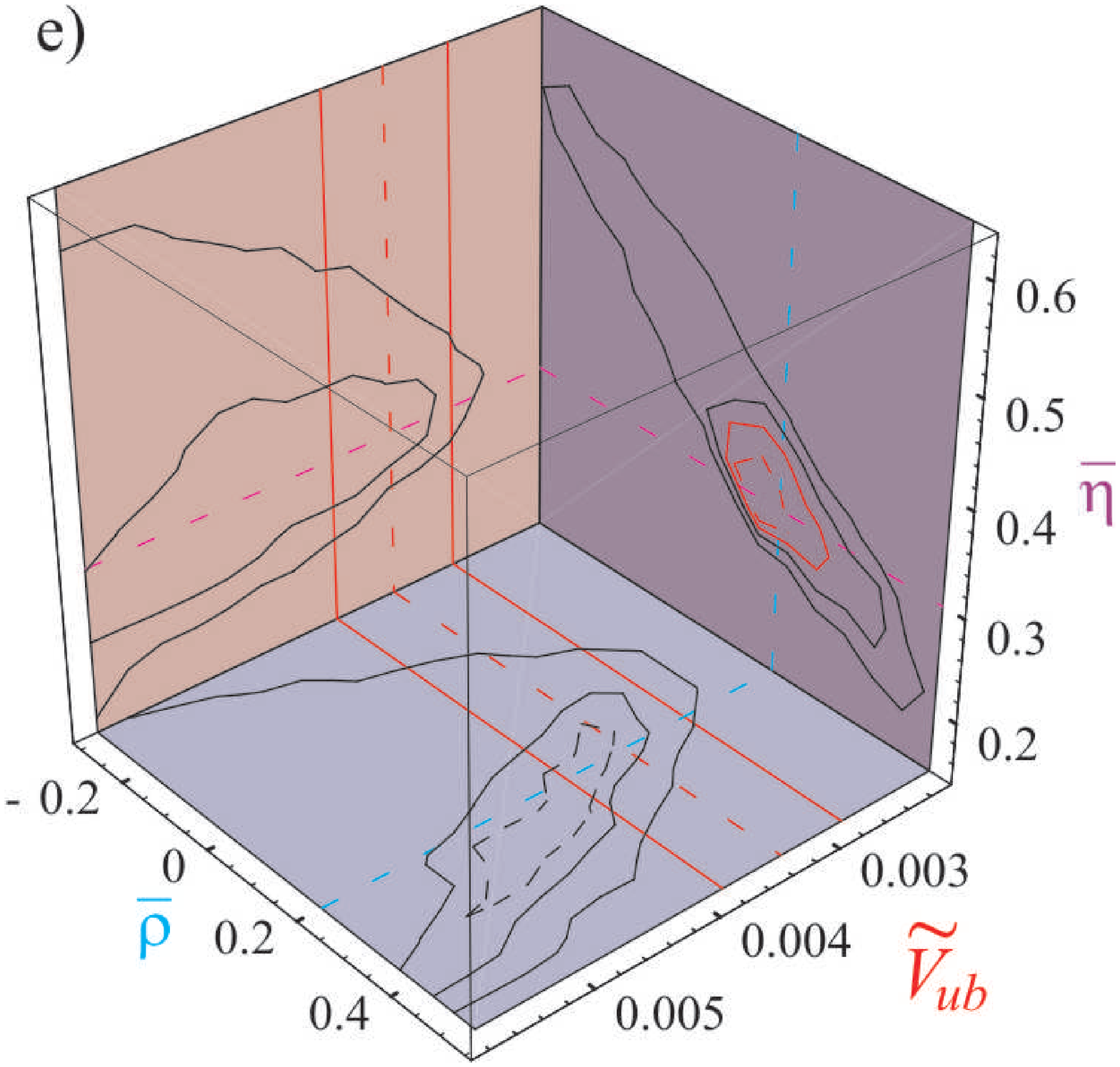,width=3.0in}  
  \hspace{0.5cm}
  \epsfig{file=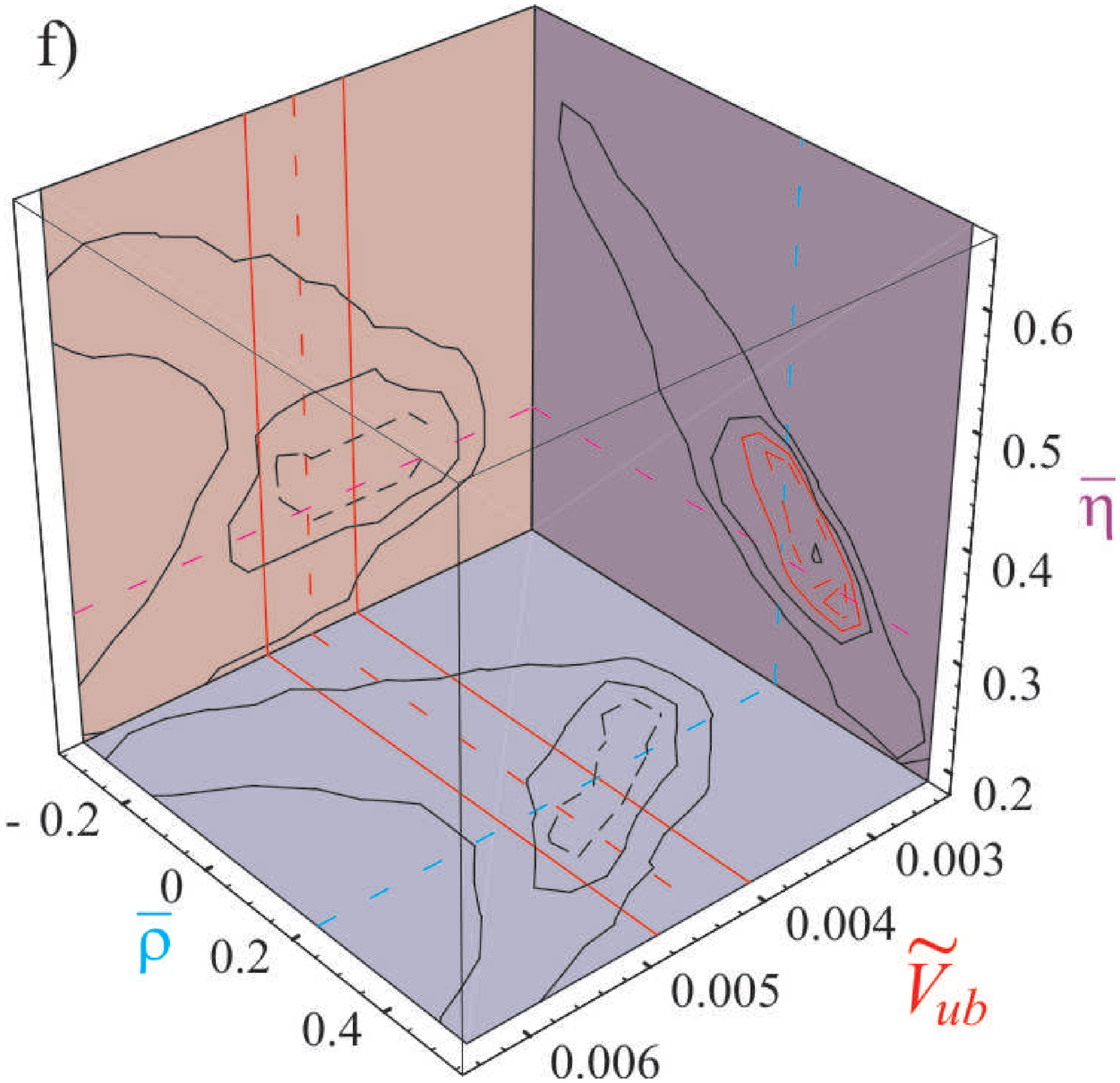,width=3.0in} 
 }
 \caption{
  ``Triplet plots'', as defined in the text, 
  showing the dependence of the fit outputs $\bar\rho$ and $\bar\eta$
  on a single scanned theoretical parameter:
  from top to bottom, $\widetilde{B_K}$, $\widetilde{V_{cb}}$,
  and $\widetilde{V_{ub}}$, 
  and from left to right, for fits using only exclusive or only inclusive
  measurements of $|V_{ub}|$ and $|V_{cb}|$.
  The dashed lines drawn for $\bar\rho$ and $\bar\eta$ are at purely
  indicative values of $0.2$ and $0.3$, respectively, and are there
  only to help guide the eye.
  The fit probability requirement remains $P(\chi^2) > 0.32$, and the
  theoretical uncertainty ranges and nominal values are those given
  elsewhere in the text.
 }
\label{fig:vis_3}
\end{figure*}

Figure~\ref{fig:vis_3} 
shows the results of applying this technique to displays of
a single theoretical parameter against the two Wolfenstein parameters
$\bar{\rho}$ and $\bar{\eta}$, with one plot shown for each of a selection
of three of the five theoretical parameters scanned, 
$\widetilde{B}_K$, $\widetilde{V}_{cb}$, and $\widetilde{V}_{ub}$.

In each of the graphs, the $\bar{\rho}$--$\bar{\eta}$
plane displayed at the right rear of the three dimensional volume
shows the effects of including only fits with $P(\chi^2)>0.32$ 
(outer black countour, Level~1), limiting the graph's four undisplayed
parameters to their theoretical bounds (inner black contour, Level~2),
and limiting the graph's displayed theoretical parameter to its bounds
(solid colored contour, Level~3),
and to its nominal central value (dashed colored contour, Level~4).
The nominal values 
used for the ``guide lines'' in the figure are 
$\bar\rho = 0.2$ and $\bar\eta = 0.3$.

\ignore{
\begin{figure*}[p]
 \centerline{
  \epsfig{file=doe_ckm_rhoeta.eps,width=7in}
 }
 \caption{
  Graphs of the kind shown in Figure~\ref{fig:ckm_threed}
  for a global CKM fit based on February 2002 experimental data,
  showing the dependence of the fit outputs for the Wolfenstein
  parameters $\bar{\rho}$ and $\bar{\eta}$ on the four theoretical
  parameters 
  ``pseudo-$V_{ub}$,'' $f_{B_d}\sqrt{B_{B_d}}$, $B_K$, and $\xi^2$.
  In each of the four graphs, the $\bar{\rho}$--$\bar{\eta}$
  plane displayed at the right rear of the three dimensional volume
  shows the effects of including only fits with $P(\chi^2)>0.32$ 
  (outer black countour, Level~1), limiting the graph's three undisplayed
  parameters to their theoretical bounds (inner black contour, Level~2),
  and limiting the graph's displayed theoretical parameter to its bounds
  (solid colored contour, Level~3),
  and to its nominal central value (dashed colored contour, Level~4).
  Purely to guide the eye, lines for nominal values $\bar\rho = 0.2$ and 
  $\bar\eta = 0.3$ are included in the figure.
 }
 \label{fig:ckm_rhoeta}
\end{figure*}
}

Strikingly clear from these figures is the power and the independence
from theoretical inputs of the $\sin 2\beta$ constraint.
In each plot in Figure~\ref{fig:vis_3} a clear radial band from 
$( \bar{\rho}, \bar{\eta} ) = (1,0)$ corresponding to this
constraint is visible, even for the outermost contour, which reflects
scanning over all theoretical ranges enlarged by a factor of five.

Also readily seen is the relative role of the precision of the predictions
for each theoretical parameter in constraining $\bar{\rho}$ and $\bar{\eta}$,
especially radially.  This can be observed by noting the difference in
size of the Level~2 (second solid black) and Level~3 (solid color)
contours.  The former shows the effect of applying our theoretical
knowledge of all but the selected parameter, and the latter that of
adding to that our knowledge of that parameter.

Thus, for instance, in comparing Figures~\ref{fig:vis_3}b, \ref{fig:vis_3}d,
and~\ref{fig:vis_3}f, based on fits to inclusive $|V_{xb}|$ data,
it is easy to see that theoretical input on
$\widetilde{B}_K$ plays the largest role in constraining 
$\bar{\rho}$ and $\bar{\eta}$, while the theoretical input for
$\widetilde{V}_{cb}$ (for inclusive measurements) has little effect
once the other theoretical parameters have been determined.
(The corresponding plots, not included here, for $\widetilde{f_B \sqrt{B_B}}$
and $\widetilde\xi$ show that they have comparatively small roles as well.)





\section{Conclusion and Future directions}

This work is already beginning to shed light on the sensitivity of
CKM fits to assumptions regarding the parameterization of theoretical
uncertainties. We believe that in the future it will become a
valuable tool for understanding the results of such fits and for 
assessing whether, in fact, inconsistencies with the standard model
may have emerged as the precision and breadth of measurements are improved.

By clearly separating the probablistic uncertainties associated with measured 
quantitities from the
{\it a priori} unknown distributions of theoretical quantities, we have
attempted to expose the effects of assumptions about the distribution
of theoretical uncertainties on the results of fits that purport to extract 
a statistic representing the degree of ``self-consistency'' of the CKM sector
of the standard model.
In particular, the clear correlation found between ``best-fit'' values for 
$\bar{\rho}$ and $\bar{\eta}$ 
and the theoretical inputs for $B_K$ and the computations of $|V_{cb}|$,
which are quite model dependent, 
should prompt a cautious approach in future attempts to ascertain the 
degree of consistency of unitarity triangle-based investigations of the 
CKM matrix.

The existence of such correlations demonstrates the dangers of constructing 
contours covering a range of values of theoretically uncertain parameters 
that may then be interpreted as contours of equal probability.

In the next decade, as lattice calculations of many of the theoretical quantities that enter
into CKM unitarity triangle studies improve and gain control of systematic uncertainties, it
will be possible to assign a probability distibution function to theoretical uncertainties
in a principled manner.  Until we reach this {\it  nirvana}, however, an explicit display
of the effect of the (unknown) distribution of theoretical errors on conclusions about the
validity of the standard model will remain important.

\end{document}